\documentclass[11pt, a4paper]{article}
\pdfoutput=1
\usepackage{jheppub}

\usepackage{comment}

\usepackage{enumitem}

\usepackage{centernot} % To do simply slashed symbols properly (with a centered slash)
\usepackage{slashed} % Alternate slash package

\usepackage[english]{babel}
\usepackage[utf8]{inputenc}  %% Needed for UNICODE text input
\usepackage[T1]{fontenc}
\usepackage[normalem]{ulem}

\usepackage{subfig}

\usepackage{amsfonts, amssymb, amsthm} %% mathtools will load amsmath
\DeclareFontFamily{U}{mathx}{\hyphenchar\font45}
\DeclareFontShape{U}{mathx}{m}{n}{
      <5> <6> <7> <8> <9> <10>
      <10.95> <12> <14.4> <17.28> <20.74> <24.88>
      mathx10
      }{}
\DeclareSymbolFont{mathx}{U}{mathx}{m}{n}
\DeclareFontSubstitution{U}{mathx}{m}{n}
\DeclareMathAccent{\widecheck}{0}{mathx}{"71}
\DeclareMathAccent{\wideparen}{0}{mathx}{"75}

\usepackage[]{mathtools}   % tbtags - Put equation labels either on top or bottom line
\numberwithin{equation}{section} % Number equations per-section
\allowdisplaybreaks              % Allows page breaks in multi-equations

\usepackage{stmaryrd}
\usepackage{bbold}     % To be able to use \mathbb{1} (in addition to \mathbb{A} for instance)
\usepackage{ifthen}
\newcommand{\dInt}[2][]{%
    \ifthenelse{\equal{#1}{}}
    {\ensuremath{\operatorname{d}{#2}\;}}
    {\ensuremath{\operatorname{d}^{#1}{#2}\;}}
}

\renewcommand{\imath}{i} % {\operatorname{i}}
\newcommand{\hc}{\text{h.c.}}

\newcommand{\Proj}[1]{{\mathbb{P}_\text{#1}}}

\DeclareMathOperator{\Tr}{Tr}
\newcommand{\sgn}[1]{\mathop{\rm sgn}\nolimits\left(#1\right)}

\makeatletter
\renewcommand*\env@matrix[1][*\c@MaxMatrixCols c]{%
	\hskip -\arraycolsep
	\let\@ifnextchar\new@ifnextchar
	\array{#1}}
\makeatother

\usepackage[capitalise]{cleveref}

\usepackage{tabularx}

\title{Dimensional Regularization and Breitenlohner-Maison/%\newline
't~Hooft-Veltman Scheme for $\gamma_5$ applied to Chiral YM Theories:\newline
Full One-Loop Counterterm and RGE Structure}

\author[a,1]{Hermès Bélusca-Maïto\note{Corresponding author.},} %% \thanks{\email{hbelusca@phy.hr}}}
\author[a]{Amon Ilakovac,} %% \thanks{\email{ailakov@phy.hr}}}
\author[a]{Marija Mađor-Božinović,} %% \thanks{\email{mmadjor@phy.hr}}}
\author[b]{and Dominik Stöckinger} %% \thanks{\email{Dominik.Stoeckinger@tu-dresden.de}}}

\affiliation[a]{Department of Physics, University of Zagreb, Bijeni\v{c}ka cesta 32, HR-10000 Zagreb, Croatia}
\affiliation[b]{Institut für Kern- und Teilchenphysik, TU Dresden, DE-01069 Dresden, Germany}

\emailAdd{hbelusca@phy.hr}
\emailAdd{ailakov@phy.hr}
\emailAdd{mmadjor@phy.hr}
\emailAdd{Dominik.Stoeckinger@tu-dresden.de}

\arxivnumber{2004.14398}
\abstract{
	We study the application of the Breitenlohner--Maison--'t~Hooft--Veltman (BMHV) scheme of Dimensional Regularization to the renormalization of chiral gauge theories, focusing on the specific counterterm structure required by the non-anticommuting Dirac $\gamma_5$ matrix and the breaking of the BRST invariance.
	Calculations are performed at the one-loop level in a massless chiral Yang-Mills theory with chiral fermions and real scalar fields.
	We discuss the setup and properties of the regularized theory in detail.
	Our central results are the full counterterm structures needed for the correct renormalization:
	the singular UV-divergent counterterms, including evanescent counterterms that have to be kept for consistency of higher-loop calculations.
	
	We find that the required singular, evanescent counterterms associated with vector and scalar fields are uniquely determined but are not gauge invariant.
	Furthermore, using the framework of algebraic renormalization, we determine the symmetry-restoring finite counterterms, that are required to restore the BRST invariance, central to the consistency of the theory.
	These are the necessary building blocks in one-loop and higher-order calculations.

	Finally, renormalization group equations are derived within this framework, and the derivation is compared with the more customary calculation in the context of symmetry-invariant regularizations.
	We explain why, at one-loop level, the extra BMHV-specific counterterms do not change the results for the RGE.
	The results we find complete those that have been obtained previously in the literature
	in the absence of scalar fields.
}

\begin{document}

\begin{flushright}
ZTF-EP-20-01 \\
\today
\end{flushright}

% {\let\newpage\relax\maketitle}
\maketitle
\flushbottom

% \begin{abstract}
% %\vspace*{2\parskip}
% {
% %\setlength{\parskip}{0pt}
% %% The abstract...
% }
% \end{abstract}

% \newpage

% % Table of contents
% \tableofcontents

\section{Introduction}
\label{sect:Intro}

The existence of chiral fermions is a fundamental fact of nature. In
quantum field theory, chiral fermions lead to the phenomenon of chiral
anomalies \cite{Adler:1969gk,Bell:1969ts} manifested e.g.\ in pion decays or
baryon number non-conservation in the Standard Model (SM). Gauge theories with
chiral fermions are only well-defined if chiral gauge anomalies are absent,
which is equivalent to the one-loop anomaly cancellation conditions
thanks to the Adler-Bardeen theorem \cite{Adler:1969er}. Technically
chiral anomalies are related to the impossibility to find a
regularization scheme preserving the chiral symmetry in question.
In practical calculations, Dimensional Regularization (DReg)
\cite{Cicuta:1972jf,Bollini:1972ui,Ashmore:1972uj,tHooft:1972tcz}
is by far the most common scheme. For a recent review of versions of
DReg and alternatives see \cite{Gnendiger:2017pys}.
Here the existence of chiral anomalies leads
to the $\gamma_5$-problem, i.e.\ the problem that $\gamma_5$ (and the
Levi-Civita symbol $\epsilon_{\mu\nu\rho\sigma}$) are tied to strictly
4 dimensions. For an extensive overview of the $\gamma_5$-problem and
references  we refer the reader to Ref.\ \cite{Jegerlehner:2000dz}.

We point out that a large set of treatments of $\gamma_5$ in DReg
has been proposed which retain the anticommutativity of $\gamma_5$ in
$d\ne4$ dimensions; these treatments are typically either defined only
for subclasses of diagrams \cite{Chanowitz:1979zu,Jegerlehner:2000dz} or
give up other properties such as cyclicity of the trace
\cite{Kreimer:1989ke,Korner:1991sx,Kreimer:1993bh}. An interesting
recent proposal was made in Ref.\ \cite{Zerf:2019ynn}, but this
proposal is so far limited to fermion traces. In practical
calculations, the anticommutative definition of $\gamma_5$ is
advantageous; however, these anticommuting schemes have not reached the
same level of mathematical rigor as the original scheme by `t Hooft
and Veltman \cite{tHooft:1972tcz} (see also the work by Akyeampong and Delbourgo,
\cite{Akyeampong:1973xi,Akyeampong:1973vk,Akyeampong:1973vj}),
for which perturbative all-order consistency with fundamental field theoretical properties has been
established by Breitenlohner and Maison
\cite{Breitenlohner:1975qe,Breitenlohner:1977hr,Breitenlohner:1975hg,Breitenlohner:1976te}.
An example of the issues which can arise at higher orders
is provided by Refs.\ \cite{Bednyakov:2015ooa,Zoller:2015tha}, which computed the four-loop
$\beta$-function for $\alpha_s$ using various prescriptions involving anticommuting
$\gamma_5$ and the reading-point prescription of
Ref.\ \cite{Korner:1991sx}, with conflicting
results. The scheme ambiguity could be resolved in
Ref.\ \cite{Poole:2019txl} only by using information external to the
regularization schemes.

In the present paper, we focus on the \emph{``Breitenlohner--Maison--'t~Hooft--Veltman''}
(BMHV) scheme. In this scheme $\gamma_5$ is
non-anticommuting in $d$ dimensions, but the scheme is rigorously
established at all orders. Gauge invariance is broken in intermediate
steps but can be restored order by order by adding suitable
counterterms.
For this reason, the usual procedure of generating
counterterms by a renormalization transformation is not
sufficient. There are in fact three additional types of counterterms:
{\it (i)} UV divergent
counterterms cancelling ``evanescent'' divergences, {\it (ii)} the finite
symmetry-restoring counterterms which restore gauge/BRST invariance,
and {\it (iii)} finite evanescent counterterms, which can optionally be
added. We remark that the existence of symmetry-restoring counterterms
follows in complete generality from the renormalizability of the
theory, which can be established e.g.\ using purely algebraic methods
\cite{Becchi:1975nq,Tyutin:1975qk,Piguet:1980nr,Piguet:1995er} (for a more
recent overview of these methods, see also \cite{Binosi:2009qm}).
Symmetry-restoring counterterms for the BMHV scheme have been considered
in the literature already for gauge theories without scalar fields \cite{Martin:1999cc},
for abelian gauge theories \cite{SanchezRuiz:2002xc},
in the evaluation of flavor-changing neutral processes at one-loop \cite{Ferrari:1994ct},
for supersymmetric QED \cite{Hollik:1999xh}, and different practical strategies for their determination have
been developed e.g.\ in Refs.\ \cite{Martin:1999cc,Grassi:1999tp,Grassi:2001zz,Fischer:2003cb}.

Our first goal is to take the BMHV scheme seriously, apply it
to general chiral gauge theories without compromises and work out its
properties in detail. In the present paper, we focus on the one-loop
level of a general gauge theory with purely right-chiral fermions and
evaluate the full counterterm structure; in
a companion paper we will present the generalization to the full
electroweak Standard Model. We expose the technical details of the
BMHV scheme and the determination of the counterterms in a
way that is close to practical calculations, with the aim that the
present paper bridges the gap between purely algebraic
approaches and  phenomenological applications.
Our study is motivated by the increasing
need for high-precision (multi-loop) electroweak calculations,
discussed e.g.\ in Ref.\ \cite{FCCeereport:1809.01830}. Our main goal
is therefore to present detailed
discussions and one-loop results which will be vital
ingredients in forthcoming, future analyses of the BMHV scheme for
multi-loop calculations in chiral gauge theories.

Before presenting the outline of this paper we mention two further
recent works on $\gamma_5$. Ref.\ \cite{Bruque:2018bmy} has considered strictly
4-dimensional schemes as alternatives to dimensional regularization,
in the hope that these schemes might offer practical advantages with respect to the
treatment of $\gamma_5$. However, this
reference showed clearly that even 4-dimensional schemes have very similar
problems for $\gamma_5$ as dimensional schemes, as long as they are
compatible with gauge invariance. Ref.\ \cite{Gnendiger:2017rfh}
considers $\gamma_5$ in various versions of dimensional schemes,
including the so-called four-dimensional formulation (FDF) of DReg
\cite{Fazio:2014xea}; this reference showed in particular that effectively FDF may
be viewed as a particularly efficient implementation of the BMHV
scheme at the one-loop level, at least for the four-dimensional
helicity version of DReg \cite{Gnendiger:2017rfh}. This is promising
in view of future practical applications of the BMHV scheme.

In the past the BMHV scheme was applied in a range of calculations and
practical procedures have been developed, see
e.g.\ \cite{Buras:1994dj,Larin:1993tq,Trueman:1995ca}; still it was
often considered
as rather impractical and less preferable than its alternatives, see
e.g.\ Refs.\ \cite{Chetyrkin:1997gb,Schubert:1993wg}. But
given the result of Ref.\ \cite{Gnendiger:2017rfh}, the general
computer-algebraic progress, and the ambiguities present in other
schemes, we believe a new thorough study of the BMHV scheme is timely
and promising.

The structure of our paper is as follows.
In \cref{sect:DimReg} we begin by collecting the relevant properties
of DReg in the BMHV scheme. In \cref{sect:RModel} we define the
chiral gauge theory we consider; we provide formulations
using Weyl spinors and using Dirac spinors; the latter is the one we
promote to $d$ dimensions. We exhibit in detail the symmetry
properties with respect to gauge invariance, BRST invariance, and
the functional form of the Slavnov-Taylor identity and its breaking in
$d$ dimensions. \cref{sect:standardrenormalizationstructure} begins
the study of renormalization in the BMHV scheme. It first collects
known results from the standard case where gauge invariance is
preserved by the regularization; then it describes the differences appearing
in the BMHV scheme.

The central new results of the present paper are presented in
\cref{sect:Rmodel1LoopSCT} and \cref{sect:BRSTrestoration}.
The UV divergent, singular counterterms are computed and discussed in
\cref{sect:Rmodel1LoopSCT}.  The symmetry-restoring counterterms are
determined in  \cref{sect:BRSTrestoration}. After describing and
assessing several possible strategies for their
determination we proceed similarly to
Ref.\ \cite{Martin:1999cc}, highlighting the logic of the overall
procedure as well as pointing out the role of technical
simplifications based on the Bonneau identities
\cite{Bonneau:1980zp,Bonneau:1979jx}.

In \cref{sect:RGE} and \cref{sect:MultRenorm} we evaluate the
one-loop RGEs and show that the obtained results are the standard, known ones.
We focus on explaining how these results are obtained in spite of the
necessity of non-standard divergent and finite counterterms. These two
sections thus provide a check of the procedure and prepare future
multi-loop applications. Both sections use different methods to derive
the $\beta$ functions, and each case leads to valuable insights on
expected issues in two-loop BMHV calculations.

Finally, we expose in \cref{sect:LModel} the changes in our main results that would appear if one wishes to use a left-handed model instead of a right-handed one. We summarize and conclude in the last section.

\section{Generalities on Dimensional Regularization}
\label{sect:DimReg}

The Dimensional Regularization (DReg) scheme allows regularizing the divergences arising from loop calculations in 4 dimensions, while explicitly preserving Lorentz covariance and in principle gauge invariance.
Schematically the procedure consists in extending the Lorentz-covariant objects --~scalar/vector and spinor fields, momenta, derivatives, and spinor matrices~-- appearing in the theory from their definition in 4 dimensions into an extended definition in a formal ``$d$''-dimensional space. Note that for supersymmetric theories this procedure breaks supersymmetry, and therefore an alternative regularization may be used instead \cite{Siegel:1979wq,Siegel:1980qs,Capper:1979ns,Stockinger:2005gx}, unless explicit supersymmetry-restoring counterterms are introduced (see e.g.\ \cite{Martin:1993yx,Mihaila:2009bn,Fischer:2003cb,Stockinger:2011gp}).
If such an extension is in principle easily implemented, problems do appear when attempting to extend the definition of genuinely intrinsically 4-dimensional objects, namely the $\gamma_5$ Dirac matrix and the Levi-Civita symbol $\epsilon_{\mu\nu\rho\sigma}$.
These two objects appear in chiral theories (of which the Standard Model is one example). Such theories usually exhibit gauge anomalies (the Adler-Bell-Jackiw anomaly) that are generated by the presence of these objects, as well as by their actual fermion content.

In this scheme, the formal $d$-dimensional space can be separated into 4-dimensional and $d-4 \equiv -2\epsilon$-dimensional subspaces as direct sums. Lorentz covariants extended into this $d$-dimensional space now possess 4-dimensional (denoted by bars: $\overline{\cdot}$\;) and $(-2\epsilon)$-dimensional (also called ``evanescent'', denoted by hats: $\widehat{\cdot}$\;) components.
Metric tensors on these subspaces are defined as
\begin{align}
    \text{$d$-dim.}:&\, g_{\mu\nu} \, , \;&
    \text{$4$-dim.}:&\, \bar{g}_{\mu\nu} \, , \;&
    \text{$(-2\epsilon)$-dim.}&:\, \hat{g}_{\mu\nu} = g_{\mu\nu} - \bar{g}_{\mu\nu} \, .
\end{align}
The existence of these objects and their inverse (with upper indices) has been shown by
explicit construction in Ref.\ \cite{Collins:1984xc}; they are
defined such that
\begin{align}
    g_{\mu\nu} g^{\nu\mu} &= d \; , \;\;&
    \bar{g}_{\mu\nu} \bar{g}^{\nu\mu} &= 4 \; , \;\;&
    \hat{g}_{\mu\nu} \hat{g}^{\nu\mu} &= d-4 \equiv -2\epsilon \; ,
\end{align}
{and}
\begin{align}
  g_{\mu\nu} g^{\nu\rho} &= g_{\mu\;}^{\;\rho} \equiv \delta_{\mu\;}^{\;\rho} \; , \;\;&
    \bar{g}_{\mu\nu} \bar{g}^{\nu\rho} &= \bar{g}_{\mu\;}^{\;\rho} = \bar{g}_{\mu\nu} g^{\nu\rho} = g_{\mu\nu} \bar{g}^{\nu\rho} \; , \;\;\\
    \hat{g}_{\mu\nu} \hat{g}^{\nu\rho} &= \hat{g}_{\mu\;}^{\;\rho} = \hat{g}_{\mu\nu} g^{\nu\rho} = g_{\mu\nu} \hat{g}^{\nu\rho} \; , &
    \bar{g}_{\mu\nu} \hat{g}^{\nu\rho} &= 0 = \hat{g}_{\mu\nu} \bar{g}^{\nu\rho} \; ,
\end{align}
expressing the fact that the quasi-$d$-dimensional space is a direct
sum of the actual $4$-dimensional space and a
quasi-$(-2\epsilon)$-dimensional space.
Our convention for the 4-dimensional metric signature is mostly minus,
i.e. $(+1, -1, -1, -1)$.
When being extended to the $d$-dimensional formalism, Lorentz indices become formal symbols that cannot take any particular value. They just obey Einstein summation convention for repeated indices, while lowering and raising indices is done using the metric tensors. We note that the metric tensors act similarly as projectors onto these different subspaces.
As an illustration for 4-vectors, the following behaviour is exhibited:
\begin{equation}\begin{gathered}
    k^\mu = g^{\mu\nu} k_\nu \; , \;\;\qquad k_\mu = g_{\mu\nu} k^\nu \; , \;\;\qquad
    \bar{k}_\mu = \bar{g}_{\mu\nu} k^\nu \; , \;\; \hat{k}_\mu = \hat{g}_{\mu\nu} k^\nu \; , \;\;\qquad
    k^2 = \bar{k}^2 + \hat{k}^2 \; , \\
    k^2 = k^\mu k_\mu = g^{\mu\nu} k_\nu k_\mu = g_{\mu\nu} k^\nu k^\mu \; , \;\;\qquad
    \bar{k}^2 = \bar{k}^\mu \bar{k}_\mu = \bar{g}^{\mu\nu} k_\nu k_\mu = \bar{g}_{\mu\nu} k^\nu k^\mu \; , \\
    \hat{k}^2 = \hat{k}^\mu \hat{k}_\mu = \hat{g}^{\mu\nu} k_\nu k_\mu = \hat{g}_{\mu\nu} k^\nu k^\mu \; , \;\;\qquad
    \bar{g}_{\mu\nu} \hat{k}^\mu = 0 \; , \;\;\qquad \hat{g}_{\mu\nu} \bar{k}^\mu = 0 \; ,
\end{gathered}\end{equation}
with similar extensions due to the fact that the different metrics, and as extension, the different contracted indices, project onto their associated subspaces.

For the usual $\gamma^\mu$ matrices extended to $d$-dimensional space, one can similarly define their 4-dimensional and $(-2\epsilon)$-dimensional versions $\bar{\gamma}^\mu$ and $\hat{\gamma}^\mu$ respectively, including the anticommutation relations between matrices of same space-time dimensionality, the anticommutation relations between matrices of different space-time dimensionalities, their contractions and their traces:
\begin{subequations}
\label{eq:GammaDreg}
\begin{align}
   \{\gamma^\mu,\gamma^\nu\} &= 2g^{\mu\nu}\mathbb{1}
   \, ,& \,
   \{\gamma^\mu,\bar{\gamma}^\nu\} &= \{\bar{\gamma}^\mu,\bar{\gamma}^\nu\} = 2\bar{g}^{\mu\nu}\mathbb{1}
   \, ,& \,
   \gamma_\mu \gamma^\mu &= d\, \mathbb{1}
   \, ,\\
   \{\bar{\gamma}^\mu,\hat{\gamma}^\nu\} &= 0
   \, ,& \,
   \{\gamma^\mu,\hat{\gamma}^\nu\} &= \{\hat{\gamma}^\mu,\hat{\gamma}^\nu\} = 2\hat{g}^{\mu\nu}\mathbb{1}
   \, ,& \,
   \gamma_\mu \bar{\gamma}^\mu &= \bar{\gamma}_\mu \bar{\gamma}^\mu = 4\, \mathbb{1}
   \, , \\
   &&
   \gamma_\mu \hat{\gamma}^\mu &= \hat{\gamma}_\mu \hat{\gamma}^\mu = (d-4) \mathbb{1}
   \, ,& \,
   \bar{\gamma}_\mu \hat{\gamma}^\mu &= 0 \, , \\
   \Tr\gamma^\mu &= 0 \, ,& \, \Tr\bar{\gamma}^\mu &= 0 \, ,& \, \Tr\hat{\gamma}^\mu &= 0 \, .
\end{align}
\end{subequations}

The real problem, of course, is how to define in DReg the Levi-Civita symbol $\epsilon$ and the $\gamma_5$ matrix, which are \emph{intrinsically $4$-dimensional quantities}. In this work we adopt the \emph{``Breitenlohner--Maison--'t~Hooft--Veltman''} (BMHV) scheme for treating $\gamma_5$ and $\epsilon_{\mu\nu\rho\sigma}$, whose consistency in perturbative renormalization has been proved by Breitenlohner and Maison
\cite{Breitenlohner:1975qe,Breitenlohner:1977hr,Breitenlohner:1975hg,Breitenlohner:1976te},
and that is able to reproduce the ABJ anomaly \cite{Akyeampong:1973xi,Akyeampong:1973vk,Akyeampong:1973vj,Marinucci:1975hx,Frampton:1978ix,Bonneau:1980yb}.
The $\epsilon$ symbol is defined by its product with the metric tensor, and the product of two $\epsilon$ symbols together,
\begin{align}
\label{eq:eps_def}
	g_{\mu\;}^{\;\mu_1} \epsilon_{\mu_1\mu_2\mu_3\mu_4} &= \epsilon_{\mu\mu_2\mu_3\mu_4} \, , \\
	\epsilon_{\mu_1\mu_2\mu_3\mu_4} \epsilon_{\nu_1\nu_2\nu_3\nu_4} &= - \sum_{\pi \in S_4} \sgn{\pi} \prod_{i=1}^4 \bar{g}_{\mu_i\nu_{\pi(i)}} \, ,
\end{align}
from which its other properties can be obtained,
\begin{equation}\begin{gathered}
	\epsilon_{\mu_1\mu_2\mu_3\mu_4} = \sgn{\pi} \epsilon_{\mu_{\pi(1)}\mu_{\pi(2)}\mu_{\pi(3)}\mu_{\pi(4)}} \, , \\
	\sum_{\pi \in S_5} \sgn{\pi} \epsilon_{\mu_{\pi(1)}\mu_{\pi(2)}\mu_{\pi(3)}\mu_{\pi(4)}} \bar{g}^{\mu_{\pi(5)}\nu} = 0 \, .
\end{gathered}\end{equation}
Here, $\pi$ is a permutation belonging to the permutation group of $n$ elements $S_n$ indicated in the corresponding expression.
In the rest of this paper we use the $\epsilon^{0123} = +1$ convention.
On the other side, the $\gamma_5$ matrix is defined to be anticommuting with Dirac matrices in the 4-dimensional subspace, and commuting in the $(-2\epsilon)$-dimensional subspace:
\begin{align}
\label{eq:Gamma5DReg}
    \{\gamma_5, \bar{\gamma}^\mu\} &= 0 \, , \;&
    [\gamma_5, \hat{\gamma}^\mu] &= 0 \, , \;&
    \{\gamma_5, \gamma^\mu\} &= \{\gamma_5, \hat{\gamma}^\mu\} = 2 \gamma_5 \hat{\gamma}^\mu \, , \;&
            [\gamma_5, \gamma^\mu] &= [\gamma_5, \bar{\gamma}^\mu] = 2 \gamma_5 \bar{\gamma}^\mu \, .
\end{align}
$\gamma_5$ otherwise keeps its usual 4-dimensional behaviour. The last of the equations \eqref{eq:Gamma5DReg} follows from the explicit definition of $\gamma_5$, and its square,
\begin{align}
\label{eq:Gamma5BMdef}
    \gamma_5 = \frac{-i}{4!} \epsilon_{\mu\nu\rho\sigma} \gamma^\mu \gamma^\nu \gamma^\rho \gamma^\sigma \, , &&
	\gamma_5^2 = \mathbb{1} \, ,
\end{align}
leading to the trace important to realize the Adler-Bell-Jackiw (ABJ) anomaly
\begin{equation}
\label{eq:ABJ}
	\Tr(\{\gamma^\alpha,\gamma_5\} \gamma_\alpha \gamma_\mu \gamma_\nu \gamma_\rho \gamma_\sigma) = 8\imath (d-4) \epsilon_{\mu\nu\rho\sigma} \, .
\end{equation}

\subsubsection*{Amplitudes in $d$ dimensions and the 4-dimensional limit}
\label{subsect:AmpsDDim}

Once an amplitude has been defined, its evaluation in $d$ dimensions is performed using standard techniques for loop calculations. Its actual Laurent expansion in $4-d = 2\epsilon$ is determined only after having completely reduced and simplified its Lorentz structures: fully evaluating Dirac $\gamma$ traces (cyclicity of the trace is valid in this scheme), fully contracting any vector, tensor and Levi-Civita symbol using the properties defined above. Any $\gamma_5$ matrix and pair of $\epsilon$ symbols can be further removed by using \cref{eq:Gamma5BMdef,eq:eps_def}. This defines a unique ``normal form'' \cite{Breitenlohner:1975qe} for the amplitude.

This allows one to define the regularized version of the amplitude via its Laurent expansion in $4-d = 2\epsilon$. From there one can define its divergent part and the associated counterterms, as well as its finite part and its evanescent part that may be neglected in the $d \to 4$ limit. The renormalized value of an amplitude is obtained after performing all the necessary subtractions of the divergences of its sub-diagrams, and the resulting finite expression is interpreted in the physical 4-dimensional space by setting all quantities to their 4-dimensional values, i.e. first taking the $d \to 4$ limit and then, setting all remaining evanescent objects to zero. This operation will be denoted by $\mathop{\text{LIM}}_{d \to 4}$ in the rest of this paper.
%%% LaTeX technicality: see also https://tex.stackexchange.com/questions/23432/how-to-create-my-own-math-operator-with-limits for the custom math-operators with limits.

\subsubsection*{Charge conjugation in $d$ dimensions for Dimensional Regularization}
\label{subsect:ChargeConjDd}

Phenomenological models may contain, for example in their Yukawa sector, fermions as well as their corresponding charge-conjugated partners. This is precisely the case in our model under study introduced in \cref{sect:RModel}. Thus the question concerning the definition of the charge-conjugation operation in the framework of dimensional regularization arises.

In usual integer dimensions the charge-conjugation operation $\widehat{\mathcal{C}}$ can always be defined, and a corresponding matrix representation $C$ explicitly constructed. For example, in 4 dimensions such a matrix, with antihermitean property, can be constructed as to be numerically equal to $C = \imath \gamma^0 \gamma^2$, and satisfies the relations:
\begin{align}
\label{eq:ChargeConj4D}
	C^{-1} \gamma^\mu C &= -{\gamma^\mu}^T \; ,& \; C^{-1} &= C^\dagger = C^T \; ,& \; C^T &= -C \; ,& \; \text{and:} \;\; C^{-1} \gamma_5 C &= \gamma_5^T \, .
\end{align}
% In even dimensions one can construct another matrix representation that provides $C^{-1} \gamma^\mu C = +{\gamma^\mu}^T$ instead, while in odd dimensions either one or the other representation can exist at the same time: for example in $d=5$ we can only construct instead $C^{-1} \gamma^\mu C = +{\gamma^\mu}^T$ since now $\gamma_5$ is part of the corresponding Clifford algebra. Note also that the sign in $C^T = -C$ does change depending on the dimensionality of the space-time considered.

% In the continuous dimensionality of the dimensional regularization such an explicit construction of a charge-conjugation matrix is not possible anymore and one needs to rely only on a definition based on its properties on the set of Dirac matrices and on its action on the $d$-dimensional spinors.

One can wonder whether in the continuous dimensionality of the dimensional regularization such a construction is still possible.
As it turns out, an explicit construction via a matrix representation has been provided in Appendix~A of \cite{Stockinger:2005gx}, based on the construction of Dirac $\gamma$ matrices in $d$ dimensions given by Collins in \cite{Collins:1984xc}.
Alternatively, one can define the charge-conjugation operation based only on its properties on the set of Dirac matrices and on its action on the $d$-dimensional spinors. For this purpose, since we work in dimension $d = 4 - 2\epsilon$ around 4, we postulate that the relations given in \cref{eq:ChargeConj4D} also hold in $d \approx 4$ (see Appendix~A of \cite{Hieda:2017sqq} for a motivation%
\footnote{
	 As an alternative definition, Appendix~A of \cite{Tsai:2009it} instead postulates a different action of the charge-conjugation operation, on a product of Dirac matrices, as being equal to minus the product of the same Dirac matrices taken in the opposite order, and not transposed. This latter definition is still satisfactory since ultimately, in most of the resulting amplitudes, the internal gamma matrices attached to loops appear inside traces.
}).
Obviously, this would not be true anymore if $d$ was to be pushed to a different integer dimension.

Our final choice for the charge-conjugation matrix in $d \approx 4$ dimension employs the same definitions as in 4 dimensions \cref{eq:ChargeConj4D}, together with the following properties:
\begin{equation}
\label{eq:ChargeConjDd}
	C^{-1} \Gamma C = \eta_\Gamma \Gamma^T \Rightarrow C \Gamma^T C^{-1} = \eta_\Gamma \Gamma \, , \;
	\text{with:} \;
	\eta_\Gamma =
		\begin{cases}
			+1 & \text{for} \; \Gamma = \mathbb{1} \, , \, \gamma_5 \, , \\
			-1 & \text{for} \; \Gamma = \gamma^\mu \, , \, \sigma^{\mu\nu} \, ,
		\end{cases}
\end{equation}
and in the presence of anticommuting fermions (see also Appendix~G.1 of \cite{Dreiner:2008tw}):
\begin{gather}
\label{eq:ChargeConjDdSpinors}
	\widehat{\mathcal{C}} \Psi \widehat{\mathcal{C}}^{-1} \equiv \Psi^C = C \overline{\Psi}^T
	\, , \qquad
	(\Psi^C)^C = \Psi \, , \qquad
	\widehat{\mathcal{C}} \overline{\Psi} \widehat{\mathcal{C}}^{-1} \equiv \overline{\Psi}^C = -\Psi^T C^{-1} = \overline{\Psi^C} \, , \\
	\overline{\Psi}^C_i \Gamma \Psi^C_j = -{\Psi}^T_i C^{-1} \Gamma C \overline{\Psi}^T_j = \overline{\Psi}_j C \Gamma^T C^{-1} {\Psi}_i = \eta_\Gamma \overline{\Psi}_j \Gamma {\Psi}_i \, .
\end{gather}
Note that employing \cref{eq:ChargeConjDd} in $d$ dimensions has an extra subtlety: while it is true that when using these definitions in 4 dimensions, we have: $C^{-1} (\gamma^\mu \gamma_5) C = +(\gamma^\mu \gamma_5)^T$, it is not so in $d$ dimensions in the BMHV scheme due to the $\gamma_5$ matrix:
\begin{equation}
	C^{-1} (\gamma^\mu \gamma_5) C = (C^{-1} \gamma^\mu C)(C^{-1} \gamma_5 C) = -(\gamma^\mu)^T \gamma_5^T = -(\gamma_5 \gamma^\mu)^T = (\overline{\gamma}^\mu \gamma_5)^T - (\widehat{\gamma}^\mu \gamma_5)^T \, ,
\end{equation}
while, of course, we have:
\begin{equation}
	C^{-1} (-\gamma_5 \gamma^\mu) C = \gamma_5^T (\gamma^\mu)^T = (\gamma^\mu \gamma_5)^T \, .
\end{equation}

\section{The Right-Handed (R) Model and its Extension to $d$ Dimensions}
\label{sect:RModel}

Let us begin the investigation of the Dirac $\gamma_5$ matrix in the BMHV scheme in a general,
massless chiral gauge theory.
In the present section we define the model first in 4 dimensions, then extend it to $d$ dimensions and provide the respective Lagrangians, BRST transformations and Slavnov-Taylor identities. The $d$-dimensional extension requires the usage of Dirac fermions instead of Weyl fermions, and requires to make a choice for the evanescent part of the fermion kinetic term and for the fermionic interaction term. We discuss several options and motivate our choice. We then analyze the breaking of BRST invariance, which in our case is caused by a single evanescent term in the tree-level action. The breaking is evaluated on the operator level and translated into Feynman rules.

\subsection{The R-model in 4 dimensions}

Our setup is similar to the one from
Refs.~\cite{Machacek:1983tz,Machacek:1983fi,Machacek:1984zw}.
The model is a gauge theory with matter fields,
based on a simple gauge Lie group\footnote{This gauge group verifies the algebraic properties exposed in \cite{vanRitbergen:1998pn}.} $\mathcal{G}$, with gauge fields $G^a_\mu$ in the adjoint representation of $\mathcal{G}$, and structure constants $f^{abc}$. The latter also define the generators ${T_G}^a_{bc} \equiv \imath f^{acb}$ of the adjoint representation.

%% For simplicity the model incorporates one irreducible representation of \emph{real massless} scalars and one irreducible representation of massless right-handed fermions.
This model incorporates \emph{real massless} scalars $\Phi^m$ and massless right-handed fermion fields described, in the 4-dimensional formulation, using Weyl spinors $\xi^i_\alpha$.
% \item Weyl fermions: for the R-model: $\xi^i_\alpha \equiv \psi_R^C$; for the L-model: $\chi^i_\alpha \equiv \psi_L$.
    % The theory can be re-expressed in terms of ``Dirac'' fermions as follows:
    % for the R-model: $\psi^i = (0, \bar{\xi}^{i\;\dot\alpha})^T$; for the L-model: $\psi^i = (\chi^i_\alpha, 0)^T$. (Note: For the two-component fermions we use the Wess \& Bagger abuse of notation: $\bar{\xi} \equiv \xi^\dagger$.)
They are both charged under the gauge group $\mathcal{G}$ and for simplicity we assume their group representations to be irreducible.
We denote their representations respectively by `S' and `R', and their associated generator matrices by $\theta^a_{mn}$ and $(T_R^a)_{ij}$. In particular the scalar representation is imaginary and antisymmetric, $\theta^a_{mn} = -\theta^a_{nm}$.%
\footnote{
  The model may be generalized to products of (semi\nobreakdash-)simple gauge groups and to reducible representations.
  In this case one needs to consider all the possible mixings for each set of
  irreducible representations that have equal quantum numbers
  (see e.g. \cite{Luo:2002ti,Schienbein:2018fsw}).
}

Before quantization, the 4-dimensional classical Lagrangian of the model can be split into four terms:
\begin{equation}\label{eq:4dimLagrangian}
	\mathcal{L}_\text{gauge} + \mathcal{L}_\text{fermions} + \mathcal{L}_\text{scalars} + \mathcal{L}_\text{Yukawa} \, ,
\end{equation}
where each piece of the Lagrangian reads:
\begin{subequations}
\begin{align}
	\mathcal{L}_\text{gauge} &= \frac{-1}{4} F^a_{\mu\nu} F^{a\,\mu\nu} \, , \\
	% \mathcal{L}_\text{fermions} &= \imath \bar{\psi}_i.\gamma^\mu.D^{ij}_\mu \psi_j
		 % = \imath \bar{\psi}_i.\Proj{L}.\gamma^\mu.\Proj{R}.D^{ij}_\mu \psi_j = \imath \xi \sigma^\mu D_\mu \bar{\xi} \; \text{(R-model), or:} \\
		% &= \imath \bar{\psi}_i.\Proj{R}.\gamma^\mu.\Proj{L}.D^{ij}_\mu \psi_j = \imath \bar{\chi} \bar{\sigma}^\mu D_\mu \chi \; \text{(L-model)} \, ,
	\mathcal{L}_\text{fermions} &= \imath \xi \sigma^\mu D_\mu \bar{\xi} \, , \\
	\mathcal{L}_\text{scalars} &= \frac{1}{2} (D_\mu \Phi^m)^2
		- \frac{\lambda^{mnop}}{4!} \Phi_m \Phi_n \Phi_o \Phi_p \, , \\ % No additional -\mu^2 |\Phi|^2 term for our work!
	% \mathcal{L}_\text{Yukawa} &= -\frac{(Y_R)^m_{ij}}{2} \Phi_m \bar{\psi}^C_i \psi_j + \hc = -\frac{(Y_R)^m_{ij}}{2} \Phi_m \bar{\xi}_i (-\imath \sigma_2) \bar{\xi}_j + \hc \; (\equiv -\frac{(Y_R)^{m\;*}_{ij}}{2} \Phi_m \xi_i (-\imath \bar{\sigma_2}) \xi_j) \\
		% &= -\frac{(Y_R)^m_{ij}}{2} {\psi_R^i}^T (\imath \sigma_2) \psi_R^j \Phi_m + \hc \;\text{(R-model), and similarly for L-model} \, .
	\mathcal{L}_\text{Yukawa} &= -\frac{(Y_R)^m_{ij}}{2} \Phi_m \bar{\xi}_i \bar{\xi}_j + \hc \, , % \; (\equiv -\frac{(Y_R)^{m\;*}_{ij}}{2} \Phi_m \xi_i \xi_j) \, ,
\end{align}
\end{subequations}
where the last equation%
\footnote{
  Note that contrary to Refs.~\cite{Machacek:1983tz,Machacek:1983fi,Machacek:1984zw} the Yukawa term has a normalisation factor $1/2$ since the two 2-component fields are identical -- the corresponding Feynman rule would generate the compensating factor 2. This is in accordance with \cite{Dreiner:2008tw,Martin:2012us}.
}
uses an index-free notation for the Lorentz invariant contraction of two Weyl spinors.

There, the covariant derivative acting on the fermion fields is defined%
\footnote{
  We choose to introduce the coupling constant $g$ in the minimal coupling term of the covariant derivative. The minus sign in front of the coupling term is part of our conventions.
}
by:
\begin{equation}
	D_{ij\;\mu} = \partial_\mu \delta_{ij} - \imath g G^a_\mu {T_R}^a_{ij} \, ,
\end{equation}
and the one for the scalar fields is similar (the ${T_R}^a_{ij}$ generator being replaced by $\theta^a_{mn}$).
% In particular for the adjoint representation, we have: $D^{ab}_\mu = \partial_\mu \delta^{ab} + (g) f^{abc} G^b_\mu$.
From the commutator of the covariant derivatives acting on a given type of field, the field strength tensor for $G$ is defined as:
\begin{equation}
	F^a_{\mu\nu} = \partial_\mu G^a_\nu - \partial_\nu G^a_\mu + g f^{abc} G^b_\mu G^c_\nu \, .
\end{equation}

Note that in $\mathcal{L}_\text{scalars}$ the scalar potential does not contain any quadratic term $\mu^2 |\Phi|^2$, because we are working in the framework of a massless theory; the scalar fields do not acquire a vacuum expectation value and the fields remain perturbatively massless.
% Note that in a purely Right or Left-handed model, Dirac-type Yukawa terms $\bar{\psi} \psi$ are forbidden because they cancel; only Majorana-type Yukawa terms $\bar{\psi}^C \psi$ do survive.
The form of the Yukawa interaction implies that the Yukawa matrix $(Y_R)^m_{ij}$ is symmetric in its fermion-group indices $i,j$.

The Weyl spinor formalism is intrinsically tied to 4-dimensional
space. As a preparation for the $d$-dimensional regularization we
replace the Weyl spinors by projections of Dirac spinors, which can be
generalized to $d$ dimensions. Specifically we promote
the right-handed Weyl fermion $\bar{\xi}$ to
\begin{align}
	\bar{\xi} \to \Proj{R}\psi \equiv \psi_R\, ,
\end{align}
where $\psi$ is a Dirac spinor whose left-handed part is understood to be
\emph{fictitious}, decoupled from the theory.
We employ here the standard right/left chirality operators (projectors) $\Proj{R} = (\mathbb{1}+\gamma_5)/2$ and $\Proj{L}=(\mathbb{1}-\gamma_5)/2$.
The fermionic contents of the theory can be rewritten as (we recall that $\overline{\psi_R} = \overline{\psi}_L \equiv \overline{\psi} \Proj{L}$):
\begin{subequations}
\label{eq:LfermionsPsi}
\begin{align}
\label{eq:LfermionsPsigauge}
	\mathcal{L}_\text{fermions} &= \imath \overline{\psi_R}_i \slashed{D}^{ij} {\psi_R}_j = \imath \overline{\psi_R}_i \slashed{\partial} {\psi_R}_i + g {T_R}^a_{ij} \overline{\psi_R}_i \slashed{G}^a {\psi_R}_j \, , \\
	\mathcal{L}_\text{Yukawa} &= -\frac{(Y_R)^m_{ij}}{2} \Phi_m \overline{\psi_R}^C_i {\psi_R}_j -\frac{(Y_R)^{m\;*}_{ij}}{2} \Phi_m \overline{\psi_R}_i {\psi_R}^C_j \, .
\end{align}
\end{subequations}
% Both of these terms are separately gauge-invariant in 4 dimensions.
We stress again that the left-handed part $\Proj{L}\psi$ entirely decouples
and does not appear at all in this Lagrangian.

\subsubsection*{Gauge-fixing}

The Lagrangian defined so far is gauge invariant. For quantization and
renormalization we promote gauge invariance to BRST invariance and a
Slavnov-Taylor identity \cite{Becchi:1975nq,Tyutin:1975qk}.
The BRST transformations of ordinary fields
are defined as infinitesimal gauge transformations, where the
transformation parameter is replaced by a Faddeev-Popov ghost field
$c^a$ (in the adjoint representation):
\begin{subequations}
\begin{align}
    s{G^a_\mu} &= D^{ab}_\mu c^b = \partial_\mu c^a + g f^{abc} G^b_\mu c^c \, , \\
    s{\psi_i} &= s{{\psi_R}_i} = \imath c^a g {T_R}^a_{ij} {\psi_R}_j \, , \\
    s{\overline{\psi}_i} &= s{\overline{\psi_R}_i} = +\imath \overline{\psi_R}_j c^a g {T_R}^a_{ji} \, , \\
    s{{\psi_L}_i} &= 0 \, , \\
    s{\overline{\psi_L}_i} &= 0 \, , \\
    s{\Phi_m} &= \imath c^a g \theta^a_{mn} \Phi_n \, .
\end{align}
\end{subequations}
Here $s$ is the generator of the BRST transformation, which acts as a
fermionic differential operator.
The BRST transformations of ghost and antighost fields $c^a$ and
$\bar{c}^a$ and the auxiliary Nakanishi-Lautrup \cite{Nakanishi:1966zz,Lautrup:1967zz} field $B^a$ are given by:
\begin{subequations}
\begin{align}
    s{c^a} &= -\frac{1}{2} g f^{abc} c^b c^c \equiv \imath g c^2 \, , \\
    s{\bar{c}^a} &= B^a \, , \\
    s{B^a} &= 0 \, .
\end{align}
\end{subequations}
One can prove that the BRST operator $s$ is nilpotent: $s^2{\phi} = 0$ for any field or linear combination of fields $\phi$.

The Lagrangian of the theory is then extended with the ghost and the gauge-fixing terms, obtained as the BRST transformation of
the expression $\bar{c}^a(\xi B^a/2+ \partial^\mu G^a_\mu)$, resulting
in (up to total derivatives)
\begin{subequations}
\begin{align}
	\mathcal{L}_\text{ghost} &= \partial^\mu \bar{c}_a \cdot D^{ab}_\mu c_b \equiv -\bar{c}_a \partial^\mu D^{ab}_\mu c_b \, , \\
	\mathcal{L}_\text{g-fix} &= \frac{\xi}{2} B^a B_a + B^a \partial^\mu G^a_\mu \, .
\end{align}
\end{subequations}
The gauge-fixing Lagrangian $\mathcal{L}_\text{g-fix}$ is equivalent to the more common form: $\mathcal{L}_\text{g-fix} = \frac{-1}{2\xi} (\partial^\mu G^a_\mu)^2$, obtained after integrating out the auxiliary $B^a$ field.
Finally, it is useful to couple the non-linear BRST transformations to
external sources (or Batalin-Vilkovsky ``anti-fields'',
\cite{Batalin:1977pb,Batalin:1981jr,Batalin:1984jr}) and add corresponding
terms to the Lagrangian (see e.g. \cite{Piguet:1995er} and references therein),
\begin{equation}
	\mathcal{L}_\text{ext} = \rho_a^\mu s{G^a_\mu} + \zeta_a s{c^a} + \bar{R}^i s{{\psi_R}_i} + R^i s{\overline{\psi_R}_i} + \mathcal{Y}^m s{\Phi_m} \, ,
\end{equation}
where the external sources do not transform under BRST transformations: $s{\mathcal{J}} = 0$ for $\mathcal{J} = \rho_a^\mu , \zeta_a , R , \bar{R} , \mathcal{Y}^m$.

The final tree-level action in 4 dimensions, which constitutes the basis for the
quantization and renormalization procedure, is then given by
\begin{equation}
\label{eq:TreeLevel4DAction}
	S_0^{(4D)} = \int \dInt[4]{x} (\mathcal{L}_\text{gauge} + \mathcal{L}_\text{fermions} + \mathcal{L}_\text{scalars} + \mathcal{L}_\text{Yukawa} + \mathcal{L}_\text{ghost} + \mathcal{L}_\text{g-fix} + \mathcal{L}_\text{ext}) \, .
\end{equation}
This tree-level action satisfies the Slavnov-Taylor identity
\begin{equation}
\label{eq:STIS0}
    \mathcal{S}(S_0^{(4D)}) = 0 \, ,
\end{equation}
where the Slavnov-Taylor operation is given for a general functional
${\cal F}$ as
\begin{equation}\begin{split}
\label{eq:SofFDefinition}
    \mathcal{S}({\cal F}) =
    \int \dInt[4]{x} \left(
        \frac{\delta {\cal F}}{\delta \rho_a^\mu} \frac{\delta {\cal F}}{\delta G^a_\mu} +
        \frac{\delta {\cal F}}{\delta \zeta_a} \frac{\delta {\cal F}}{\delta c^a} +
        \frac{\delta {\cal F}}{\delta \mathcal{Y}^m} \frac{\delta {\cal F}}{\delta \Phi_m} +
        \frac{\delta {\cal F}}{\delta \bar{R}^i} \frac{\delta {\cal F}}{\delta \psi_i} +
        \frac{\delta {\cal F}}{\delta R^i} \frac{\delta {\cal F}}{\delta \overline{\psi}_i} +
        B^a \frac{\delta {\cal F}}{\delta \bar{c}_a} \right)\,.
\end{split}\end{equation}
The Slavnov-Taylor identity is the basic, defining symmetry property
of the theory. We will require that the Slavnov-Taylor identity ${\cal
  S}(\Gamma)=0$ is satisfied for the fully renormalized, finite
effective action $\Gamma$ (which incorporates the tree-level action,
loop corrections and counterterm contributions).
On the level of the 4-dimensional tree-level action, the Slavnov-Taylor identity
summarizes three properties: {\it (i)} the gauge invariance of the physical part
of the Lagrangian, {\it (ii)} the BRST invariance of the gauge-fixing and ghost
Lagrangian, and {\it (iii)} the nilpotency of the BRST transformations.

\subsubsection*{Quantum Numbers and Constraints from Gauge-invariance}

We summarize in \cref{tbl:fields_quantum_numbers} the list of quantum numbers (mass dimension, ghost number and (anti)commutativity) of the fields and the external sources (BV ``anti-fields'') of the theory, that are necessary for building the whole set of all possible renormalizable mass-dimension $\leq 4$ field-monomial operators with a given ghost number.

\begin{table}[h]
\renewcommand*{\arraystretch}{1.2} %% Change the default height of the rows
\centering
\begin{tabular}{r|*{10}{c}|cc}
    \hline
    %\cline{2-13}
    \multicolumn{1}{r}{}
              & $G^a_\mu$ & $\bar{\psi}_i$, $\psi_i$ & $\Phi_m$ & $c^a$ & $\bar{c}^a$ & $B^a$ & $\rho_a^\mu$ & $\zeta_a$ & $R^i$, $\bar{R}^i$ & $\mathcal{Y}^m$ & $\partial_\mu$ & $s$ \\
    \hline
    mass dim. &  1        &  3/2                     &  1       &  0    &  2          &  2    &  3           &  4        &  5/2               &  3              &  1             &  0  \\
    ghost \#  &  0        &  0                       &  0       &  1    & -1          &  0    & -1           & -2        & -1                 & -1              &  0             &  1  \\
    comm.     & +1        & -1                       & +1       & -1    & -1          & +1    & -1           & +1        & +1                 & -1              & +1             & -1  \\
    \hline
\end{tabular}
\caption{List of fields, external sources and operators, and their quantum numbers.}
\label{tbl:fields_quantum_numbers}
\end{table}

Concerning the gauge transformations under the group $\mathcal{G}$, the mentioned gauge invariance of the terms in \cref{eq:4dimLagrangian} implies two consequences\footnote{They can be proved alternatively by imposing their BRST invariance.} for the fermionic and scalar sectors:
\begin{itemize}[leftmargin=*]
    \item imposing gauge-invariance of the Yukawa interaction implies that the Yukawa matrices satisfy the constraint:
        \begin{subequations}
        \begin{equation}
        \label{eq:GaugeInvarYuk}
            (Y_R)^n_{ij} \theta^a_{nm} + (Y_R)^m_{ik} {T_R}^a_{kj} - {T_{\overline{R}}}^a_{ik} (Y_R)^m_{kj} = 0 \, ,
        \end{equation}
        which is a more explicit version of Eq.~(A.15) from \cite{Machacek:1983tz}. % 1st article from Machacek \& Vaughn. Specialized to the right fermions only, by identification $T_L^a = T_{\overline{R}}^a$.
        The generators ${T_R}^a$ verify ${T_R}^{a\;\dagger} = {T_R}^a$, and from them the conjugate representation $\overline{R}$ is defined with generators ${T_{\overline{R}}}^a \equiv -{T_R}^{a\;T} = -{T_R}^{a\;*}$.
        The complex-conjugate counterpart of this equation is
        \begin{equation}
            (Y_R)^{n\;*}_{ij} \theta^a_{nm} + (Y_R)^{m\;*}_{ik} {T_{\overline{R}}}^a_{kj} - {T_R}^a_{ik} (Y_R)^{m\;*}_{kj} = 0 \, ;
        \end{equation}
        \end{subequations}
    \item imposing gauge-invariance of the scalar self-coupling interaction implies that the scalar quartic coupling matrix $\lambda$ satisfies the constraint:
        \begin{equation}
        \label{eq:GaugeInvarLamb}
            \theta^a_{mq} \lambda^{qnop} + \theta^a_{nq} \lambda^{mqop} + \theta^a_{oq} \lambda^{mnqp} + \theta^a_{pq} \lambda^{mnoq} = 0 \, ,
        \end{equation}
        which agrees with Eq.~(2.7) of \cite{Machacek:1984zw}. % 3rd article from Machacek \& Vaughn.
\end{itemize}
In case the gauge group representations of the quantum fields are reducible and contain two different, but group theoretically identical irreducible representations, the mixings between group theoretically
identical irreducible representations might appear through Yukawa couplings, see \cite{Luo:2002ti,Schienbein:2018fsw}. For that reason, in the following, we consider only irreducible gauge boson, fermion and scalar group representations, if not stated otherwise.

\subsubsection*{Group invariants}

In this section, we summarize the different group invariants that are employed in all of our calculations.
Recall that the right-handed fermions are in an irreducible representation $R$ of the gauge group $\mathcal{G}$ with corresponding hermitian group generators ${T_R}^a$, and the real scalar fields are in an irreducible representation $S$ of $\mathcal{G}$ with imaginary generators $\theta^a$. The adjoint representation of the gauge group $\mathcal{G}$ is denoted by $G$ and its Casimir index is $C_2(G)$.

We define the Casimir and Dynkin indices for these representations, as well as some invariants built out of the Yukawa matrices:
\begin{align}
	C_2(R) \mathbb{1} &= T_R^a T_R^a \, ,& \quad
    C_2(S) \mathbb{1} &= \theta^a \theta^a \, , \\
    S_2(R)\delta^{ab} &= \Tr(T_R^a T_R^b) \, ,& \quad
	S_2(S)\delta^{ab} &= \Tr(\theta^a \theta^b) \, , \\
    Y_2(R)_{ij} &= (Y_R^m Y_R^{m\,\dagger})_{ij} \equiv Y_2(R)\delta_{ij} \, , \hspace{-8cm} \\
    Y_2(S)^{mn} &= \frac{1}{2} \Tr(Y_R^m Y_R^{n\,\dagger} + Y_R^{m\,\dagger} Y_R^n) \equiv Y_2(S)\delta^{mn} \, . \hspace{-8cm}
\end{align}

Due to the presence of charge-conjugated fermions (or, when mapping a left-handed model to its corresponding right-handed model by interpreting left-handed fermions as charge-conjugated right-handed fermions, as presented in \cref{sect:LModel}), we also introduce the corresponding complex-conjugate fermion representation $\overline{R}$ associated with group generators ${T_{\overline{R}}}^a \equiv -{T_R}^{a\;*} = -{T_R}^{a\;T}$, since the generators themselves are hermitian: ${T_R}^{a\;\dagger} = {T_R}^a$. Defining the Yukawa matrices for the conjugate representation as: $Y_{\overline{R}}^m \equiv (Y_R^m)^\dagger = (Y_R^m)^*$ since the Yukawa matrix $(Y_R)^m_{ij}$ is symmetric in its fermion-group indices $i,j$.
We then obtain the group invariants for this $\overline{R}$ representation:
\begin{align}
	C_2(\overline{R}) \mathbb{1} &= {T_{\overline{R}}}^a {T_{\overline{R}}}^a = (-{T_R}^{a\,T})(-{T_R}^{a\,T}) = {T_R}^a {T_R}^a = C_2(R) \mathbb{1} \, , \\
	S_2(\overline{R}) \delta^{ab} &= \Tr({T_{\overline{R}}}^b {T_{\overline{R}}}^a) = \Tr((-{T_R}^{b\,T})(-{T_R}^{a\,T})) = \Tr({T_R}^a {T_R}^b) = S_2(R) \delta^{ab} \, , \\
	Y_2(\overline{R})_{ij} &= ((Y_{\overline{R}})^m (Y_{\overline{R}})^{m\,\dagger})_{ij} = (Y_R^{m\,\dagger} Y_R^m)_{ij} = (Y_R^m Y_R^{m\,\dagger})_{ji} = Y_2(R)_{ji} \equiv Y_2(R)_{ij} \, .
\end{align}
Also, it can be shown, using \cref{eq:GaugeInvarYuk}, that:
\begin{equation}
    \Tr( Y_R^m {T_R}^a Y_R^{n\,\dagger} ) = \Tr( Y_R^{m\,\dagger} {T_{\overline{R}}}^a Y_R^n ) = \frac{Y_2(S)}{2} \theta^a_{mn} \, .
\end{equation}

\subsection{Promoting the R-model to $d$ dimensions}
\label{subsect:RModelDReg}

We now proceed to extend the R-model to $d$ dimensions. While it is straightforward to do so for the bosonic fields, the fermionic fields need some care, even if we start from the version \cref{eq:LfermionsPsi}
of the Lagrangian in terms of Dirac spinors.

The first difficulty is associated with the fermion-gauge interaction term in
\cref{eq:LfermionsPsigauge}, which involves the right-handed
chiral current
$\overline{\psi}_i \gamma^\mu {\psi_R}_j$ in 4 dimensions. The
following are
three \emph{inequivalent choices} for the $d$-dimensional versions of
this term:
\begin{align}
  \overline{\psi}_i \gamma^\mu \Proj{R} \psi_j \, , &&
  \overline{\psi}_i \Proj{L} \gamma^\mu \psi_j \, , &&
  \overline{\psi}_i \Proj{L} \gamma^\mu \Proj{R} \psi_j \, .
\label{eq:inequivalentchoices}
\end{align}
They are different because $\Proj{L} \gamma^\mu \neq \gamma^\mu \Proj{R}$ in $d$ dimensions, see \cref{eq:Gamma5DReg}.
Each of these does lead to valid $d$-dimensional extensions of the model that are perfectly renormalizable using dimensional regularization and the BMHV scheme. However, the intermediate calculations and the final $d$-dimensional results will differ, depending on the choice for this interaction term.

Our choice for the rest of this work is to use the third option, which is equal to
\begin{equation}
	\overline{\psi} \Proj{L} \gamma^\mu \Proj{R} \psi = \overline{\psi} \Proj{L} \overline{\gamma}^\mu \Proj{R} \psi = \overline{\psi_R} \overline{\gamma}^\mu \psi_R \, ,
\end{equation}
is the most symmetric one, and leads to the simplest expressions (see also the discussions in
Refs.~\cite{Martin:1999cc,Jegerlehner:2000dz}). %% == JegerlehnerGamma5
One should note that it is actually the most straightforward choice as it carries the information that right-handed fermions were originally present on the left and on the right sides of the interaction term.

The second, more critical problem, is that as it stands the pure fermionic kinetic term
$\imath \overline{\psi_R}_i \slashed{\partial} {\psi_R}_i = \imath \overline{\psi}_i \Proj{L} \slashed{\partial} \Proj{R} \psi_i$
projects only the purely $4$-dimensional derivative, leading to
a purely 4-dimensional propagator%
\footnote{
	Indeed, the corresponding propagator is $\Delta(p) = \Proj{R} \imath \slashed{p} \Proj{L} / \bar{p}^2$. Expressing the Fourier-transformed kinetic term as $\widetilde{\overline{\psi}}_i \mathcal{K}(p) \widetilde{\psi}_i = \widetilde{\overline{\psi}}_i \Proj{L} \slashed{p} \Proj{R} \widetilde{\psi}_i$, the expression for the propagator $\Delta(p)$ is the only possibility such that: $\Delta(p) \mathcal{K}(p) = \Proj{R}$ and $\mathcal{K}(p) \Delta(p) = \Proj{L}$. The problematic term is then the $\bar{p}^2$, i.e. the 4-dimensional scalar product in the denominator, which cancels a similar term coming from the Dirac matrices contractions sandwiched between the projectors, according to \cref{eq:Gamma5DReg}.
}
and to unregularized loop diagrams.
We are thus led to consider the full Dirac fermion $\psi$ in its entirety and use
instead the fully $d$ dimensional covariant kinetic term
$\imath \overline{\psi}_i \slashed{\partial} {\psi}_i$.
% It can be re-expressed in terms of projectors as follows:
% $\imath \overline{\psi}_i \slashed{\partial} {\psi}_i = \imath \overline{\psi}_i \overline{\slashed{\partial}} {\psi}_i + \imath \overline{\psi}_i \widehat{\slashed{\partial}} {\psi}_i = \imath (\overline{\psi}_i \Proj{L} \slashed{\partial} \Proj{R} {\psi}_i + \overline{\psi}_i \Proj{R} \slashed{\partial} \Proj{L} {\psi}_i) + \imath (\overline{\psi}_i \Proj{L} \slashed{\partial} \Proj{L} {\psi}_i + \overline{\psi}_i \Proj{R} \slashed{\partial} \Proj{R} {\psi}_i)$.
The fictitious left-chiral field $\psi_L$ is thus introduced, which appears only within the kinetic term and nowhere else (it does not couple in particular to the gauge bosons of the theory), and we enforce it to be invariant under gauge transformations.

Hence, our final choice for the $d$-dimensionally regularized fermionic kinetic and gauge interaction terms is:
\begin{equation}
\label{eq:Lfermions}
	\mathcal{L}_\text{fermions} = \imath \overline{\psi}_i \slashed{\partial} {\psi}_i + g {T_R}^a_{ij} \overline{\psi_R}_i \slashed{G}^a {\psi_R}_j \, .
\end{equation}
Since this is a crucial ingredient of our analysis we rewrite it in
several ways, first as a sum of a purely 4-dimensional, gauge invariant part and a
purely evanescent term
\begin{align}
\label{eq:Lfermionsplit}
	\mathcal{L}_\text{fermions}
		&=\mathcal{L}_\text{fermions,inv} + \mathcal{L}_\text{fermions,evan} \, , \\
	\mathcal{L}_\text{fermions,inv}
		&= \imath \overline{\psi}_i \overline{\slashed{\partial}} {\psi}_i
		 + g {T_R}^a_{ij} \overline{\psi_R}_i \slashed{G}^a {\psi_R}_j \, , \\
	\mathcal{L}_\text{fermions,evan}
		&= \imath \overline{\psi}_i \widehat{\slashed{\partial}} {\psi}_i
	\, .
\end{align}
Here the first term contains purely
4-dimensional derivatives and gauge fields. It is gauge and BRST-invariant since
the fictitious left-chiral field $\psi_L$ is a gauge singlet.
This invariant term can also be written as a sum of purely left-chiral and
purely right-chiral terms involving the 4-dimensional covariant derivative as
\begin{align}
	\mathcal{L}_\text{fermions,inv}
		&= \imath \overline{\psi_L}_i \overline{\slashed{\partial}} {\psi_L}_i
		 + \imath \overline{\psi_R}_i \overline{\slashed{\partial}} {\psi_R}_i
		 + g {T_R}^a_{ij} \overline{\psi_R}_i \slashed{G}^a {\psi_R}_j
		\\
		&= \imath \overline{\psi_L}_i \overline{\slashed{\partial}} {\psi_L}_i
		 + \imath \overline{\psi_R}_i \overline{\slashed{D}} {\psi_R}_i
		\, ,
\end{align}
which highlights its gauge invariance.
The second term in \cref{eq:Lfermionsplit} is purely evanescent, i.e.\ it vanishes
in 4-dimensions. The evanescent term can be rewritten as
\begin{align}
	\mathcal{L}_\text{fermions,evan} &=
		\imath \overline{\psi_L}_i \widehat{\slashed{\partial}} {\psi_R}_i
		+
		\imath \overline{\psi_R}_i \widehat{\slashed{\partial}} {\psi_L}_i
	\, ,
\end{align}
which highlights the fact that it mixes left- and right-chiral fields
which have different gauge transformation properties. This causes the
breaking of gauge and BRST invariance --- the central difficulty of
the BMHV scheme.

The rest of the model is straightforwardly extended to $d$ dimensions: we define the $d$-dimensional BRST transformations on the fields formally exactly in the same way as in 4 dimensions:
\begin{subequations}
\begin{align}
    s_d{G^a_\mu} &= D^{ab}_\mu c^b = \partial_\mu c^a + g f^{abc} G^b_\mu c^c \, , \\
    s_d{\psi_i} &= s_d{{\psi_R}_i} = \imath c^a g {T_R}^a_{ij} {\psi_R}_j \, , \\
    s_d{\overline{\psi}_i} &= s_d{\overline{\psi_R}_i} = +\imath \overline{\psi_R}_j c^a g {T_R}^a_{ji} \, , \\
    s_d{{\psi_L}_i} &= 0 \, , \\
    s_d{\overline{\psi_L}_i} &= 0 \, , \\
    s_d{\Phi_m} &= \imath c^a g \theta^a_{mn} \Phi_n \, , \\
    s_d{c^a} &= -\frac{1}{2} g f^{abc} c^b c^c \equiv \imath g c^2 \, , \\
    s_d{\bar{c}^a} &= B^a \, , \\
    s_d{B^a} &= 0 \, ,
\end{align}
\end{subequations}
and again the external sources are invariant under BRST transformations.
This version of the BRST operator $s_d$ is nilpotent, like its
4-dimensional counterpart. Furthermore, we note that the right-hand
sides of these equations contain no $d$-dependent prefactors or
evanescent objects.

The full $d$-dimensional tree-level action $S_0$ of the model thus reads:
\begin{equation}
\label{eq:S0Def1}
	S_0 = \int \dInt[d]{x} (\mathcal{L}_\text{gauge} + \mathcal{L}_\text{fermions} + \mathcal{L}_\text{scalars} + \mathcal{L}_\text{Yukawa} + \mathcal{L}_\text{ghost} + \mathcal{L}_\text{g-fix} + \mathcal{L}_\text{ext}) \, ,
\end{equation}
where all terms except $\mathcal{L}_\text{fermions}$ remain formally exactly as before
(and with all Lorentz indices interpreted in $d$ dimensions).

\subsubsection*{Properties and expansion of the $d$-dimensional tree-level action}

We now provide two ways to rewrite the $d$-dimensional
classical action, which will be very useful in the discussion of
higher orders and renormalization. First, we note that we can naturally
decompose $S_0$ according to the split of the fermion Lagrangian
\eqref{eq:Lfermionsplit} into
\begin{subequations}
\begin{equation}
\label{eq:S0Def2}
  S_0 = S_{0,\text{inv}}+S_{0,\text{evan}}
\end{equation}
i.e. into a BRST-invariant and a purely evanescent part, with
\begin{align}
\begin{split}
  S_{0,\text{inv}} &=
  \int \dInt[d]{x} \big(
  \mathcal{L}_\text{gauge} +
        \mathcal{L}_\text{fermions,inv} + \mathcal{L}_\text{scalars} +
        \mathcal{L}_\text{Yukawa}
        \\
        &{}\qquad\qquad +
        \mathcal{L}_\text{ghost} + \mathcal{L}_\text{g-fix} +
        \mathcal{L}_\text{ext}
        \big)
	\, ,
\end{split}
	\\
	S_{0,\text{evan}} &= \int \dInt[d]{x} \mathcal{L}_\text{fermions,evan}
	\, .
\end{align}
\end{subequations}
Here, the first part of the action
contains everything except the
evanescent part of the $d$-dimensional fermion kinetic term.
It is clearly BRST-invariant since the 4-dimensional part of the
fermion covariant derivative term is gauge and BRST-invariant and all
other sectors of the theory are insensitive to the transition from 4
to $d$ dimensions.

Second, we write the $d$-dimensional action of the model as a sum of integrated
field monomials and introduce notations for each field monomial, for later usage
(and where we used the condensed notation $\int_x \equiv \int \dInt[d]{x}$):
\begin{subequations}
\label{eq:S0Def3}
\begin{equation}\label{eq:RModelDReg_Action}
\begin{split}
    S_0 =\;&
        (S_{GG} + S_{GGG} + S_{GGGG})
         + (S_{\overline{\psi}\psi} + \overline{S_{\overline{\psi} G \psi_R}})
         + (S_{\Phi\Phi} + S_{\Phi G \Phi} + S_{\Phi GG \Phi}) \\
        &+ ((Y_R)^m_{ij} S_{\overline{\psi_R}^C_i \Phi^m {\psi_R}_j} + \hc) + \lambda_{mnop} S_{\Phi^4_{mnop}} \\
        &+ S_\text{g-fix} + (S_{\bar{c} c} + S_{\bar{c} G c})
         + (S_{\rho c} + S_{\rho G c}) + S_{\zeta c c} + S_{\bar{R} c \psi_R} + S_{R c \overline{\psi_R}} + S_{\mathcal{Y} c \Phi}
        \, ,
\end{split}\end{equation}
with the gauge kinetic and self-interaction terms
\begin{equation}\begin{split}
	\int_x \frac{-1}{4} F^a_{\mu\nu} F^{a\,\mu\nu} &= S_{GG} + S_{GGG} + S_{GGGG} \, , \; \text{with:} \\
		S_{GG}   &= \int_x \frac{1}{2} G^a_\mu (g^{\mu\nu} \partial^2 - \partial^\mu \partial^\nu) G^a_\nu \, , \\
		S_{GGG}  &= \int_x (-g) f^{abc} (\partial_\mu G^a_\nu) G^{b\,\mu} G^{c\,\nu} \, , \\
		S_{GGGG} &= \int_x \frac{-g^2}{4} f^{eac} f^{ebd} G^a_\mu G^{b\,\mu} G^c_\nu G^{d\,\nu} \, ,
\end{split}\end{equation}
the fermion kinetic and interaction terms, using the notation $A \overset{\leftrightarrow}{\partial} B \equiv A (\partial B) - (\partial A) B$
\begin{equation}\begin{split}
	S_{\overline{\psi}\psi} &= \int_x \imath \overline{\psi}_i \slashed{\partial} \psi_i \equiv \int_x \frac{\imath}{2} \overline{\psi}_i \overset{\leftrightarrow}{\slashed{\partial}} \psi_i \, , \\
	\overline{S_{\overline{\psi} G \psi_R}} &= \int_x g {T_R}^a_{ij} \overline{\psi}_i \Proj{L} \slashed{G}^a \Proj{R} \psi_j = \int_x g {T_R}^a_{ij} \overline{\psi}_i \overline{\slashed{G}^a} \Proj{R} \psi_j \, ,
\end{split}\end{equation}
the scalar kinetic and interaction terms
\begin{equation}\begin{split}
	\int_x \frac{1}{2} (D_\mu \Phi^m)^2 &= S_{\Phi\Phi} + S_{\Phi G \Phi} + S_{\Phi GG \Phi} \, , \; \text{with:} \\
		S_{\Phi\Phi}     &= \int_x \frac{1}{2} (\partial_\mu \Phi^m)^2 \equiv \int_x \frac{-1}{2} \Phi^m \partial^2 \Phi^m \, , \\
		S_{\Phi G \Phi}  &= \int_x -\imath g \theta^a_{mn} (\partial^\mu \Phi^m) G^a_\mu \Phi^n \, , \\
		S_{\Phi GG \Phi} &= \int_x \frac{g^2}{2} (\theta^a \theta^b)_{mn} \Phi^m G^a_\mu G^{b\,\mu} \Phi^n \, ,
\end{split}\end{equation}
the Yukawa and the scalar quartic self-coupling terms
\begin{equation}\begin{split}
	(Y_R)^m_{ij} S_{\overline{\psi_R}^C_i \Phi^m {\psi_R}_j} + \hc &= \int_x \left( -\frac{(Y_R)^m_{ij}}{2} \Phi_m \overline{\psi_R}^C_i {\psi_R}_j -\frac{(Y_R)^{m\;*}_{ij}}{2} \Phi_m \overline{\psi_R}_i {\psi_R}^C_j \right) \, , \\
	\lambda_{mnop} S_{\Phi^4_{mnop}} &= \int_x \frac{-\lambda_{mnop}}{4!} \Phi^m \Phi^n \Phi^o \Phi^p \, ,
\end{split}\end{equation}
the gauge-fixing terms
\begin{equation}
	S_\text{g-fix} = \int_x \frac{\xi}{2} B^a B_a + B^a \partial^\mu G^a_\mu \, ,
\end{equation}
the ghost kinetic and interaction terms
\begin{equation}\begin{split}
	\int_x (\partial^\mu \bar{c}_a) (D_\mu c_a) &= S_{\bar{c} c} + S_{\bar{c} G c} \, , \; \text{with:} \\
		S_{\bar{c} c}   &= \int_x (\partial^\mu \bar{c}_a) (\partial_\mu c_a) \equiv \int_x -\bar{c}_a \partial^2 c_a \, , \\
		S_{\bar{c} G c} &= \int_x g f^{abc} (\partial^\mu \bar{c}_a) G^b_\mu c_c \, ,
\end{split}\end{equation}
and the external BRST source terms
\begin{equation}\begin{split}
	\int_x \rho_a^\mu s_d{G^a_\mu} &= \int_x \rho_a^\mu D^{ab}_\mu c^b = S_{\rho c} + S_{\rho G c} \, , \; \text{with:} \\
		S_{\rho c}   &= \int_x \rho_a^\mu (\partial_\mu c_a) \, , \\
		S_{\rho G c} &= \int_x g f^{abc} \rho_a^\mu G^b_\mu c_c \, ,
\end{split}\end{equation}
and
\begin{equation}\begin{split}
	S_{\zeta c c} &= \int_x \zeta_a s_d{c^a} = \int_x \frac{-1}{2} g f^{abc} \zeta_a c^b c^c \, , \\
	S_{\bar{R} c \psi_R} &= \int_x \bar{R}^i s_d{\psi_i} = \int_x \imath g \bar{R}^i c^a {T_R}^a_{ij} {\psi_R}_j \equiv \int_x \imath g \bar{R}^i c^a {T_R}^a_{ij} \Proj{R} \psi_j \, , \\
	S_{R c \overline{\psi_R}} &= \int_x R^i s_d{\overline{\psi}_i} \equiv \int_x s_d{\overline{\psi}_i} R^i = \int_x \imath g \overline{\psi_R}_j c^a {T_R}^a_{ji} R^i \equiv \int_x \imath g \overline{\psi}_j \Proj{L} c^a {T_R}^a_{ji} R^i \, , \\
	S_{\mathcal{Y} c \Phi} &= \int_x \mathcal{Y}^m s_d{\Phi_m} = \int_x \imath g \mathcal{Y}^m c^a \theta^a_{mn} \Phi_n
	\, .
\end{split}\end{equation}
\end{subequations}

\subsection{BRST breaking of the R-model in $d$ dimensions}
\label{subsect:BRST_breaking_RModel}

% Using the functional form\footnote{Alternatively, one can act with $s_d$ on each individual field in the action. The obtained results are identical.} of the BRST transformation acting on the classical action:
% \begin{equation}
    % s_d = \int \dInt[d]{x} \left(
        % \frac{\delta S_0}{\delta \rho_a^\mu} \frac{\delta}{\delta G^a_\mu} + \frac{\delta S_0}{\delta \zeta_a} \frac{\delta}{\delta c^a} +
        % \frac{\delta S_0}{\delta \mathcal{Y}^m} \frac{\delta}{\delta \Phi_m} +
        % \frac{\delta S_0}{\delta \bar{R}^i} \frac{\delta}{\delta \psi_i} + \frac{\delta S_0}{\delta R^i} \frac{\delta}{\delta \overline{\psi}_i} +
        % B^a \frac{\delta}{\delta \bar{c}_a} \right)
        % \, ,
% \end{equation}

Our next step is to determine
to what extent our choice of the $d$-dimensional action
$S_0$ given in
\cref{eq:S0Def1,eq:S0Def2,eq:S0Def3} breaks the defining
BRST invariance and the Slavnov-Taylor identity.
As already mentioned in \cref{subsect:RModelDReg} the
$d$-dimensional action can be split into a BRST-invariant and an
evanescent term. It is easy to see that the part $S_{0,\text{inv}}$
on its own satisfies
\begin{align}
  s_d S_{0,\text{inv}} &= 0
\end{align}
and hence, due to the Quantum Action Principle, the $d$-dimensional Slavnov-Taylor identity
\begin{align}
 \mathcal{S}_d(S_{0,\text{inv}}) &=0\, ,
\end{align}
where the Slavnov-Taylor operator $\mathcal{S}_d$ is given in the same way as its
4-dimensional version in \cref{eq:SofFDefinition} except for
replacing all 4-dimensional objects by $d$-dimensional ones.
However, the evanescent part of the action $S_{0,\text{evan}}$  is not
BRST-invariant since it couples left- and
right-chiral fermions with different gauge transformation
properties.
This breaking of BRST invariance leads to a breaking of the
Slavnov-Taylor identity
in the form
\begin{subequations}
  \label{eq:WidehatDeltaDefinition}
\begin{align}
  s_d S_0 &=  s_d S_{0,\text{evan}}
  \equiv \widehat{\Delta} \, ,\\
 \mathcal{S}_d(S_0) &= \widehat{\Delta} \, ,
\end{align}
\end{subequations}
with the same non-vanishing integrated breaking term $\widehat{\Delta}$ appearing
in both equations. The breaking is given by
\begin{equation}
\label{eq:BRSTTreeBreaking}
    \widehat{\Delta} = \int \dInt[d]{x}
        (g {T_R}^a_{ij}) c^a \left\{
            \overline{\psi}_i \left(\overset{\leftarrow}{\widehat{\slashed{\partial}}} \Proj{R} + \overset{\rightarrow}{\widehat{\slashed{\partial}}} \Proj{L}\right) \psi_j
        \right\}
    \equiv \int \dInt[d]{x} \widehat{\Delta}(x)
% = \int \dInt[d]{x}
        % \left(\frac{g}{2} {T_R}^a_{ij}\right) c^a \left\{
            % \partial_\mu (\overline{\psi}_i \widehat{\gamma}^\mu \psi_j)
            % - \overline{\psi}_i \overset{\leftrightarrow}{\widehat{\slashed{\partial}}} \gamma_5 \psi_j
        % \right\}
    \, .
\end{equation}
% where we employed cancellation due to contraction between symmetric and anti-symmetric tensors, as well as the relations imposed by gauge invariance relating the $\theta$ generators with the Yukawa matrices, \cref{eq:GaugeInvarYuk} and its conjugate, and the totally symmetric scalar self-coupling $\lambda$, \cref{eq:GaugeInvarLamb}.

For the purpose of restoring the BRST symmetry, as we will see in \cref{sect:BRSTrestoration}, the evaluation of Feynman diagrams with an insertion of this breaking $\widehat{\Delta}$ will be required. This breaking generates an interaction vertex whose Feynman rule (with all momenta incoming) is:
\\
% c^a(-p_1-p_2) \bar{\psi}_{i,\alpha}(p_1) \psi_{j,\beta}(p_2)
\begin{equation}
\begin{tabular}{rl}
	\raisebox{-40pt}{\includegraphics[scale=0.6]{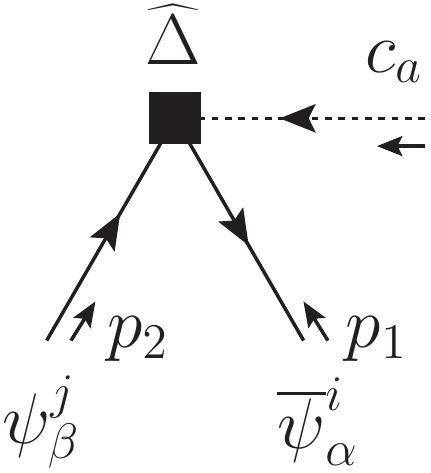}} &
	$\begin{aligned}
		&= \frac{g}{2} {T_R}^a_{ij} \left((\widehat{\slashed{p_1}} + \widehat{\slashed{p_2}}) + (\widehat{\slashed{p_1}} - \widehat{\slashed{p_2}}) \gamma_5 \right)_{\alpha\beta} \\
		&= g {T_R}^a_{ij} \left(\widehat{\slashed{p_1}} \Proj{R} + \widehat{\slashed{p_2}} \Proj{L} \right)_{\alpha\beta}
		\, .
	\end{aligned}$
\end{tabular}
\end{equation}
It is useful to provide as well the Feynman rule corresponding to the charge-conjugated fermions,
since the Yukawa couplings contain occurrences of these, and to applying flipping rules as in
\cite{Denner:1992vza,Denner:1992me}. % == DennerHahnKueblbeck
The breaking can be equivalently written as
\begin{equation}
    \widehat{\Delta} = \int \dInt[d]{x}
        (g {T_{\overline{R}}}^a_{ij}) c^a \left\{
        \overline{\psi^C}_i \left(\overset{\leftarrow}{\widehat{\slashed{\partial}}} \Proj{L} + \overset{\rightarrow}{\widehat{\slashed{\partial}}} \Proj{R}\right) \psi^C_j
        \right\}
% = \int \dInt[d]{x}
        % \left(\frac{g}{2} {T_{\overline{R}}}^a_{ij}\right) c^a \left\{
        % \partial_\mu (\overline{\psi^C}_i \widehat{\gamma}^\mu \psi^C_j)
        % + \overline{\psi^C}_i \overset{\leftrightarrow}{\widehat{\slashed{\partial}}} \gamma_5 \psi^C_j
        % \right\}
    \, ,
\end{equation}
generating the Feynman rule:
\\
% c^a(-p_1-p_2) \overline{\psi^C}_{i,\alpha}(p_1) \psi^C_{j,\beta}(p_2)
\begin{equation}
\begin{tabular}{rl}
	\raisebox{-40pt}{\includegraphics[scale=0.6]{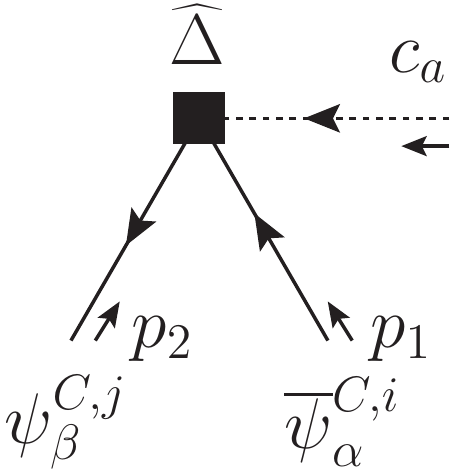}} &
	$\begin{aligned}
		&= \frac{g}{2} {T_{\overline{R}}}^a_{ij} \left((\widehat{\slashed{p_1}} + \widehat{\slashed{p_2}}) - (\widehat{\slashed{p_1}} - \widehat{\slashed{p_2}}) \gamma_5 \right)_{\alpha\beta} \\
		&= g {T_{\overline{R}}}^a_{ij} \left(\widehat{\slashed{p_1}} \Proj{L} + \widehat{\slashed{p_2}} \Proj{R} \right)_{\alpha\beta}
		\, ,
	\end{aligned}$
\end{tabular}
\end{equation}
where the difference with the previous result is in the appearance of the generator ${T_{\overline{R}}}^a$ for the fermionic conjugate representation $R$.

At this point it is natural to introduce the so-called linearized
Slavnov-Taylor operator $b_d$. In our later applications we will require
the Slavnov-Taylor identity at higher orders in the form $\mathcal{S}(S_0+{\cal F})$,
where the functional ${\cal F}$ might be the 1-loop
regularized or renormalized effective action or the 1-loop counterterm
action. We can then write to first order in ${\cal F}$,
\begin{equation}
    \mathcal{S}_d(S_0+{\cal F}) = \mathcal{S}_d(S_0) + b_d{\cal F} + {\cal O}({\cal F}^2)
\end{equation}
where $b_d$ can be written in functional form as
\begin{equation}\begin{split}
\label{eq:bdDefinition}
  b_d =
  \int \dInt[d]{x} & \left(
        \frac{\delta S_0}{\delta \rho_a^\mu} \frac{\delta}{\delta G^a_\mu} + \frac{\delta S_0}{\delta G^a_\mu} \frac{\delta}{\delta \rho_a^\mu} + \frac{\delta S_0}{\delta \zeta_a} \frac{\delta}{\delta c^a} + \frac{\delta S_0}{\delta c^a} \frac{\delta}{\delta \zeta_a} + \frac{\delta S_0}{\delta \mathcal{Y}^m} \frac{\delta}{\delta \Phi_m} + \frac{\delta S_0}{\delta \Phi_m} \frac{\delta}{\delta \mathcal{Y}^m} \right.\\
        &\hspace{0.5cm}\left. + \frac{\delta S_0}{\delta \bar{R}^i} \frac{\delta}{\delta \psi_i} + \frac{\delta S_0}{\delta \psi_i} \frac{\delta}{\delta \bar{R}^i} + \frac{\delta S_0}{\delta R^i} \frac{\delta}{\delta \overline{\psi}_i} + \frac{\delta S_0}{\delta \overline{\psi}_i} \frac{\delta}{\delta R^i} + B^a \frac{\delta}{\delta \bar{c}_a} \right)
        \, .
\end{split}\end{equation}
The linearized Slavnov-Taylor operator is an extension of the BRST
transformations in the sense that
\begin{equation}
    b_d = s_d +
    \int \dInt[d]{x} \left(
        \frac{\delta S_0}{\delta G^a_\mu} \frac{\delta}{\delta \rho_a^\mu}
        + \frac{\delta S_0}{\delta c^a} \frac{\delta}{\delta \zeta_a}
        + \frac{\delta S_0}{\delta \Phi_m} \frac{\delta}{\delta \mathcal{Y}^m}
        +
        \frac{\delta S_0}{\delta \psi_i} \frac{\delta}{\delta \bar{R}^i}
        + \frac{\delta S_0}{\delta \overline{\psi}_i} \frac{\delta}{\delta R^i} \right)
    \, ,
\end{equation}
i.e. $b_d$ and $s_d$ act in the same way on fields but only $b_d$
acts in a non-trivial way on the sources.
A subtlety, compared to the standard situation with symmetry-preserving
regularization, is that $b_d$ is not nilpotent, ${b_d}^2 \neq 0$. The
reason is that the $d$-dimensional action $S_0$ is not BRST-invariant%
\footnote{
	We might have defined a nilpotent object
	$b_d^{\text{nilpotent}}$ by using the invariant action
	$S_{0,\text{inv}}$ in place of $S_0$ in the definition of
	$b_d$. However, it is our choice of $b_d$ which will appear in the
	later analysis.
}.

For later usage it is advantageous to also define the 4-dimensional linearized
Slavnov-Taylor operator, $b$, as the restriction to 4 dimensions of $d$-dimensional
operator $b_d$, based on the Slavnov-Taylor operation \cref{eq:SofFDefinition}
and on the 4-dimensional action $S_0^{(4D)}$. Its functional form is then:
\begin{equation}
    b = s +
    \int \dInt[4]{x} \left(
        \frac{\delta S_0^{(4D)}}{\delta G^a_\mu} \frac{\delta}{\delta \rho_a^\mu}
        + \frac{\delta S_0^{(4D)}}{\delta c^a} \frac{\delta}{\delta \zeta_a}
        + \frac{\delta S_0^{(4D)}}{\delta \Phi_m} \frac{\delta}{\delta \mathcal{Y}^m}
        +
        \frac{\delta S_0^{(4D)}}{\delta \psi_i} \frac{\delta}{\delta \bar{R}^i}
        + \frac{\delta S_0^{(4D)}}{\delta \overline{\psi}_i} \frac{\delta}{\delta R^i} \right)
    \, .
\end{equation}
Contrary to its $d$-dimensional counterpart $b_d$, the operator $b$ is nilpotent:
$b^2 = 0$, because the 4-dimensional action $S_0^{(4D)}$ is BRST-invariant \cite{Piguet:1980nr}.

\section{Standard Renormalization Transformation versus General Counterterm Structure}
\label{sect:standardrenormalizationstructure}

In the majority of practical loop calculations in gauge theories, a
regularization is assumed which preserves gauge and BRST invariance of
the theory. In such cases, the necessary counterterm structure can
simply be obtained from the classical Lagrangian by applying a
renormalization transformation. We briefly recall the structure of the
required renormalization transformation here; this will provide a
useful benchmark against which the counterterm structure in the
BMHV scheme can be compared.

% \dspar{citation for renormalization transformation? Lee/Zinn-Justin
  % for SSB gauge theory but for unbroken theory? Weinberg/Ch. 17, other
  % standard textbooks? Textbook by Denner? Cite algebraic
  % renormalization like Piguet/Sorella? E.g. (6.26) in Piguet/Rouet
  % goes exactly in the direction we need.}

% \dspar{why are the field renormalization constants in section 8
  % diagonal in the field indices? Do you assume irreducible matter
  % representations?}

The renormalization transformation consists of renormalization of
physical parameters%
\footnote{We employ additive renormalization for the physical parameters since
multiplicative renormalization for them would not be sufficient in general.},
\begin{subequations}
  \label{eq:rentransform}
  \begin{align}
  g & \to g + \delta g \, , \\
  (Y_R)^m_{ij} & \to (Y_R)^m_{ij} + \delta (Y_R)^m_{ij} \, , \\
  \lambda^{mnop} & \to \lambda^{mnop} + \delta \lambda^{mnop} \, ,
\end{align}
and fields, using multiplicative renormalization,
\begin{align}
  G^a_\mu &\to \sqrt{Z_G} G^a_\mu  \, , \\
  ({\psi_R}_i , \overline{\psi_R}_i) &\to \sqrt{Z_{\psi_R}} ({\psi_R}_i , \overline{\psi_R}_i) \, , \\
  ({\psi_L}_i , \overline{\psi_L}_i) &\to \quad ({\psi_L}_i , \overline{\psi_L}_i) \, , \\
  \Phi_m &\to \sqrt{Z_\Phi} \Phi_m \, , \\
  c^a &\to \sqrt{Z_c} c^a \, .
\end{align}
Here the fictitious left-chiral fermion field does not renormalize,
and we have used a ghost field renormalization that is different from the antighost field one.
The remaining fields, sources and the gauge parameter renormalize in a
dependent way, as
\begin{align}
  \left\{B^a, \bar{c}^a, \xi\right\} & \to
    \left\{ \sqrt{Z_G}^{-1} B^a, \sqrt{Z_G}^{-1} \bar{c}^a, Z_G \xi \right\}
  \, , \\
  \rho_a^\mu & \to {\sqrt{Z_G}}^{-1} \rho_a^\mu\, , \\
  \zeta_a & \to {\sqrt{Z_c}}^{-1} \zeta_a \, , \\
  (R^i , \bar{R}^i) & \to  {\sqrt{Z_{\psi_R}}}^{-1} (R^i , \bar{R}^i) \, , \\
  \mathcal{Y}^m & \to  {\sqrt{Z_\Phi}}^{-1} \mathcal{Y}^m  \, .
\end{align}
\end{subequations}
If this renormalization transformation is applied on the BRST
invariant part of the tree-level action we obtain an invariant
counterterm action $S_{\text{ct,inv}}$,
\begin{equation}
\label{eq:Sctinv}
	S_{0,\text{inv}}
	\stackrel{\text{Eqs.\ (\ref{eq:rentransform})}}{\longrightarrow}
	S_{0,\text{inv}}+S_{\text{ct,inv}} \, .
\end{equation}
This is invariant in the sense that the Slavnov-Taylor identity
\begin{equation}
\label{eq:STIctinv}
	\mathcal{S}_d(S_{0,\text{inv}}+S_{\text{ct,inv}}) = 0
\end{equation}
holds.

This structure can be compared later to the actual counterterm
structure needed in the BMHV scheme. As a preview, we
note that the following general counterterm structure can be expected:
\begin{equation}
\label{eq:CT_structure}
  S_{\text{sct,inv}}
+  S_{\text{sct,evan}}
+  S_{\text{fct,inv}}
+  S_{\text{fct,restore}}
+  S_{\text{fct,evan}} \, ,
\end{equation}
where
\begin{itemize}[leftmargin=*]
\item
  $S_{\text{sct,inv}}$ and $S_{\text{fct,inv}}$ correspond to the
  invariant counterterms generated by a renormalization transformation
  as in \cref{eq:Sctinv}. The
  subscripts ``sct'' and ``fct'' denote singular parts
  (i.e.\ involving $1/\epsilon$ poles) and finite parts,
  respectively.
\item
  $S_{\text{sct,evan}}$ corresponds to additional singular
  counterterms needed to cancel additional $1/\epsilon$ poles of loop
  diagrams. We will see that these counterterms are purely
  evanescent. Similarly, evanescent divergent counterterms are also
  familiar from computations using regularization by dimensional
  reduction (see \cite{Gnendiger:2017pys} for a recent review). % == todornottod
  There, such counterterms are needed to establish scheme equivalence
  \cite{Jack:1993ws,Jack:1994bn}, to ensure unitarity,
  finiteness, and consistency with infrared factorization in higher-order computations
  \cite{Harlander:2006rj, Harlander:2006xq,
    Kilgore:2011ta, Kilgore:2012tb,Broggio:2015ata,Broggio:2015dga}.
\item
  $S_{\text{fct,restore}}$ corresponds to finite counterterms needed
  to restore the symmetry. Determining these counterterms is one of
  the central goals of the present paper, and is presented in \cref{sect:BRSTrestoration}.
\item
  $S_{\text{fct,evan}}$ corresponds to additional counterterms which
  are both finite and evanescent. Adding or changing such counterterms
  can swap e.g.\ between different options as in
  \cref{eq:inequivalentchoices}; these counterterms vanish in the
  4-dimensional limit, but they can affect calculations at higher
  orders.
\end{itemize}

Let us present for further use a more detailed analysis of the
structure of the invariant counterterms. We focus on the
counterterms arising in first order of
the renormalization constants $\delta g$, $\delta Y$, $\delta
\lambda$ and $\delta Z_\varphi\equiv Z_\varphi-1$. At first order in these
quantities we can express the invariant counterterm action as a
linear combination of basis functionals $L_\varphi$,
\begin{align}
\label{eq:Sctinvstructure}
\begin{split}
    S_{\text{ct,inv}} =\;&
    \frac{\delta Z_G}{2} L_G +
    \frac{\delta Z_{\psi_R}}{2} \overline{L_{\psi_R}} +
    \frac{\delta Z_\Phi}{2} L_\Phi +
    \frac{\delta Z_c}{2} L_c
    \\
    &+
    \frac{\delta g}{g} L_g +
    \big( \delta (Y_R)^m_{ij} {L_{Y_R}}^m_{ij} + \hc \big) +
    \delta \lambda^{mnop} L_{\lambda^{mnop}} \, ,
\end{split}
\end{align}
and in the following we collect the properties of these functionals.
Introducing the field-numbering operators:
\begin{subequations}
\begin{align}
	N_\varphi &= \int \dInt[d]{x} \varphi_i(x) \frac{\delta}{\delta \varphi_i(x)} \, , \; \text{for $\varphi_i \in \{G^a_\mu, \Phi^m, c_a, \bar{c}_a, B^a, \rho_a^\mu, \zeta_a, R^i, \bar{R}^i, \mathcal{Y}^m$\} } \, , \\
	N_\psi^{R/L} &= \int \dInt[d]{x} (\Proj{R/L} \psi_i(x))_s \frac{\delta}{\delta \psi_i(x)_s} \, , \\
	N_{\overline{\psi}}^{L/R} &= \int \dInt[d]{x} (\overline{\psi}_i(x) \Proj{L/R})^s \frac{\delta}{\delta \overline{\psi}_i(x)^s} \, ,
\end{align}
\end{subequations}
(and summing over repeated generic group index $i$ and spinor index $s$),
we can first write the functionals $L_\varphi$ as derivatives of the tree-level action:
\begin{equation}\label{eq:Lfuncts_fields}
\begin{aligned}
	L_G &= ( N_G - N_{\bar{c}} - N_B - N_\rho + 2 \xi \frac{\partial}{\partial \xi} ) S_0 \equiv \mathcal{N}_G S_0
	\, , \\
	L_c &= \left( N_c - N_\zeta \right) S_0 \equiv \mathcal{N}_c S_0
	\, , \\
	L_\Phi &= \left( N_\Phi - N_\mathcal{Y} \right) S_0 \equiv \mathcal{N}_{\Phi} S_0
	\, , \\
	\overline{L_{\psi_R}} &= -(N_\psi^R + N_{\overline{\psi}}^L - N_{\bar{R}} - N_R) S_{0,\text{inv}} \equiv \mathcal{N}_{\psi}^R S_{0,\text{inv}}
	\, , \\
	L_{\psi_R} &= -(N_\psi^R + N_{\overline{\psi}}^L - N_{\bar{R}} - N_R) S_0 \equiv \mathcal{N}_{\psi}^R S_0 \\
		&= \overline{L_{\psi_R}} + S_{0,\text{evan}}
	\, ,
\end{aligned}\end{equation}
and
\begin{align}\label{eq:Lfuncts_couplings}
	L_g &\equiv g \frac{\partial S_0}{\partial g} \, , &&
	{L_{Y_R}}^m_{ij} &\equiv \frac{\partial S_0}{\partial (Y_R)^m_{ij}} \, , &&
	L_{\lambda_{mnop}} &\equiv \frac{\partial S_0}{\partial \lambda_{mnop}} \, .
\end{align}
In most of these equations the result does not change if we replace
$S_0$ by its invariant part $S_{0,\text{inv}}$, excepting for
$L_{\psi_R}$ and $\overline{L_{\psi_R}}$ where we have given both
expressions and expressed the difference in terms of the evanescent term
$S_{0,\text{evan}}$. It is the latter quantity $\overline{L_{\psi_R}}$
that appears in the renormalization transformation \cref{eq:Sctinvstructure}.

The $L_\varphi$ functionals corresponding to field renormalization can be
written as a total $b_d$-variation and in terms of the monomials of
\cref{subsect:RModelDReg} as
\begin{align}
    \begin{split}
    L_G &= b_d \int \dInt[d]{x} \widetilde{\rho}_a^\mu G^a_\mu \\
        &= 2 S_{GG} + 3 S_{GGG} + 4 S_{GGGG} + \overline{S_{\overline{\psi} G \psi_R}} + S_{\Phi G \Phi} + 2 S_{\Phi GG \Phi} - S_{\bar{c} c} - S_{\rho c}
        \, ,
    \end{split} \\
	\intertext{where $\widetilde{\rho}_a^\mu = \rho_a^\mu + \partial^\mu \bar{c}_a$ is the natural combination arising from the \emph{ghost equation} (see third equation in \eqref{eq:STIInvarianceEqs});}
    \begin{split}
    L_c &= -b_d \int \dInt[d]{x} \zeta_a c^a \\
        &= S_{\bar{c} c} + S_{\bar{c} G c} + S_{\rho c} + S_{\rho G c} + S_{\zeta c c} + S_{\bar{R} c \psi_R} + S_{R c \overline{\psi_R}} + S_{\mathcal{Y} c \Phi}
        \, ,
    \end{split} \\
    \begin{split}
    L_\Phi &= b_d \int \dInt[d]{x} \mathcal{Y}^m \Phi_m \\
           &= 2 \left( S_{\Phi\Phi} + S_{\Phi G \Phi} + S_{\Phi GG \Phi} \right) + 4 \lambda_{mnop} S_{\Phi^4_{mnop}} + ( (Y_R)^m_{ij} S_{\overline{\psi_R}^C_i \Phi^m {\psi_R}_j} + \hc )
        \, ,
    \end{split} \\
    \begin{split}
    L_{\psi_R} &= - b_d \int \dInt[d]{x} (\bar{R}^i \Proj{R} \psi_i + \overline{\psi}_i \Proj{L} R^i) \\
        \footnotemark&=
        \left( 2 \int \dInt[d]{x} \frac{\imath}{2} \overline{\psi}_i (\slashed{\partial} \Proj{R} + \Proj{L} \slashed{\partial}) \psi_i \right) + 2 \overline{S_{\overline{\psi} G \psi_R}} + 2 ((Y_R)^m_{ij} S_{\overline{\psi_R}^C_i \Phi^m {\psi_R}_j} + \hc)
        % \, ,
    % \end{split}
    % \begin{split}
    % \overline{L_{\psi_R}} &= L_{\psi_R} - \int \dInt[d]{x} \imath \overline{\psi}_i \widehat{\slashed{\partial}} \psi_i \\
        % &=
        % \left( 2 \int \dInt[d]{x} \imath \overline{\psi}_i \overline{\slashed{\partial}} \Proj{R} \psi_i \right) + 2 \overline{S_{\overline{\psi} G \psi_R}} + 2 ((Y_R)^m_{ij} S_{\overline{\psi_R}^C_i \Phi^m {\psi_R}_j} + \hc)
        \, ,
    \end{split}
\end{align}
\footnotetext{
	Observing that $\imath \overline{\psi}_i (\slashed{\partial} \Proj{R} + \Proj{L} \slashed{\partial}) \psi_i = 2 \imath \overline{\psi}_i \overline{\slashed{\partial}} \Proj{R} \psi_i + \imath \overline{\psi}_i \widehat{\slashed{\partial}} \psi_i$, we note that there exists a difference between this calculation and the result given in \cite{Martin:1999cc}, amounting to: $L_{\psi_R}^\text{CPM} - L_{\psi_R}^\text{ours} = \imath \int \dInt[d]{x} \overline{\psi}_i \widehat{\slashed{\partial}} \gamma_5 \psi_i$.
}
while the $L_\varphi$ functionals corresponding to renormalization of physical couplings
can be expressed in terms of the monomials of
\cref{subsect:RModelDReg} as
\begin{align}
    \begin{split}
    L_g =\;& S_{GGG} + 2 S_{GGGG} + S_{\Phi G \Phi} + 2 S_{\Phi G G \Phi} + \overline{S_{\overline{\psi} G \psi_R}} \\
        & + S_{\bar{c} G c} + S_{\rho G c} + S_{\zeta c c} + S_{\bar{R} c \psi_R} +S_{R c \overline{\psi_R}} + S_{\mathcal{Y} c \Phi}
    \, ,
    \end{split} \\
    {L_{Y_R}}^m_{ij} =\;& S_{\overline{\psi_R}^C_i \Phi^m {\psi_R}_j} \, , \\
    L_{\lambda_{mnop}} =\;& S_{\Phi^4_{mnop}} \, .
\end{align}

Despite the non-nilpotency of $b_d$, several of the $L_\varphi$ are actually
$b_d$-invariant in the following sense:
\begin{align}
  b_d L_\varphi&=0 \quad \text{for} \quad \varphi=G,\Phi \, , \\
  b_d \overline{L_{\psi_R}} &= 0 \, , \\
  b_d \left[ \delta (Y_R)^m_{ij} {L_{Y_R}}^m_{ij} \right] &=0 \, , \\
  b_d \left[ \delta \lambda^{mnop} L_{\lambda^{mnop}} \right] &= 0 \, ,
\end{align}
where the last two equations hold provided that the renormalization
constants $\delta (Y_R)$ and $\delta \lambda$ satisfy the analogous
gauge invariance constraints as
\cref{eq:GaugeInvarYuk,eq:GaugeInvarLamb}.
In contrast, the functional $L_c$ is not $b_d$-invariant in this sense%
\footnote{
	This fact appears to be in contradiction with a claim made in \cite{Martin:1999cc}.
};
instead, it is easy to see that
\begin{align}
	b_d L_c &= \widehat{\Delta}
\end{align}
with the same breaking as in \cref{eq:BRSTTreeBreaking}. As a
result, also $L_g$, corresponding to gauge coupling renormalization,
is not $b_d$-invariant. However, one may define the quantity $L_{F^2}$
corresponding to the field strength tensor; this quantity has the
useful properties
\begin{align}
  L_{F^2} &=
  \frac{-1}{4} \int \dInt[d]{x} F^a_{\mu\nu} F^{a\,\mu\nu} = S_{GG} + S_{GGG} + S_{GGGG}
 \, ,\\
 b_d L_{F^2} &= 0
 \, ,\\
  	L_g 	& = L_c + L_G - 2 L_{F^2} \, .
\end{align}
Note, however, that in the limit $d \to 4$ and evanescent terms vanishing, all the
$L_\varphi$ functionals presented here become invariant under the linear $b$
transformation in 4 dimensions.

\section{Evaluation of the One-Loop Singular Counterterm Action $S_\text{sct}^{(1)}$ in the R-Model}
\label{sect:Rmodel1LoopSCT}

In this section, we evaluate the one-loop (order $\hbar^1$) contributions that define the singular counterterm action $S_\text{sct}^{(1)}$.
The calculations are performed in $d = 4 - 2\epsilon$ dimensions.
% We define: $1/\bar\epsilon \equiv 1/\epsilon - \gamma_E + \ln(4\pi)$.
Since the tree-level action $S_0$ also contains vertex terms $K_\phi s_d{\phi}$ with BRST sources $K_\phi$, their loop corrections have to be computed as well.
Together with the tree-level action $S_0$, the singular counterterm action participates in the definition of the dimensionally-regularized effective action $\Gamma_\text{DReg}$. This action may not yet be BRST-invariant, and thus additional \emph{finite counterterms} will be necessary to restore the BRST symmetry, \emph{up to non-spurious (and finite) anomalous terms}, thus completing the definition of $\Gamma_\text{DReg}$.
Supposing anomalous terms have been properly cancelled so as BRST symmetry is restored, the renormalized effective action $\Gamma_\text{Ren}$ is then defined from $\Gamma_\text{DReg}$ at the loop-order of interest by taking the renormalized limit, i.e. the limit $d \to 4$ and remaining evanescent terms vanishing.

Here and in the rest of the paper, the amplitudes of the necessary Feynman diagrams have been computed using the \verb|Mathematica| packages \verb|FeynArts|~\cite{Hahn:2000kx} and \verb|FeynCalc|~\cite{Mertig:1990an,Shtabovenko:2016sxi,Shtabovenko:2020gxv}; the $\epsilon$-expansion of the amplitudes has been cross-checked using the \verb|FeynCalc|'s interface \verb|FeynHelpers|~\cite{Shtabovenko:2016whf} to \verb|Package-X|~\cite{Patel:2016fam}.
The group-structure invariants are defined the same way as in the articles from Machacek \& Vaughn \cite{Machacek:1983tz,Machacek:1983fi,Machacek:1984zw}.

\subsection{Notational conventions for the quantum effective action and Green's functions}
\label{subsect:NotatConvGreenFcts}

Before continuing, we define in this section some notations adopted in the rest of this paper.
The quantum effective action (see e.g. Chapter~16 in \cite{Weinberg:1996kr} for a review) $\Gamma[\Phi]$ is the generating functional in the interacting theory for the one-particle-irreducible (1PI, or ``proper'') truncated correlation functions, incorporating all the quantum corrections. It is defined as the Legendre transform of the vacuum energy functional (i.e. the sum of all connected vacuum-vacuum amplitudes, itself defined from the partition function $Z$). As such $\Gamma[\Phi]$ is a functional of ``classical fields'' defined as the vacuum expectation values of their corresponding field operators in presence of suitable external currents. It can be expanded in generic $d$-dimensional coordinate space:
\begin{equation}
    \Gamma[\Phi] = \sum_{n \geq 2} \frac{1}{|n|!} \int \left(\prod_{i=1}^{n} \dInt[d]{x_i} \phi_i(x_i)\right) \Gamma_{\phi_n \cdots \phi_1}(x_1, \dots, x_n) \, ,
\end{equation}
where $|n|! \equiv \prod_j n_j!$, with $n_j$ the number of fields of a given type $j$, spanning all the different types of fields in the given 1PI function, and $n$ the total number of fields in it. The condition $n \geq 2$ is present because tadpoles can be eliminated (see e.g.\ \cite{Peskin:1995ev,Srednicki:2007qs}) by adjusting the external sources $J_{\phi_i}$ that couple linearly to the fields $\phi_i$ entering in the definition of the generating functional $Z[J]$. The coefficients $\Gamma_{\phi_n \cdots \phi_1}(x_1, \dots, x_n)$ designate the correlation (Green's) functions defined by:
\begin{equation}\begin{split}
    \Gamma_{\phi_n \cdots \phi_1}(x_1, \dots, x_n) &= \left.\frac{\delta^n \Gamma[\Phi]}{\delta \phi_n(x_n) \cdots \delta \phi_1(x_1)}\right|_{\phi_i=0} = -\imath \langle \phi_n(x_n) \cdots \phi_1(x_1) \rangle^\text{\,1PI}
	% \\ &\equiv -\imath \langle \Omega \vert \mathbb{T}[\phi_n(x_n) \cdots \phi_1(x_1)] \vert \Omega \rangle^\text{\,1PI} \, ,
    \, .
\end{split}\end{equation}
% where $\lvert \Omega \rangle$ is the interacting vacuum, and $\mathbb{T}[\cdot]$ is the time-ordered product.
Note that in a renormalized version of the quantum effective action, the coefficients $\Gamma_{\phi_n \cdots \phi_1}$ (thus, the associated 1PI correlation functions) would be finite. Note also that the order of the fields in the functional derivative matters in the case of anticommuting fields, so that $\Gamma_{\phi_n \cdots \phi_{i+1} \phi_i \cdots \phi_1}(x_1, \dots, x_n) = -\Gamma_{\phi_n \cdots \phi_i \phi_{i+1} \cdots \phi_1}(x_1, \dots, x_n)$ if $\phi_i$ anticommutes with $\phi_{i+1}$.

These formulae can be re-expressed in momentum space, via Fourier transform:
\begin{equation}
    \Gamma[\Phi] = \sum_{n \geq 2} \frac{1}{|n|!} \int \left(\prod_{i=1}^{n} \frac{\dInt[d]{p_i}}{(2\pi)^d} \widetilde{\phi_i}(p_i)\right) \widetilde{\Gamma}_{\phi_n \cdots \phi_1}(p_1, \dots, p_n) (2\pi)^d \delta^d({\textstyle \sum_{j=1}^{n} p_j}) \, ,
\end{equation}
where the tilde over the fields indicate that they have been Fourier-transformed. The coefficients $\widetilde{\Gamma}_{\phi_n \cdots \phi_1}(p_1, \dots, p_n)$ are the Green's functions in momentum space, with all the momenta taken to be incoming:
\begin{equation}\begin{gathered}
    \widetilde{\Gamma}_{\phi_n \cdots \phi_1}(p_1, \dots, p_n) (2\pi)^d \delta^d({\textstyle \sum_{j=1}^{n} p_j}) = (2\pi)^{d \times n} \left.\frac{\delta^n \Gamma[\Phi]}{\delta \widetilde{\phi_n}(p_n) \cdots \delta \widetilde{\phi_1}(p_1)}\right|_{\widetilde{\phi_i}=0}
    \, , \\
    \widetilde{\Gamma}_{\phi_n \cdots \phi_1}(p_1, \dots, p_n) \equiv -\imath \langle \widetilde{\phi_n}(p_n) \cdots \widetilde{\phi_1}(p_1) \rangle^\text{\,1PI}
    \, ,
\end{gathered}\end{equation}
and the delta-distribution ensures momentum conservation for these Green's functions (originating from their invariance under spatial translations, in coordinate space).
When there is no ambiguity, we adopt the shortened notation $\widetilde{\Gamma}_{\phi_n \cdots \phi_1}$ in place of $\widetilde{\Gamma}_{\phi_n \cdots \phi_1}(p_1, \dots, p_n)$.
Under these definitions, the evaluation of $\langle \phi_n \cdots \phi_1 \rangle^\text{\,1PI}$ is done using the standard diagrammatic method, and the Feynman rules for the vertex with ordered fields $\phi_1 \cdots \phi_n$ are given by the value of $\imath \widetilde{\Gamma}_{\phi_n \cdots \phi_1} = \langle \phi_n \cdots \phi_1 \rangle^\text{\,1PI}$.

An \emph{insertion} of a local field-operator $\mathcal{O}(x)$ in $\Gamma$, denoted by $\mathcal{O}(x) \cdot \Gamma$, is defined by the set of all Feynman diagrams where $\mathcal{O}(x)$ is inserted as an ``interaction vertex'', or equivalently by the generating functional (see Ref.\ \cite{Piguet:1980nr})
\begin{equation}
    \mathcal{O}(x) \cdot \Gamma[\Phi] = \sum_{n \geq 2} \frac{-\imath }{|n|!} \int \left(\prod_{i=1}^{n} \dInt[d]{x_i} \phi_i(x_i)\right) \langle \mathcal{O}(x) \phi_n(x_n) \cdots \phi_1(x_1) \rangle^\text{\,1PI} \, .
\end{equation}
The \emph{integrated insertion} $\mathcal{O} \cdot \Gamma$ is defined by
\begin{equation}
	\mathcal{O} \cdot \Gamma[\Phi] = \int \dInt[d]{x} \mathcal{O}(x) \cdot \Gamma[\Phi] \, ,
\end{equation}
and thus invariance under spatial translations will ensure momentum conservation at the ``vertex'' $\mathcal{O}$ in momentum space.

All the above relations are generic and may be interpreted both for
the theory with or without counterterms. Now we introduce specific
notation for regularized and (partially or fully) renormalized
quantities. In the context of DReg, the effective action is first
defined for $d\ne4$ and obtained from genuine loop diagrams and
diagrams involving counterterm insertions. At the 1-loop level we
use the notation $\Gamma^{(1)}$ for the effective action including
tree-level and genuine 1-loop contributions, but no counterterms; the
object $\Gamma^{(1)}_{\text{DReg}}$ contains also 1-loop
counterterms. Hence, we can write
\begin{subequations}
  \label{eq:definitionGammaDReg}
  \begin{align}
  \Gamma^{(1)} &=
  S_0+(\text{genuine 1PI 1-loop diagrams})
  \, ,\\
  \Gamma_\text{DReg}^{(1)} &=\Gamma^{(1)} + S_{\text{ct}}
  \, ,
\end{align}
\end{subequations}
where $S_0$ and $S_{\text{ct}}$ denote the tree-level and the 1-loop
counterterm action, respectively, and where the argument $[\Phi]$ is dropped.
All these quantities are still $\epsilon$-dependent and contain
evanescent objects. The quantity $\Gamma_\text{DReg}^{(1)}$ contains
counterterms, which by construction must cancel the UV $1/\epsilon$
divergences; hence this quantity allows the limit $\epsilon \to 0$.

The final, fully renormalized effective action at the 1-loop level is
then defined by taking the operation $\mathop{\text{LIM}}_{d \to 4}$
described in \cref{sect:DimReg}, i.e.\ by setting
$\epsilon = 0$ and neglecting all the evanescent objects:
\begin{equation}
\label{eq:definitionGammaRen}
  \Gamma_{\text{Ren}}^{(1)}[\varphi,\Phi,K_{\Phi},g_i,\xi,\mu] =
  \mathop{\text{LIM}}_{d \to 4}\Gamma_\text{DReg}^{(1)}[\varphi,\Phi,K_{\Phi},g_i,\xi,\mu] \, ,
\end{equation}
where in this equation we emphasised the fact that the effective action, both in the dimensional-regularized and the renormalized cases, depends on the fields, the external fields, the coupling constants of the theory, the gauge fixing parameter $\xi$ and the renormalization scale $\mu$.

\subsection{Calculation of the One-Loop Divergent Terms}

We present in this section the results of the divergent parts of the self-energies and vertices of the theory, evaluated at one-loop order.
In the following calculations, all momenta are taken incoming. The blobs shown in the diagrams represent the collection of the one-loop corrections not explicitly shown, that can be easily obtained diagrammatically via the standard methods.

\subsubsection{Self-energies}

\noindent
\uline{Scalar field}:
\raisebox{-20pt}{\includegraphics[scale=0.6]{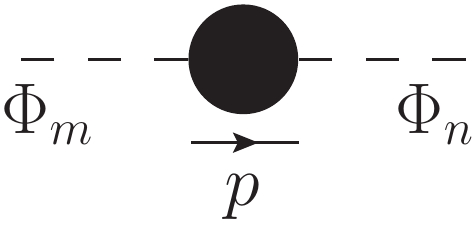}}
\begin{equation}
    \imath \widetilde{\Gamma}_{\Phi\Phi}^{nm}(p,-p)|^{(1)}_\text{div}
        = -\frac{\imath \hbar}{16 \pi^2 \epsilon} \left\{ (g^2 (3-\xi) C_2(S)) \delta^{mn} p^2 - Y_2(S) \delta^{mn} \overline{p}^2 - \frac{2 Y_2(S)}{3} \delta^{mn} \widehat{p}^2 \right\} \, .
\end{equation}

\noindent
\uline{Fermion field}: % -- Same in the L-model, with $\Proj{R} \to \Proj{L}$:
\raisebox{-20pt}{\includegraphics[scale=0.6]{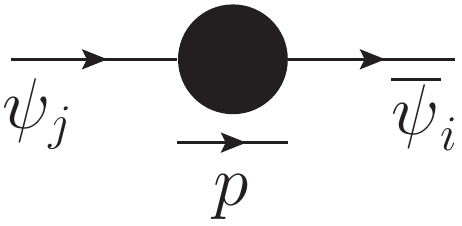}}
\begin{equation}
    \imath \widetilde{\Gamma}_{\psi \bar{\psi}}^{ji}(-p,p)|^{(1)}_\text{div} = \frac{\imath \hbar}{16 \pi^2 \epsilon} \left(g^2 \xi C_2(R) + \frac{Y_2(R)}{2}\right) \delta^{ij} \;\overline{\centernot{p}}\; \Proj{R} \, ,
\end{equation}
and for the charge-conjugated fermion field:
\begin{equation}
	\imath \widetilde{\Gamma}_{\psi^C \bar{\psi}^C}^{ji}(-p,p)|^{(1)}_\text{div} = \frac{\imath \hbar}{16 \pi^2 \epsilon} \left(g^2 \xi C_2(R) + \frac{Y_2(R)}{2}\right) \delta^{ij} \;\overline{\centernot{p}}\; \Proj{L} \, .
\end{equation}

\noindent
\uline{Gauge boson}: % -- Same in the L-model, with $R \to L$:
\raisebox{-20pt}{\includegraphics[scale=0.6]{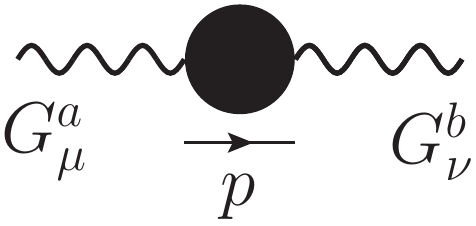}}
\begin{equation}\begin{split}
\label{eq:QGBSE}
    \imath \widetilde{\Gamma}_{GG}^{ba,\nu\mu}(p,-p)|^{(1)}_\text{div} =\;&
          -\frac{\imath \hbar g^2}{16 \pi^2 \epsilon} \frac{(13-3\xi) C_2(G) - S_2(S)}{6} \delta^{ab} (p^\mu p^\nu - p^2 g^{\mu\nu}) \\
        & + \frac{\imath \hbar g^2}{16 \pi^2 \epsilon} \frac{2 S_2(R)}{3} \delta^{ab} (\overline{p}^\mu \overline{p}^\nu - \overline{p}^2 \overline{g}^{\mu\nu}) - \frac{\imath \hbar g^2}{16 \pi^2 \epsilon} \frac{S_2(R)}{3} \delta^{ab} \widehat{p}^2 \overline{g}^{\mu\nu}
        \, .
\end{split}\end{equation}

\noindent
\uline{Ghost field}:
% independent of the choice of $\Proj{L} \gamma^\mu \Proj{R}$ or $\gamma^\mu \Proj{R}$, since it does not have any fermion loop with fermion-gauge boson vertex
\raisebox{-20pt}{\includegraphics[scale=0.6]{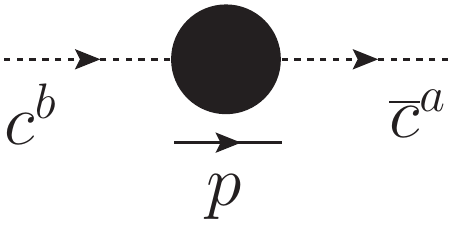}}
\begin{equation}
    \imath \widetilde{\Gamma}_{c\bar{c}}^{ba}(-p,p)|^{(1)}_\text{div} = -\frac{\imath \hbar g^2}{16 \pi^2 \epsilon} \frac{3-\xi}{4} C_2(G) \delta^{ab} p^2 \, .
\end{equation}

\subsubsection{Standard Vertices}

\noindent
\uline{Yukawa vertex}:\\
% independent of the choice of $\Proj{L} \gamma^\mu \Proj{R}$ or $\gamma^\mu \Proj{R}$:\\
\raisebox{-20pt}{\includegraphics[scale=0.6]{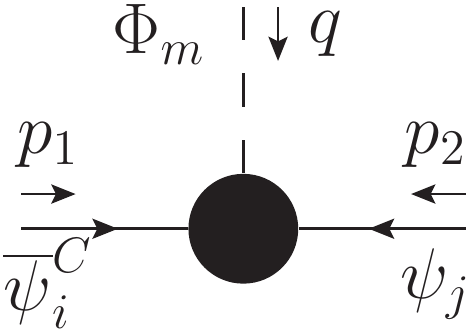}}
% $\xi_i(p_1) \to \xi_j(p_2) \Phi_m(q=-p_1-p_2)$
\begin{equation}\begin{split}
    \imath \widetilde{\Gamma}_{\psi \overline{\psi^C} \Phi}^{ji,m}|^{(1)}_\text{div}
        &= \frac{\imath \hbar}{16 \pi^2 \epsilon} \left( Y_R^n (Y_R^m)^* Y_R^n - g^2 \xi C_2(S) Y_R^m - g^2 (3+\xi) {T_{\overline{R}}}^a Y_R^m {T_R}^a \right)_{ij} \Proj{R} \\
        &= \frac{\imath \hbar}{16 \pi^2 \epsilon} \left( Y_R^n (Y_R^m)^* Y_R^n - g^2 \frac{2 C_2(R) (3+\xi) - C_2(S) (3-\xi)}{2} Y_R^m \right)_{ij} \Proj{R} \, ,
\end{split}\end{equation}
where the last line is obtained by evaluating $({T_{\overline{R}}}^a Y_R^m {T_R}^a)_{ij}$, using \cref{eq:GaugeInvarYuk}: $({T_{\overline{R}}}^a Y_R^m {T_R}^a)_{ij} = (C_2(R) - C_2(S)/2) (Y_R)^m_{ij}$.
\\
\\
\noindent
\raisebox{-20pt}{\includegraphics[scale=0.6]{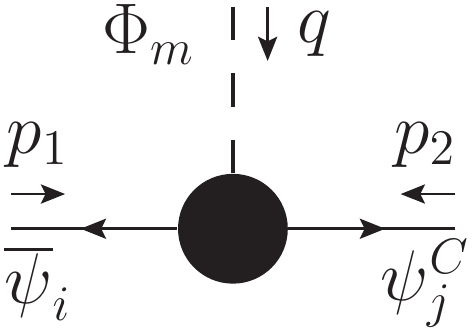}}
% $\bar{\xi}_i(p_1) \to \bar{\xi}_j(p_2) \Phi_m(q=-p_1-p_2)$
\begin{equation}
    \imath \widetilde{\Gamma}_{\psi^C \bar{\psi} \Phi}^{ji,m}|^{(1)}_\text{div}
        = \frac{\imath \hbar}{16 \pi^2 \epsilon} \left( (Y_R^n)^* Y_R^m (Y_R^n)^* - g^2 \frac{2 C_2(R) (3+\xi) - C_2(S) (3-\xi)}{2} (Y_R^m)^* \right)_{ij} \Proj{L} \, .
\end{equation}

\noindent
\uline{Fermion-gauge boson interaction}:\\
\raisebox{-20pt}{\includegraphics[scale=0.6]{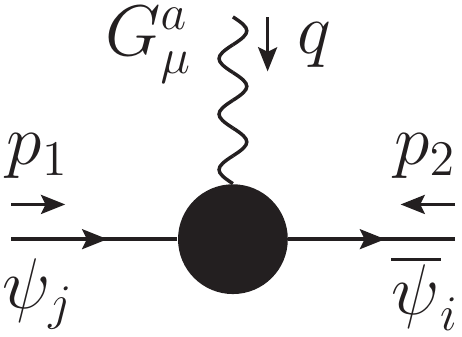}}
% $\xi_i(p_1) \to \xi_j(p_2) G^a_\mu(q=-p_1-p_2)$
\begin{equation}
    \imath \widetilde{\Gamma}_{\psi \bar{\psi} G}^{ji,a,\mu}|^{(1)}_\text{div} = \frac{\imath \hbar g}{16 \pi^2 \epsilon} \left(g^2 \frac{(3+\xi) C_2(G) + 4 \xi C_2(R)}{4} + \frac{Y_2(R)}{2}\right) {T_R}^a_{ij} \overline{\gamma}^\mu \Proj{R} \, .
\end{equation}

\noindent
\uline{$\Phi\Phi G$ Scalar-gauge boson interaction}:\\
\begin{minipage}{0.3\textwidth}
	\centering
	\includegraphics[scale=0.6]{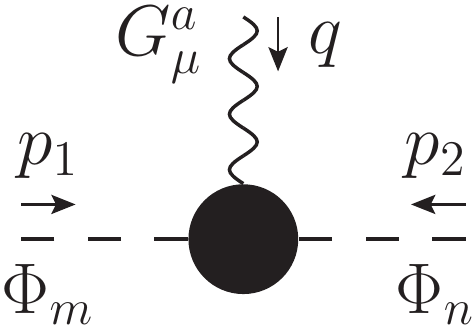}
\end{minipage}
\hfill
\begin{minipage}{0.7\textwidth}
	$+ (p_1,m) \leftrightarrow (p_2,n)$ permutation.
\end{minipage}
% $\Phi_m(p_1) \to \Phi_n(p_2) G^a_\mu(q=-p_1-p_2)$
\begin{multline}
	\imath \widetilde{\Gamma}_{\Phi\Phi G}^{nm,a,\mu}(q=-p_1-p_2,p_1,p_2)|^{(1)}_\text{div} = \\
		\frac{\imath \hbar g^3}{16 \pi^2 \epsilon} \left(\frac{3+\xi}{4} C_2(G) - (3-\xi) C_2(S)\right) \theta^a_{nm} (p_1-p_2)^\mu
		+ \frac{\imath \hbar g}{16 \pi^2 \epsilon} Y_2(S) \theta^a_{nm} \overline{(p_1-p_2)}^\mu
	\, .
\end{multline}

\noindent
\uline{Ghost-gauge boson interaction}:\\ % independent of the choice of $\Proj{L} \gamma^\mu \Proj{R}$ or $\gamma^\mu \Proj{R}$:\\
\raisebox{-20pt}{\includegraphics[scale=0.6]{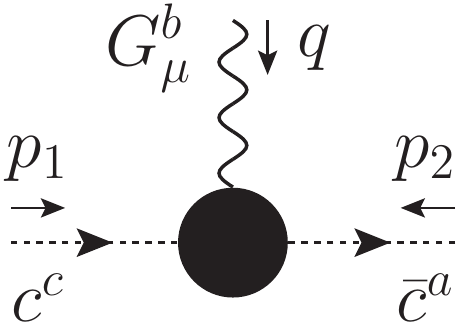}}
% $c_c(p_1) \to \bar{c}_a(p_2) G^b_\mu(q=-p_1-p_2)$
\begin{equation}
    \imath \widetilde{\Gamma}_{c G \bar{c}}^{cba}(p_2,q=-p_1-p_2,p_1)|^{(1)}_\text{div} = \frac{\hbar g^3}{16 \pi^2 \epsilon} \frac{\xi C_2(G)}{2} f^{abc} p_2^\mu \, .
\end{equation}

\noindent
\uline{Triple gauge boson vertex}:\\
\begin{minipage}{0.3\textwidth}
	\centering
	\includegraphics[scale=0.6]{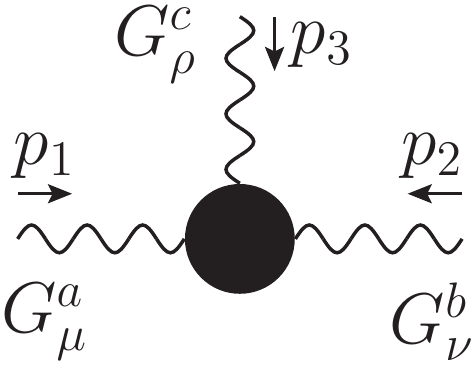}
\end{minipage}
\hfill
\begin{minipage}{0.7\textwidth}
	$+ \{ (p_1,\mu,a) \, , \, (p_2,\nu,b) \, , \, (p_3,\rho,c) \}$ permutations.
\end{minipage}
% $G^a_\mu(p_1) \to G^b_\nu(p_2) G^c_\rho(p_3=-p_1-p_2)$
\begin{multline}
	\imath \widetilde{\Gamma}_{GGG}^{cba,\rho\nu\mu}(p_1,p_2,p_3=-p_1-p_2)|^{(1)}_\text{div} = \\
		\frac{-\hbar g^3}{16 \pi^2 \epsilon} f^{abc} \frac{(17-9\xi) C_2(G) - 2 S_2(S)}{12}((p_2-p_3)^\mu g^{\nu\rho} + (p_3-p_1)^\nu g^{\mu\rho} + (p_1-p_2)^\rho g^{\mu\nu})
		\\
        + \frac{\hbar g^3}{16 \pi^2 \epsilon} f^{abc} \frac{2 S_2(R)}{3} (\overline{(p_2-p_3)}^\mu \overline{g}^{\nu\rho} + \overline{(p_3-p_1)}^\nu \overline{g}^{\mu\rho} + \overline{(p_1-p_2)}^\rho \overline{g}^{\mu\nu})
        \, .
\end{multline}

\noindent
\uline{Quartic gauge boson vertex}:\\
\begin{minipage}{0.3\textwidth}
	\centering
	\includegraphics[scale=0.6]{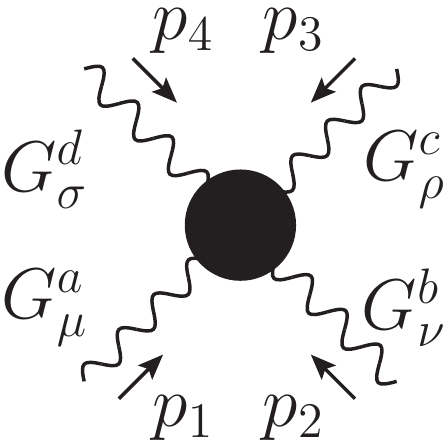}
\end{minipage}
\hfill
\begin{minipage}{0.7\textwidth}
	$+ \{ (p_1,\mu,a) \, , \, (p_2,\nu,b) \, , \, (p_3,\rho,c) \, , \, (p_4,\sigma,d) \}$ permutations.
\end{minipage}
% $G^a_\mu(p_1) G^b_\nu(p_2) \to G^c_\rho(p_3) G^d_\sigma(p_4=-p_1-p_2-p_3)$
\begin{multline}
    \imath \widetilde{\Gamma}_{GGGG}^{abcd,\mu\nu\rho\sigma}|^{(1)}_\text{div} = \\
		\frac{\imath \hbar g^4}{16 \pi^2 \epsilon} \frac{2 (2-3\xi) C_2(G) - S_2(S)}{6}
			\begin{pmatrix}[c*{2}{!{,}c}] g_{\mu\nu} g_{\rho\sigma} & g_{\mu\rho} g_{\nu\sigma} & g_{\mu\sigma} g_{\nu\rho} \end{pmatrix} \cdot
			\begin{pmatrix} f^{eac} f^{ebd} + f^{ead} f^{ebc} \\ f^{eab} f^{ecd} + f^{ead} f^{ecb} \\ f^{eab} f^{edc} + f^{eac} f^{edb} \end{pmatrix}
		\\
		- \frac{\imath \hbar g^4}{16 \pi^2 \epsilon} \frac{2 S_2(R)}{3}
			\begin{pmatrix}[c*{2}{!{,}c}] \overline{g}_{\mu\nu} \overline{g}_{\rho\sigma} & \overline{g}_{\mu\rho} \overline{g}_{\nu\sigma} & \overline{g}_{\mu\sigma} \overline{g}_{\nu\rho} \end{pmatrix} \cdot
			\begin{pmatrix} f^{eac} f^{ebd} + f^{ead} f^{ebc} \\ f^{eab} f^{ecd} + f^{ead} f^{ecb} \\ f^{eab} f^{edc} + f^{eac} f^{edb} \end{pmatrix}
		\, .
\end{multline}
We employed here a matrix-like ``scalar product'' to express in a compact form the result and to indicate how the Lorentz tensors are associated with the corresponding group structures.

\noindent
\uline{Tadpoles, and interactions with an odd number of scalar fields:} For triple scalar vertex, scalar-gauge boson vertices with one or three scalar fields, at one-loop the only possibility is that all the scalar fields are connected to a single internal fermion loop; since we are studying a massless theory these contributions vanish. The same reason also apply for tadpoles in Dimensional Regularization.

\noindent
\uline{$\Phi\Phi GG$ Scalar-gauge boson interaction}:\\
\begin{minipage}{0.3\textwidth}
	\centering
	\includegraphics[scale=0.6]{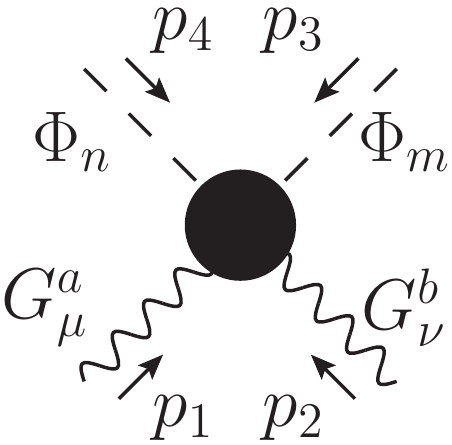}
\end{minipage}
\hfill
\begin{minipage}{0.7\textwidth}
	$+ \{ (p_1,\mu,a) \, , \, (p_2,\nu,b) \}$ and $\{ (p_3,m) \, , \, (p_4,n) \}$ permutations.
\end{minipage}
%%% $\Phi_m(p_1) \Phi_n(p_2) \to G^a_\mu(p_3) G^b_\nu(p_4=-p_1-p_2-p_3)$
% $G^a_\mu(p_1) G^b_\nu(p_2) \to \Phi_m(p_3) \Phi_n(p_4=-p_1-p_2-p_3)$
\begin{equation}\begin{split}
    \imath \widetilde{\Gamma}_{\Phi\Phi GG}^{mnab,\mu\nu}|^{(1)}_\text{div} =\;& \frac{\imath \hbar g^4}{16 \pi^2 \epsilon} \left( \frac{3+\xi}{2} C_2(G) - (3-\xi) C_2(S) \right) \{\theta^a, \theta^b\}_{mn} g_{\mu\nu} \\
        & + \frac{\imath \hbar}{16 \pi^2 \epsilon} Y_2(S) g^2 \{\theta^a, \theta^b\}_{mn} \overline{g}_{\mu\nu}
    \, .
\end{split}\end{equation}

\noindent
\uline{Quartic scalar vertex}:\\
\begin{minipage}{0.3\textwidth}
	\centering
	\includegraphics[scale=0.6]{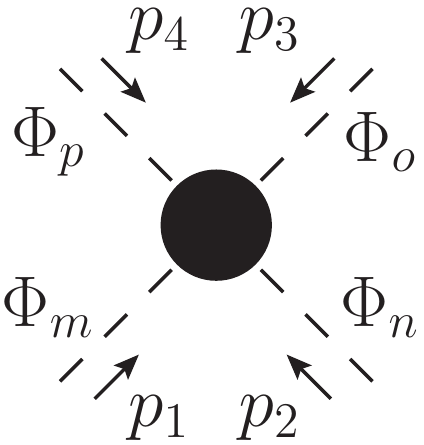}
\end{minipage}
\hfill
\begin{minipage}{0.7\textwidth}
	$+ \{ (p_1,m) \, , \, (p_2,n) \, , \, (p_3,o) \, , \, (p_4,p) \}$ permutations.
\end{minipage}
% $\Phi_m(p_1) \Phi_n(p_2) \to \Phi_o(p_3) \Phi_p(p_4=-p_1-p_2-p_3)$
\begin{equation}
    \imath \widetilde{\Gamma}_{\Phi\Phi\Phi\Phi}^{mnop}|^{(1)}_\text{div} =
    \frac{\imath \hbar}{16 \pi^2 \epsilon} \frac{1}{2} (3 g^4 A - g^2 \xi \Lambda^S - 4 H + \Lambda^2)_{mnop}
    \, ,
\end{equation}
using the following group invariants, as defined by Eqs.~(2.16), (2.17), (2.18) and (2.19) in \cite{Machacek:1984zw} % 3rd article from Machacek \& Vaughn
and employing the same conventions:
\begin{equation}
\label{eq:quartic_scalar_invariants}
\begin{aligned}
    A_{mnop} &= \frac{1}{8} \sum_\text{perms} \{\theta^a, \theta^b\}_{mn} \{\theta^a, \theta^b\}_{op} \, , &
    H_{mnop} &= \frac{1}{4} \sum_\text{perms} \Tr{Y_R^m Y_R^{\dagger\;n} Y_R^o Y_R^{\dagger\;p}} \, , \\
    \Lambda^2_{mnop} &= \frac{1}{8} \sum_\text{perms} \lambda_{mnqr} \lambda_{qrop} \, , &
    \Lambda^S_{mnop} &= \lambda_{mnop} \sum_{k=m,n,o,p} C_2(k) \, ,
\end{aligned}
\end{equation}
where in the definition of $\Lambda^S_{mnop}$ the sum is performed on each scalar line represented by the index $k$, and $C_2(k)$ is the eigenvalue of the Casimir operator $(\theta^a \theta^a)_{mn}$ for the scalar representation of line $k$. In our case the scalar fields are in the same scalar (and irreducible) representation, therefore we have $\Lambda^S_{mnop} = 4 C_2(S) \lambda_{mnop}$.

\subsubsection{Vertices with External BRST Sources}

We provide here the explicit list of Feynman diagrams necessary to evaluate the Green's functions at one-loop, since these are not conventional ones as they contain BRST-source-vertex insertions necessary for this formalism.

\noindent
\uline{From $\rho^a_\mu s_d{G^a_\mu}$:} there exist two different Green's functions involving this insertion, whose divergent parts are:\\
\begin{minipage}{0.3\textwidth}
	\centering
	\includegraphics[scale=0.6]{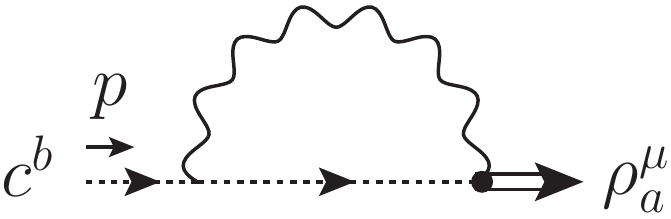}
\end{minipage}
\hfill
\begin{minipage}{0.7\textwidth}
% $c_b(p) \to \rho^a_\mu(-p)$
\begin{equation}
    \imath \widetilde{\Gamma}_{c \rho}^{ba,\mu}(-p,p)|^{(1)}_\text{div} = -\frac{\hbar g^2}{16 \pi^2 \epsilon} \frac{3-\xi}{4} C_2(G) \delta^{ab} p^\mu \, ,
\end{equation}
\end{minipage}

\noindent
\begin{minipage}{0.3\textwidth}
	\centering
	\includegraphics[scale=0.6]{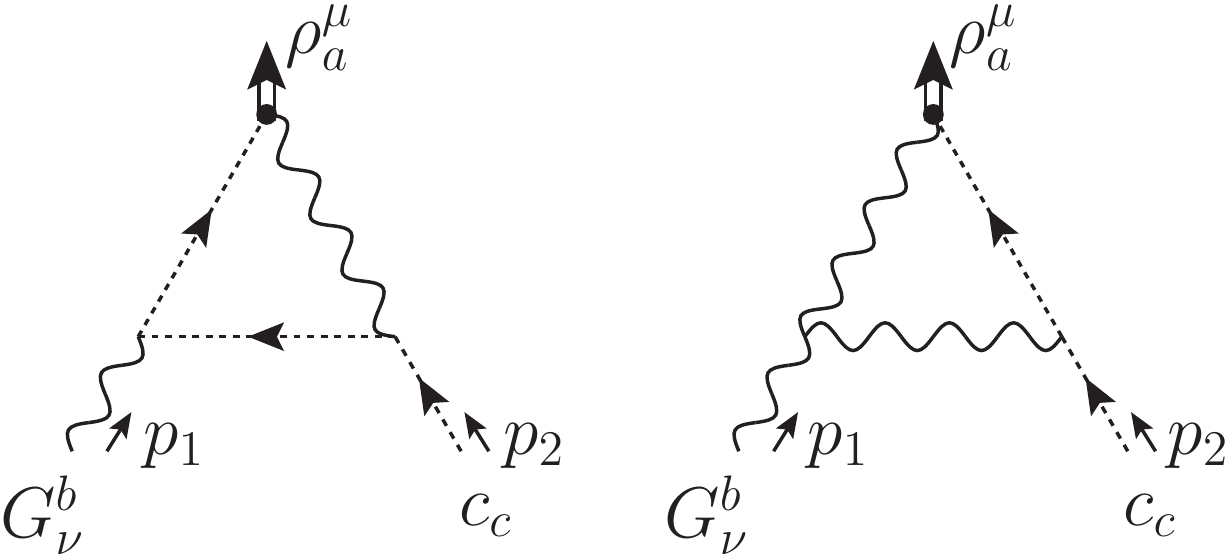}
\end{minipage}
\hfill
\begin{minipage}{0.7\textwidth}
% $G^b_\mu(p) c_c(-p-q) \to \rho^a_\mu(q)$
\begin{equation}
    \imath \widetilde{\Gamma}_{c G \rho}^{cba,\nu\mu}|^{(1)}_\text{div} = \frac{\imath \hbar g^3}{16 \pi^2 \epsilon} \frac{\xi C_2(G)}{2} f^{abc} g^{\mu\nu} \, .
\end{equation}
\end{minipage}

\noindent
\uline{From $\zeta^a s_d{c^a}$:}\\
\begin{minipage}{0.3\textwidth}
	\centering
	\includegraphics[scale=0.6]{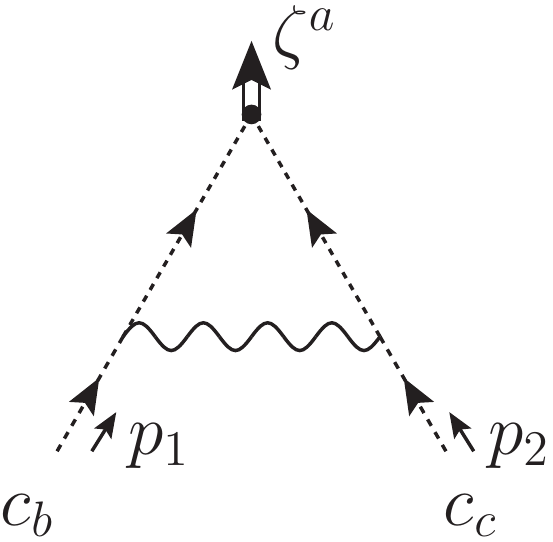}
\end{minipage}
\hfill
\begin{minipage}{0.7\textwidth}
% $c_b(p) c_c(-p-q) \to \zeta^a(q)$
\begin{equation}
    \imath \widetilde{\Gamma}_{c c \zeta}^{cba}|^{(1)}_\text{div} = -\frac{\imath \hbar g^3}{16 \pi^2 \epsilon} \frac{\xi C_2(G)}{2} f^{abc} \, ,
\end{equation}
\end{minipage}
where we accounted for the diagram's symmetry factor $= 2$ due to the fact there are two interchangeable vertices -- the $(\bar{c} G c)$ vertices -- leaving the diagram invariant.

\noindent
\uline{From $\bar{R}_i s_d{\psi_i}$:}\\
\begin{minipage}{0.3\textwidth}
	\centering
	\includegraphics[scale=0.6]{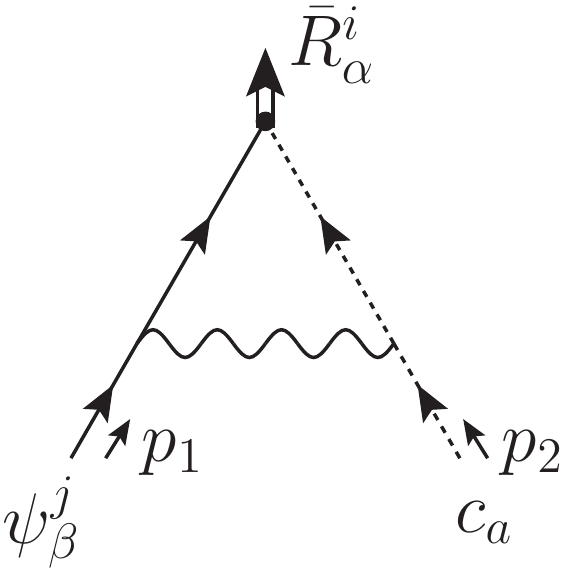}
\end{minipage}
\hfill
\begin{minipage}{0.7\textwidth}
% $\psi_{j,\beta}(p) c_a(-p-q) \to \bar{R}_{i,\alpha}(q)$
\begin{equation}
    \imath \widetilde{\Gamma}_{\psi c \bar{R}}^{jai,\beta\alpha}|^{(1)}_\text{div} = -\frac{\hbar g^3}{16 \pi^2 \epsilon} \frac{\xi C_2(G)}{2} {T_R}^a_{ij} \Proj{R}_{\alpha\beta} \, .
\end{equation}
\end{minipage}

\noindent
\uline{From $s_d{\bar{\psi}_i} R_i \equiv R_i s_d{\bar{\psi}_i}$:}\\
\begin{minipage}{0.3\textwidth}
	\centering
	\includegraphics[scale=0.6]{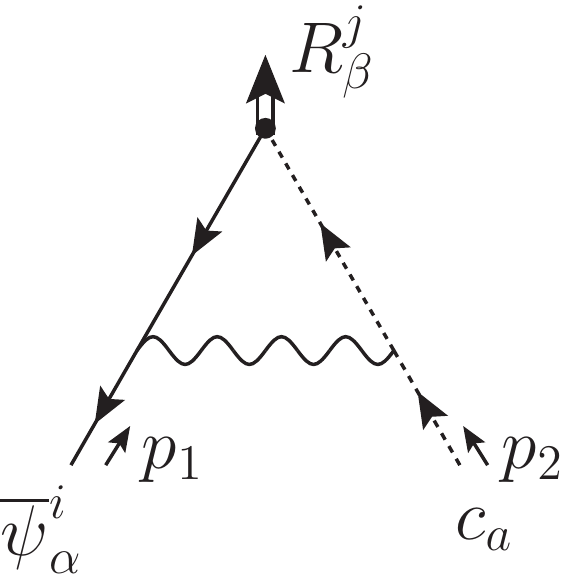}
\end{minipage}
\hfill
\begin{minipage}{0.7\textwidth}
% $\bar{\psi}_{i,\alpha}(p) c_a(-p-q) \to R_{j,\beta}(q)$
\begin{equation}
    \imath \widetilde{\Gamma}_{R c \bar{\psi}}^{jai,\beta\alpha}|^{(1)}_\text{div} = -\frac{\hbar g^3}{16 \pi^2 \epsilon} \frac{\xi C_2(G)}{2} {T_R}^a_{ij} \Proj{L}_{\alpha\beta} \, .
\end{equation}
\end{minipage}

\noindent
\uline{From $\mathcal{Y}_m s_d{\Phi_m}$:}\\
\begin{minipage}{0.3\textwidth}
	\centering
	\includegraphics[scale=0.6]{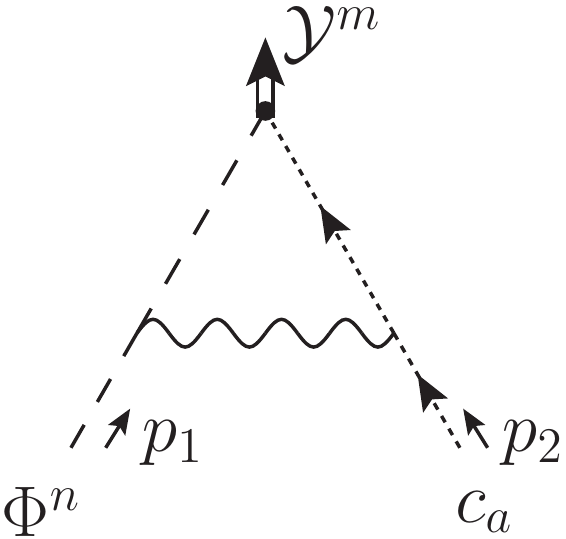}
\end{minipage}
\hfill
\begin{minipage}{0.7\textwidth}
% $\Phi_n(p) c_a(-p-q) \to \mathcal{Y}_m(q)$
\begin{equation}
    \imath \widetilde{\Gamma}_{\Phi c \mathcal{Y}}^{nam}|^{(1)}_\text{div} = -\frac{\hbar g^3}{16 \pi^2 \epsilon} \frac{\xi C_2(G)}{2} \theta^a_{mn} \, .
\end{equation}
\end{minipage}

\subsection{The One-Loop Singular Counterterm Action $S_\text{sct}^{(1)}$}

After computing all UV divergent one-loop Feynman diagrams, we can determine
the singular one-loop counterterm action.
It is defined such that the divergent parts of the one-loop vertices cancel:
\begin{equation}
	S_\text{sct}^{(1)} = -\Gamma|^{(1)}_\text{div} \, .
\end{equation}
Since it is the first main result of the present paper we present it
in two different ways. First, we provide the contributions with and
without scalar fields separately,
\begin{equation}
\label{eq:SingularCT1Loop}
	S_\text{sct}^{(1)} = S_\text{sct}^{(1)\,\substack{\text{No Scalar} \\ \text{contrib.}}} + S_\text{sct}^{(1)\,\substack{\text{Scalar} \\ \text{contrib.}}}
	\, ,
\end{equation}
where $S_\text{sct}^{(1)\,\substack{\text{No Scalar} \\ \text{contrib.}}}$ represents the terms without any contribution from the scalar fields, and agrees with Eq.~(37) of \cite{Martin:1999cc}, and reads:
\begin{equation}
\label{eq:SingularCT1Loop_NoScalarContrib}
\begin{split}
	S_\text{sct}^{(1)\,\substack{\text{No Scalar} \\ \text{contrib.}}} =\;&
		\frac{\hbar g^2}{16 \pi^2 \epsilon} \left\{
		\frac{13-3\xi}{6} C_2(G) S_{GG} + \frac{17-9\xi}{12} C_2(G) S_{GGG} + \frac{2-3\xi}{3} C_2(G) S_{GGGG} \right.\\
		&\left. - \frac{2 S_2(R)}{3} (\overline{S_{GG}} + \overline{S_{GGG}} + \overline{S_{GGGG}})
		- \xi C_2(R) (\overline{S_{\bar{\psi}\psi_R}} + \overline{S_{\overline{\psi} G \psi_R}})
		- \frac{3+\xi}{4} C_2(G) \overline{S_{\overline{\psi} G \psi_R}} \right.\\
		&\left. + \frac{3-\xi}{4} C_2(G) \left( S_{\bar{c} c} + S_{\rho c} \right)
		        - \frac{\xi C_2(G)}{2} \left( S_{\bar{c} G c} + S_{\rho G c} + S_{\zeta c c} + S_{\bar{R} c \psi_R} + S_{R c \overline{\psi_R}} \right) \right\} \\
		&- \frac{\hbar g^2}{16 \pi^2 \epsilon} \frac{S_2(R)}{3} \int \dInt[d]{x} \frac{1}{2} \bar{G}^{a\,\mu} \widehat{\partial}^2 \bar{G}^a_\mu
	\, .
\end{split}\end{equation}
The counterterm action $S_\text{sct}^{(1)\,\substack{\text{Scalar} \\ \text{contrib.}}}$
represents the terms generated from the scalar contributions, and reads:
\begin{equation}
\label{eq:SingularCT1Loop_ScalarContrib}
\begin{split}
	S_\text{sct}^{(1)\,\substack{\text{Scalar} \\ \text{contrib.}}} =\;&
		\frac{\hbar}{16 \pi^2 \epsilon} \left\{
		- g^2 \frac{S_2(S)}{6} (S_{GG} + S_{GGG} + S_{GGGG})
		- \frac{Y_2(R)}{2} \left( \overline{S_{\bar{\psi}\psi_R}} + \overline{S_{\overline{\psi} G \psi_R}} \right) \right.\\
		&\left. + g^2 (3-\xi) C_2(S) \left( S_{\Phi\Phi} + S_{\Phi G \Phi} + S_{\Phi G G \Phi} \right) - g^2 \frac{3+\xi}{4} C_2(G) \left( S_{\Phi G \Phi} + 2 S_{\Phi G G \Phi} \right) \right.\\
		&\left. - Y_2(S) \left( \overline{S_{\Phi\Phi}} + \overline{S_{\Phi G \Phi}} + \overline{S_{\Phi G G \Phi}} \right)
		+ \frac{1}{2} (3 g^4 A - g^2 \xi \Lambda^S - 4 H + \Lambda^2)_{mnop} S_{\Phi^4_{mnop}} \right.\\
		&\left. + \left( Y_R^n (Y_R^m)^* Y_R^n - g^2 \frac{2 C_2(R) (3+\xi) - C_2(S) (3-\xi)}{2} Y_R^m \right)_{ij} S_{\overline{\psi_R}^C_i \Phi^m {\psi_R}_j} + \hc \right.\\
		&\left. - g^2 \frac{\xi C_2(G)}{2} S_{\mathcal{Y} c \Phi} \right\}
		- \frac{\hbar}{16 \pi^2 \epsilon} \frac{2 Y_2(S)}{3} \widehat{S_{\Phi\Phi}}
	\, .
\end{split}\end{equation}
It contains both additional contributions to the operators without
scalar fields and contributions to additional operators involving
scalar fields.
In both equations the monomials introduced in \cref{eq:RModelDReg_Action} %\cref{eq:S0Def3}
have been used; a bar such as in $\overline{S_{GG}}$ corresponds to taking all Lorentz
indices in the respective monomial only in purely 4 dimensions; a hat
such as in $\widehat{S_{\Phi\Phi}}$ corresponds to taking all Lorentz
indices purely in $d-4$ dimensions.
Using again the condensed notation $\int_x \equiv \int \dInt[d]{x}$, the new object $\overline{S_{\bar{\psi}\psi_R}} = \int_x \imath \overline{\psi}_i \overline{\slashed{\partial}} \Proj{R} \psi_i \equiv \int_x \frac{\imath}{2} \overline{\psi}_i \overset{\leftrightarrow}{\overline{\slashed{\partial}}} \Proj{R} \psi_i$
corresponds to the 4-dimensional kinetic term of the purely right-handed fermion.
It \emph{differs from its $d$-dimensional equivalent} $S_{\overline{\psi}\psi}$.
Its appearance can be interpreted as the fact that \emph{only} the right-handed fermion component renormalizes, while the fictitious left-handed component required to properly extend the 4-dimensional chiral fermion kinetic term to $d$ dimensions, see \cref{subsect:RModelDReg}, does not renormalize. This is understandable since all fermion interaction vertices in the model are explicitly chiral (contain the right-handed projector $\Proj{R}$), thus any fermion propagator connecting such vertices get their extra left-handed component projected out. Any loop correction to a fermion propagator contains at least one such vertex connected to the fermion line, therefore such correction will only contribute to the renormalization of the right-handed part of the fermion kinetic term.

In addition to the explicit evanescent operator in the last line of
\cref{eq:SingularCT1Loop_NoScalarContrib}, generating the Feynman rule $-\imath \widehat{p}^2 \overline{g}_{\mu\nu} \delta^{ab}$, we obtain an additional evanescent operator $\widehat{S_{\Phi\Phi}} = -1/2 \int_x \Phi^m \widehat{\partial}^2 \Phi^m$ from the scalar sector, generating the Feynman rule $\imath \widehat{p}^2 \delta^{mn}$.
We observe that, should we have used instead another $d$-dimensional choice for the fermion-gauge interaction term with a $\gamma^\mu \Proj{R}$, we would have obtained \emph{many more} evanescent operators.

We can re-express the result for the singular counterterms in the
structure announced in \cref{sect:standardrenormalizationstructure}
and make contact to the usual renormalization transformation.
The sum of the singular counterterms can be written as
\begin{equation}
\label{eq:Ssctinvevan}
	S_\text{sct}^{(1)} = S_{\text{sct,inv}}^{(1)} + S_{\text{sct,evan}}^{(1)} \, ,
\end{equation}
where the first term arises from renormalization transformation as in \cref{eq:Sctinv}
and is given by \cref{eq:Sctinvstructure}:
\begin{align*}
%\label{eq:Sctinvstructure}
\begin{split}
    S_{\text{ct,inv}} =\;&
    \frac{\delta Z_G}{2} L_G +
    \frac{\delta Z_{\psi_R}}{2} \overline{L_{\psi_R}} +
    \frac{\delta Z_\Phi}{2} L_\Phi +
    \frac{\delta Z_c}{2} L_c
    \\
    &+
    \frac{\delta g}{g} L_g +
    \big( \delta (Y_R)^m_{ij} {L_{Y_R}}^m_{ij} + \hc \big) +
    \delta \lambda^{mnop} L_{\lambda^{mnop}} \, ,
\end{split}
\end{align*}
while the second term contains purely evanescent quantities.
The renormalization constants needed in \cref{eq:rentransform} agree with the
usual ones (see e.g. \cite{Machacek:1983tz,Machacek:1983fi,Machacek:1984zw})
and read
\begin{align}
\label{eq:RenConstFirst}
\delta Z_G^{(1)} &=
    \frac{\hbar}{16 \pi^2 \epsilon} g^2 \frac{(13-3\xi) C_2(G) - 4 S_2(R) - S_2(S)}{6}
\, ,
\\
\delta Z_{\psi_R}^{(1)} &=
    \frac{-\hbar}{16 \pi^2 \epsilon} \left(g^2 \xi C_2(R) + \frac{Y_2(R)}{2}\right)
\, ,
\\
\delta Z_{\Phi}^{(1)} &=
    \frac{\hbar}{16 \pi^2 \epsilon} \left(g^2 (3-\xi) C_2(S) - Y_2(S)\right)
\, ,
\\
\delta Z_c^{(1)} &= 2 \delta Z_{\rho c}^{(1)} + \delta Z_G^{(1)}
    = \frac{\hbar}{16 \pi^2 \epsilon} g^2 \frac{(22-6\xi) C_2(G) - 4 S_2(R) - S_2(S)}{6} \, ,
\intertext{where $\delta Z_{\rho c}^{(1)}$ is the coefficient of $S_{\rho c}$ in $S_\text{sct}^{(1)}$:}
    \delta Z_{\rho c}^{(1)} &\equiv \frac{\hbar}{16 \pi^2 \epsilon} g^2 \frac{3-\xi}{4} C_2(G) \, ; \nonumber
\\
\delta g^{(1)}/g &=
    \frac{-\hbar}{16 \pi^2 \epsilon} g^2 \frac{22 C_2(G) - 4 S_2(R) - S_2(S)}{12}
\, ,
\\
\delta (Y_R)^{m,(1)}_{ij} &=
    \delta Z_{Y,ij}^{m,(1)} - (\delta Z_{\psi_R}^{(1)} + \delta Z_{\Phi}^{(1)}/2) (Y_R)^m_{ij} \, ,
\intertext{where $\delta Z_{Y,ij}^{m,(1)}$ is the coefficient of $S_{\overline{\psi_R}^C_i \Phi^m {\psi_R}_j}$ in $S_\text{sct}^{(1)}$:}
    \delta Z_{Y,ij}^{m,(1)} &\equiv \frac{\hbar}{16 \pi^2 \epsilon} \left( (Y_R^n (Y_R^m)^* Y_R^n) -g^2 \frac{2 C_2(R) (3+\xi) - C_2(S) (3-\xi)}{2} Y_R^m \right)_{ij}
    \, ; \nonumber
\\
\label{eq:RenConstLast}
\delta \lambda_{mnop}^{(1)} &=
    \delta Z_{4\Phi,mnop}^{(1)} - 2 \delta Z_{\Phi}^{(1)} \lambda_{mnop} \, ,
\intertext{where $\delta Z_{4\Phi,mnop}^{(1)}$ is the coefficient of $S_{\Phi^4_{mnop}}$ in $S_\text{sct}^{(1)}$:}
    \delta Z_{4\Phi,mnop}^{(1)} &\equiv \frac{\hbar}{16 \pi^2 \epsilon} \frac{1}{2} (3 g^4 A - g^2 \xi \Lambda^S - 4 H + \Lambda^2)_{mnop}
    \, . \nonumber
\end{align}
The evanescent counterterms appearing in \cref{eq:Ssctinvevan} can be written as
\begin{equation}
\label{eq:Ssctevan}
\begin{split}
    S_{\text{sct,evan}}^{(1)} =\;&
        \frac{-\hbar}{16 \pi^2 \epsilon} \left\{ g^2 \frac{S_2(R)}{3} \left( 2 (\widetilde{S}_{GG} + \widetilde{S}_{GGG} + \widetilde{S}_{GGGG}) + \int \dInt[d]{x} \frac{1}{2} \bar{G}^{a\,\mu} \widehat{\partial}^2 \bar{G}^a_\mu \right) \right.\\
		&\left. + Y_2(S) \left( (\widetilde{S}_{\Phi\Phi} + \widetilde{S}_{\Phi G \Phi} + \widetilde{S}_{\Phi GG \Phi}) + \frac{2}{3} \widehat{S_{\Phi\Phi}} \right) \right\}
    \, ,
\end{split}\end{equation}
with
\begin{equation}
\label{eq:SsctevanTildeOps}
	\widetilde{S}_\mathcal{O} = \overline{S}_\mathcal{O} - S_\mathcal{O} \quad
	\text{for} \;
	\mathcal{O} = GG, GGG, GGGG, \Phi\Phi, \Phi G \Phi, \Phi GG \Phi \, .
\end{equation}

We close this section with the following remarks:
\begin{itemize}[leftmargin=*]
\item The renormalization transformation as usual provides most of the
   counterterms. It must be applied to the invariant part of the
   tree-level action, not to the evanescent part which contains the
   $d$-dimensional extension of the fermion kinetic term. As a result
   the counterterms $S_{\text{sct,inv}}^{(1)}$ contain only purely
   4-dimensional fermion terms.

\item The remaining evanescent counterterms are specific to the BMHV
   scheme. They involve all vertices of scalars and vectors with up to
   4 legs. The evanescent terms of the form
   $\widetilde{S}^{(1)}_{\mathcal{O}}$ are still gauge invariant,
   despite being evanescent; the two additional evanescent terms
   present in \cref{eq:Ssctevan}, contributions to the gauge boson and
   scalar two-point function counterterms, are not gauge invariant.

\item The corresponding result for a gauge theory without scalars has
   already been obtained in Ref.\ \cite{Martin:1999cc}. The scalars
   contribute in two ways: they provide additional contributions to the
   invariant counterterms $S_{\text{sct,inv}}^{(1)}$ and thus to the
   renormalization constants in \crefrange{eq:RenConstFirst}{eq:RenConstLast}.
   These contributions
   are standard and equal to the case without the BMHV scheme.
   Second, there is an explicit evanescent scalar operator present in
   \cref{eq:Ssctevan}. It originates from fermion loop contributions to
   the scalar self-energy.

\item The result presented here is specific to our choice of the
   regularized, $d$-dimensional theory \cref{eq:S0Def1}, based on
   \cref{eq:Lfermions}. In particular, this choice does not generate
   an extra evanescent counterterm to the fermion two-point function.
   Had we used another choice out of the options
   indicated in \cref{eq:inequivalentchoices}, the result would have been
   different. As an illustration we provide here the results for the
   self-energies corresponding to replacing the object
   $\Proj{L} \gamma_\mu \Proj{R}$ by $\gamma_\mu \Proj{R}$ (choice designated by ``Alt'')
   in the fermion-gauge boson interaction. The scalar self-energy does not
   change, but the fermion and gauge boson self-energies change as
\begin{align}
     \imath \widetilde{\Gamma}_{\psi \bar{\psi}}^{ji}(p)|^{\text{Alt},(1)}_\text{div} &=
	 \imath \widetilde{\Gamma}_{\psi \bar{\psi}}^{ji}(p)|^{(1)}_\text{div} -
     \frac{\imath \hbar g^2}{16 \pi^2 \epsilon} C_2(R) \delta^{ij}
     \;\widehat{\centernot{p}}\; \Proj{R} \, ,
     \\
     \imath \widetilde{\Gamma}_{GG}^{ba,\nu\mu}(p)|^{\text{Alt},(1)}_\text{div} &=
	 \imath \widetilde{\Gamma}_{GG}^{ba,\nu\mu}(p)|^{(1)}_\text{div}
     + \frac{\imath \hbar g^2}{16 \pi^2 \epsilon} \frac{S_2(R)}{3}
     \delta^{ab} (\overline{p}^\mu \widehat{p}^\nu + 2 \widehat{p}^\mu
     \widehat{p}^\nu + \widehat{p}^\mu \overline{p}^\nu +
     \overline{p}^2 \widehat{g}^{\mu\nu}) \, .
\end{align}
We see that both self-energies receive additional evanescent contributions and
the structure of the resulting $S_{\text{sct,evan}}^{(1)}$ will become
considerably more complicated. In particular, a new evanescent counterterm to
the fermion two-point function would have appeared,
$S_{\text{sct,evan}}^{(1)} \supset \hbar/(16 \pi^2 \epsilon) g^2 C_2(R) \int_x \imath \overline{\psi}_i \widehat{\slashed{\partial}} \Proj{R} \psi_i$.
\end{itemize}

\section{BRST Symmetry Breaking and its Restoration; Bonneau Identities}
\label{sect:BRSTrestoration}

Here we turn to the central point of our study --- the determination
of the symmetry-restoring counterterms required in the
BMHV scheme.
We begin this section with a brief general overview of the situation
and then describe the actual evaluation.

The basic requirement is that after renormalization, the finite
effective action $\Gamma_{\text{Ren}}$ satisfies the Slavnov-Taylor identity,
\begin{align}
  {\cal S} (\Gamma_{\text{Ren}}) &= 0\, .
\end{align}
In the previous \cref{sect:Rmodel1LoopSCT}
we have determined the singular counterterms which render the theory
finite at the one-loop level. Including finite counterterms to be
determined below, the one-loop effective action in $d$ dimensions can
be written following \cref{eq:definitionGammaDReg} as
\begin{align}
  \Gamma_\text{DReg}^{(1)} &= \Gamma^{(1)} +
  S_{\text{sct}}^{(1)} + S_{\text{fct}}^{(1)} \, ,
\end{align}
where $\Gamma^{(1)}$ denotes the effective action from tree-level and
genuine 1-loop diagrams (without counterterms). The limit $d \to 4$ exists, and
the renormalized one-loop effective action is obtained by taking the
$\mathop{\text{LIM}}_{d \to 4}\Gamma_\text{DReg}^{(1)}$, as defined in
\cref{subsect:AmpsDDim} and \cref{eq:definitionGammaRen}.
The Slavnov-Taylor identity in $d$ dimensions can be written at the one-loop level as
\begin{align}
  {\cal S}_d (\Gamma_\text{DReg}^{(1)}) &=
  {\cal S}_d (\Gamma^{(1)})
  + b_d S_{\text{sct}}^{(1)} + b_d S_{\text{fct}}^{(1)}
  \, ;
\label{eq:STIoneloopstructure}
\end{align}
here the linearized operator $b_d$ of \cref{eq:bdDefinition} has been
used and terms of higher loop order have been neglected.

The first
term on the right-hand side of equation \eqref{eq:STIoneloopstructure} is
expected to be nonzero. It
corresponds to the breaking of the Slavnov-Taylor identity by one-loop
regularized Green's functions. The second term by construction cancels
any UV divergences present in the first term. The last term contains
the finite counterterms to be discussed in the present section. These
finite counterterms must be chosen such that the finite parts of the
previous terms are cancelled (at least in the $\mathop{\text{LIM}}_{d \to 4}$).

The determination of the symmetry-restoring finite counterterms thus
requires two technical steps:
\begin{enumerate}
\item
  Evaluate the symmetry breaking caused by the genuine one-loop
  diagrams and the required singular counterterms, i.e.\ evaluate
  $  {\cal S}_d (\Gamma^{(1)})$ and
  $b_d S_{\text{sct}}^{(1)}$.
\item
  Find the symmetry-restoring counterterms $ S_{\text{fct}}^{(1)}$,
  whose $b_d$-variation cancels the symmetry breaking.
\end{enumerate}
Before presenting these calculations in detail we provide several
remarks on these steps.
\begin{itemize}[leftmargin=*]
\item
  \emph{Remarks on the structure of finite counterterms.} The
  symmetry-restoring finite counterterms are not unique. In general,
  the finite counterterms can always be written as (see also
  \cref{sect:standardrenormalizationstructure})
  \begin{equation}
  \label{eq:SFct_Struct}
     S_{\text{fct}}^{(1)} =
     S_{\text{fct,inv}}^{(1)}
     + S_{\text{fct,restore}}^{(1)}
     + S_{\text{fct,evan}}^{(1)} \, .
  \end{equation}
  Here $S_{\text{fct,inv}}^{(1)}$ originates from the renormalization
  transformation \eqref{eq:Sctinv} and is symmetry invariant in the sense
  of \eqref{eq:STIctinv}; the evanescent counterterms
  $S_{\text{fct,evan}}^{(1)}$ vanish in the $\mathop{\text{LIM}}_{d \to 4}$
  by definition and are therefore irrelevant for symmetry
  restoration at the one-loop level%
  \footnote{The choice of one-loop
    evanescent counterterms will have an impact on two- and higher-loop calculations.
  }.
  Therefore, the actual symmetry-restoring one-loop counterterms are
  given by $S_{\text{fct,restore}}^{(1)}$. They are clearly only
  unambiguous up to shifting around terms obtained by
  renormalization transformation and/or evanescent terms. What we will
  provide in the present section is one particular representative
  choice for these symmetry-restoring counterterms.

\item
  \emph{Remarks on the technical evaluation of the symmetry breaking caused
  by the first and second terms on the r.h.s.\ of
  \eqref{eq:STIoneloopstructure}.} There are several methods to determine
  the breaking of the symmetry. An obvious one is to directly compute
  all the required Green's functions and plug them into the Slavnov-Taylor
  identity. Such a direct approach was used e.g.\ in
  Ref.\ \cite{Ferrari:1994ct} for comparing the BMHV \emph{vs.} the
  naive $\gamma_5$ schemes in flavor-changing neutral processes,
  in Refs.\ \cite{Grassi:1999tp,Grassi:2001zz} in the study of chiral gauge
  theories and e.g.\ in
  Refs.\ \cite{Hollik:1999xh,Hollik:2001cz,Fischer:2003cb} in similar applications 
  on supersymmetric gauge theories. An advantage of this method is the
  direct connection to Green's functions appearing in physical processes
  and the explicit control over the symmetry breaking.

%% See also Section~7 of \cite{Binosi:2009qm} for an example of such
%% application in the context of the ``Pinch Technique''.

  A second, more indirect method is based on the \emph{regularized quantum
  action principle}, established for dimensional regularization in
  Ref.\ \cite{Breitenlohner:1977hr}. This regularized quantum action
  principle implies
  \begin{equation}
  \label{eq:QAPforSTI}
    {\cal S}_d (\Gamma^{(1)}) = \widehat{\Delta} \cdot \Gamma^{(1)} \, ,
    %% = [\widehat{\Delta} \cdot \Gamma_\text{DReg}]^{(1)} = [\widehat{\Delta} \cdot \Gamma_\text{DReg}]^{(1)}_\text{div} + [\widehat{\Delta} \cdot \Gamma_\text{DReg}]^{(1)}_\text{fin}
  \end{equation}
  where $\widehat{\Delta} = s_d S_0$ is the original tree-level BRST symmetry
  breaking \cref{eq:WidehatDeltaDefinition}, %% \cref{eq:BRSTTreeBreaking}
  while the full
  r.h.s.\ denotes the generating functional of one-loop regularized
  Green's functions with one insertion corresponding%
  \footnote{
    The r.h.s.\ of \cref{eq:QAPforSTI} also
    contains the tree-level result \cref{eq:WidehatDeltaDefinition}, but
    this tree-level result will be irrelevant in the following when we
    take only the UV divergent part and/or the
    $\mathop{\text{LIM}}_{d \to 4}$ of \cref{eq:QAPforSTI}.
  }
  to $\widehat{\Delta}$.
  Using this relation, the computation is
  simplified since the r.h.s.\ involves far fewer, and simpler Feynman
  diagrams than the left-hand side. Furthermore, it does not involve the
  evaluation of products of 1PI Green's functions, as would be the case in
  the direct approach.
  This indirect method has been applied in the literature,
  e.g.\ in Ref.\ \cite{Breitenlohner:1977hr} to scale
  invariance, in \cite{Martin:1999cc,SanchezRuiz:2002xc} to chiral
  non-abelian and abelian gauge theories at the one-loop level, and in
  Refs.\ \cite{Stockinger:2005gx,Hollik:2005nn,Stockinger:2018oxe} in a
  similar way to supersymmetric theories at the 2- and 3-loop level.

\end{itemize}
In this work we will apply the second method, that we find more advantageous.
\cref{subsect:Deltahatbreaking} will also present additional reasons why it is so.

In view of these remarks, the condition that the Slavnov-Taylor
identity is satisfied at the one-loop level in the 4-dimensional limit
can be written as
\begin{align}
\label{eq:defsymmetryrestore} %% eq:STERenorm, eq:STERenormRHS
  0 &= \mathop{\text{LIM}}_{d \to 4} \left( [\widehat{\Delta} \cdot \Gamma^{(1)}]_\text{div} + b_d{S_\text{sct}^{(1)}} + [\widehat{\Delta} \cdot \Gamma^{(1)}]_\text{fin} + b_d{S_{\text{fct,restore}}^{(1)}}\right)
  \, ,
\end{align}
% \begin{equation}
% \label{eq:STERenormRHS}
    % 0 \stackrel{n=1}{=} \mathop{\text{LIM}}_{d \to 4} \left( [\widehat{\Delta} \cdot \Gamma_\text{DReg}]^{(1)}_\text{div} + b_d{S_\text{sct}^{(1)}} \right) + [N[\widehat{\Delta}] \cdot \Gamma_\text{Ren}]^{(1)} + b{S_\text{fct,restore}^{(1)}}
    % \, ,
% \end{equation}
% with $b_d$ (resp. $b$) the linear BRST operation in $d$ (resp. 4) dimensions, see \cref{eq:LinearBRST} in \cref{app:LinearBRST}.
where the subscripts ``div''/``fin'' denote the $1/\epsilon$ and finite
parts, respectively.
This is the defining condition for the one-loop symmetry-restoring
counterterms.
The following \cref{subsect:bd_Ssct} will
present the evaluation of the divergent quantities
$[\widehat{\Delta} \cdot \Gamma^{(1)}]_\text{div}$ and
$b_d{S_\text{sct}^{(1)}}$,  and \cref{subsect:Deltahatbreaking} will present
the evaluation of the
finite parts of $[\widehat{\Delta} \cdot \Gamma^{(1)}]_\text{fin}$.
In \cref{subsect:fincts} we will determine and present
the required finite, symmetry-restoring counterterms.

\newcommand{\OLDDReg}{{}}%{\text{DReg}}

\subsection{Evaluation of $[\widehat{\Delta} \cdot \Gamma^{(1)}]_\text{div}$ and comparison with $b_d{S_\text{sct}^{(1)}}$}
\label{subsect:bd_Ssct}

In this subsection we present the evaluation of the divergent
quantities appearing in \cref{eq:defsymmetryrestore}, i.e.\
$[\widehat{\Delta} \cdot \Gamma^{(1)}]_\text{div}$ and
$b_d{S_\text{sct}^{(1)}}$. By construction, it is clear that these two
quantities must add up to something finite; however, we will show in
the following that they actually add up to zero. The basic reason is that both
quantities are pure divergences, and no terms of the form $\epsilon/\epsilon$
are generated from combining evanescent terms with UV singularities.

We start by evaluating $b_d{S_\text{sct}^{(1)}}$. First, as explained
in \cref{sect:standardrenormalizationstructure}, all the $L_\phi$ terms
present in the invariant part of the singular counterterms in
\cref{eq:Ssctinvevan,eq:Sctinvstructure} %% \cref{eq:SingularCT1Loop_Lphi}
are $b_d$-invariant, except for $L_c$ and $L_g$ where $b_d L_{c,g} = \widehat{\Delta}$.
Several of the evanescent terms specified in \cref{eq:Ssctevan} are $b_d$-invariant as well.

We therefore need to evaluate $b_d ((Y_R^n (Y_R^m)^* Y_R^n)_{ij} S_{\overline{\psi_R}^C_i \Phi^m {\psi_R}_j} + \hc)$ and $b_d ((3 g^4 A - 4 H + \Lambda^2)_{mnop} S_{\Phi^4_{mnop}})$.
In the first term, the action of $b_d$ generates a group structure that can be simplified using the gauge-invariance property \cref{eq:GaugeInvarYuk}. After this simplification, we end up with a structure $\propto \theta^a_{no} (Y_R^n (Y_R^m)^* Y_R^o + Y_R^o (Y_R^m)^* Y_R^n)_{ij}$ that cancels due to the antisymmetry of $\theta^a$.
Let us now turn to the second term:
\begin{equation}\begin{split}
    b_d ((3 g^4 A - 4 H + \Lambda^2)_{mnop} S_{\Phi^4_{mnop}}) = 4 (3 g^4 A - 4 H + \Lambda^2)_{qnop} \theta^a_{qm} \frac{\imath g}{2} \int \dInt[d]{x} c_a S_{\Phi^4_{mnop}}
    \, .
\end{split}\end{equation}
The group factor is completely symmetric in its indices, much like the tree-level scalar self-coupling $\lambda_{mnop}$, and its contraction with $\theta^a_{qm}$ can be rewritten similarly to \cref{eq:GaugeInvarLamb}.
For each term involved: $A_{qnop} \theta^a_{qm} S_{\Phi^4_{mnop}}$, $\Lambda^2_{qnop} \theta^a_{qm} S_{\Phi^4_{mnop}}$ and $H_{qnop} \theta^a_{qm} S_{\Phi^4_{mnop}}$, we throughly exploit the allowed symmetrizations in group indices so as to exhibit contractions between symmetric and antisymmetric symbols or internal cancellations, leading to the complete cancellation of these three terms. The last term in $H_{qnop}$ furthermore requires the usage of \cref{eq:GaugeInvarYuk}.

All in all, we obtain:
\begin{equation}
\label{eq:bdSingularCT1Loop}
\begin{split}
	b_d{S_\text{sct}^{(1)}} =\;& \frac{-\hbar}{16 \pi^2 \epsilon} \left\{ g^2 \frac{\xi C_2(G)}{2} \widehat{\Delta}
		+ g^2 \frac{S_2(R)}{3} b_d \int \dInt[d]{x} \frac{1}{2} \bar{G}^{a\,\mu} \widehat{\partial}^2 \bar{G}^a_\mu
		+ \frac{2 Y_2(S)}{3} b_d \widehat{S_{\Phi\Phi}} \right\}
	\, ,
\end{split}
\end{equation}
where, in the last two terms, $b_d$ actually acts like the BRST
transformation, leading to:
\begin{subequations}
\begin{align}
    b_d \int \dInt[d]{x} \frac{1}{2} \bar{G}^{a\,\mu} \widehat{\partial}^2 \bar{G}^a_\mu = \int \dInt[d]{x} (s_d{\bar{G}^{a\,\mu}}) \widehat{\partial}^2 \bar{G}^a_\mu = \int \dInt[d]{x} (\overline{\partial}^\mu c_a + g f^{abc} \bar{G}^{b\,\mu} c_c) \widehat{\partial}^2 \bar{G}^a_\mu \, , \\
    b_d \widehat{S_{\Phi\Phi}} = b_d \int \dInt[d]{x} \frac{-1}{2} \Phi_m \widehat{\partial}^2 \Phi_m = - \int \dInt[d]{x} (s_d{\Phi_m}) \widehat{\partial}^2 \Phi_m = \int \dInt[d]{x} \imath g \theta^a_{mn} c^a \Phi_m \widehat{\partial}^2 \Phi_n \, .
\end{align}
\end{subequations}
We note that the breaking terms are organized according to the field sectors: one for the fermions (proportional to the tree-level breaking $\widehat{\Delta}$), one for the gauge bosons and one for the scalars.
We further note that, as announced, \cref{eq:bdSingularCT1Loop} is a
pure $1/\epsilon$ singular term; no finite terms are generated by
applying the $d$-dimensional operator $b_d$ onto the singular
counterterm action.

For evaluating $[\widehat{\Delta} \cdot \Gamma^{(1)}]_\text{div}$ we calculate the one-loop vertex corrections with insertion of the $\widehat{\Delta}$ evanescent operator.
All momenta are incoming and all the results use $d = 4 - 2\epsilon$.
Below is the list of all diagrams with a $\widehat{\Delta}$ insertion that have a non-vanishing divergent part:
\begin{table}[h]
\centering
\begin{tabular}{*{3}{>{\centering\arraybackslash}m{0.3\textwidth}}}
	\uline{$\widehat{\Delta} c^a G^b_\mu$}: &
	\uline{$\widehat{\Delta} c^a G^b_\mu G^c_\nu$}: &
	\uline{$\widehat{\Delta} c^a \Phi^m \Phi^n$}: \\
			\includegraphics[scale=0.6]{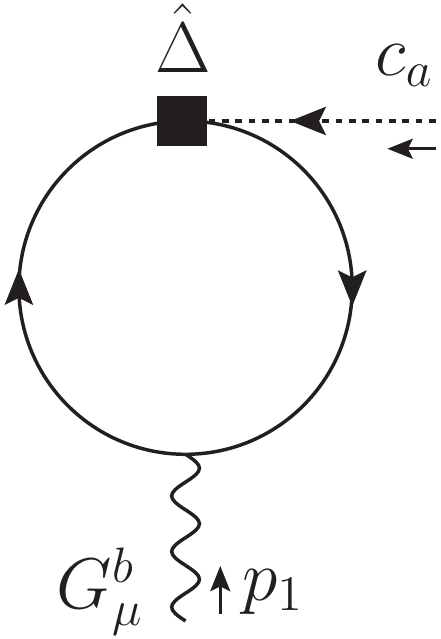}
		&
			\includegraphics[scale=0.6]{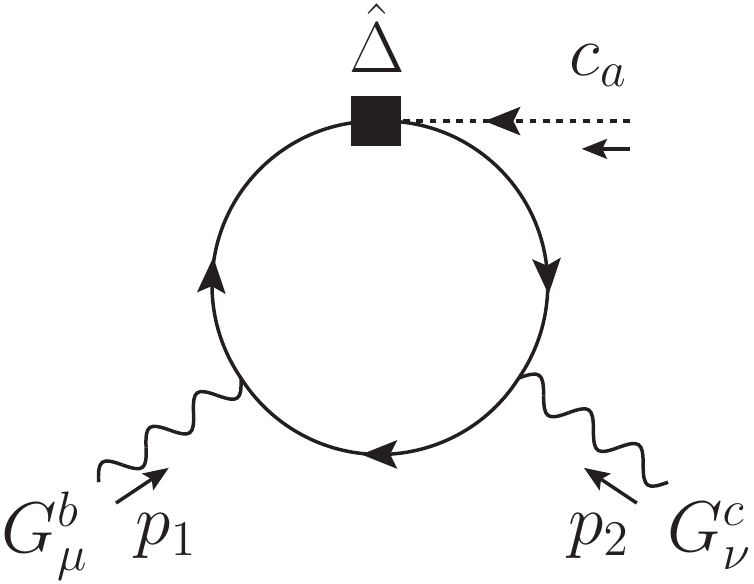} \newline
			$+ (p_1,\mu,b) \leftrightarrow (p_2,\nu,c)$ permutation.
		&
			\includegraphics[scale=0.6]{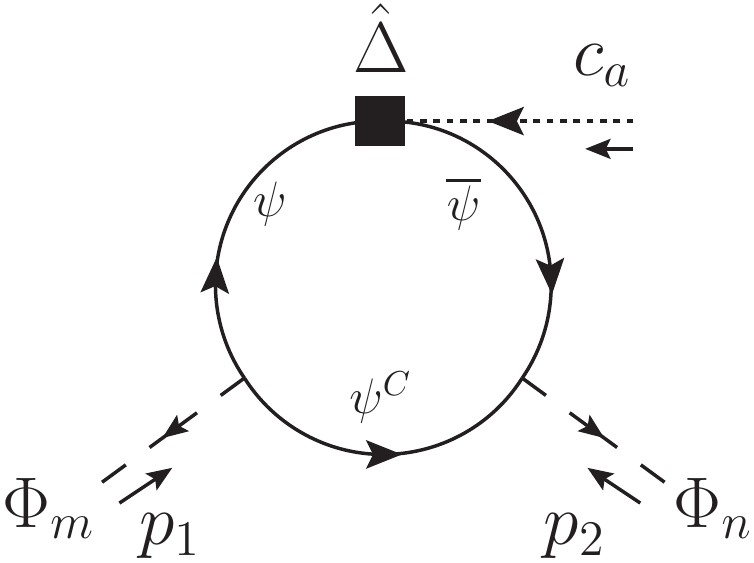} \newline
			$+ (p_1,m) \leftrightarrow (p_2,n)$ permutation.
	\\
	\multicolumn{3}{l}{\uline{$\widehat{\Delta} c^a \bar{\psi}_{i,\alpha} \psi_{j,\beta}$}:} \\
	\multicolumn{3}{c}{
		%\begin{figure}[h!]
			\centering
			\subfloat[Vanishing diagrams.]{%
				\includegraphics[scale=0.55]{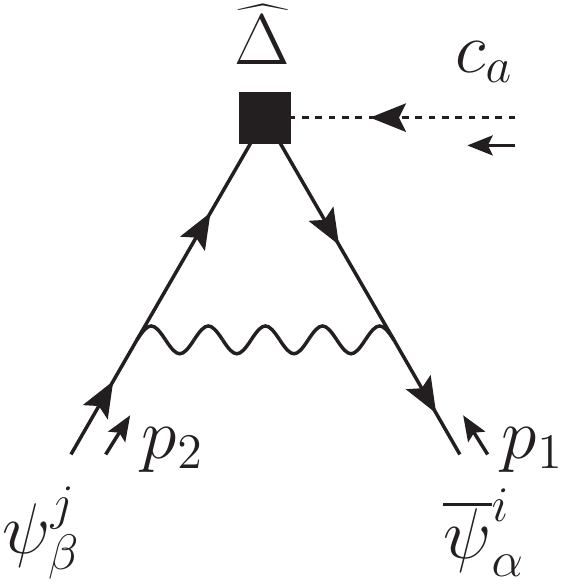}
				\includegraphics[scale=0.55]{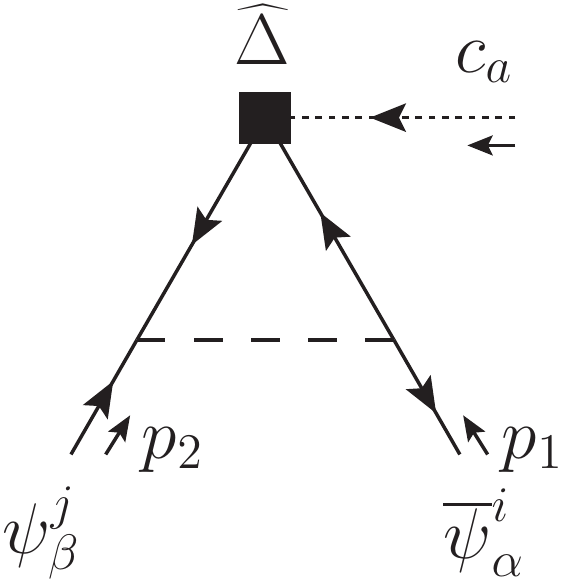}
			} \hfill
			\subfloat[Diagrams giving the $\Proj{R}$ and $\Proj{L}$ contributions \newline respectively.]{%
				\includegraphics[scale=0.55]{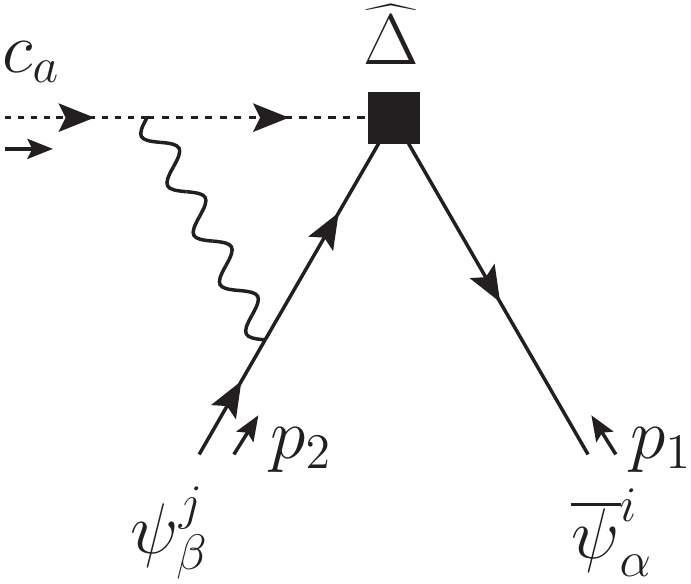}
				\includegraphics[scale=0.55]{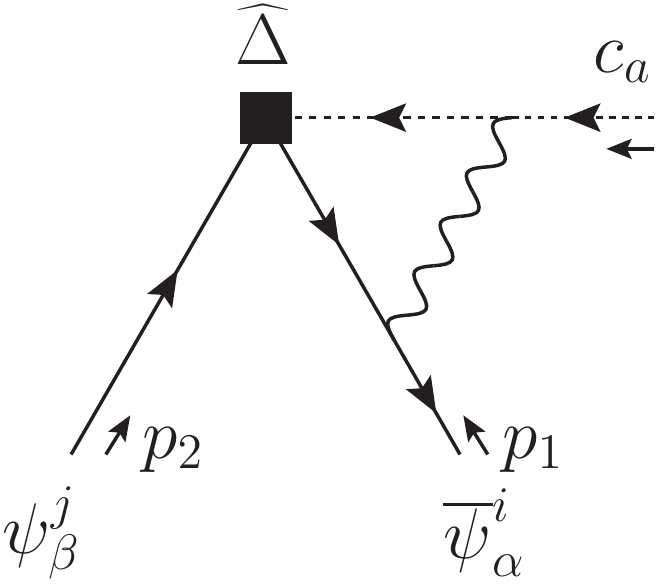}
			}
		%\end{figure}
	}
\end{tabular}
\end{table}

\begin{subequations}
\begin{align}
% $\widehat{\Delta}(q=0) c^a(-p_1) \to G^b_\mu(p_1)$
    \imath [\widehat{\Delta} \cdot \Gamma_\OLDDReg {}_{Gc}^{ba,\mu}]^{(1)}_\text{div} &= \frac{\hbar g^2}{16 \pi^2 \epsilon} \frac{S_2(R)}{3} \delta^{ab} \widehat{p_1}^2 \overline{p_1}^\mu
    \, , \\
% $\widehat{\Delta}(q=0) c^a(-p_1-p_2) \to G^b_\mu(p_1) G^c_\nu(p_2)$
    \imath [\widehat{\Delta} \cdot \Gamma_\OLDDReg {}_{GGc}^{cba,\nu\mu}]^{(1)}_\text{div} &= \frac{-\imath \hbar g^3}{16 \pi^2 \epsilon} \frac{S_2(R)}{3} f^{abc} (\widehat{p_1}^2 - \widehat{p_2}^2) \overline{g}^{\mu\nu}
    \, , \\
% $\widehat{\Delta}(q=0) c^a(-p_1-p_2) \to \Phi^m(p_1) \Phi^n(p_2)$
    \imath [\widehat{\Delta} \cdot \Gamma_\OLDDReg {}_{\Phi\Phi c}^{nm,a}]^{(1)}_\text{div} &= \frac{-\hbar g}{16 \pi^2 \epsilon} \frac{2 Y_2(S)}{3} \theta^a_{mn} (\widehat{p_1}^2 - \widehat{p_2}^2)
    \, , \\
% $\widehat{\Delta}(q=0) c^a(-p_1-p_2) \to \bar{\psi}_{i,\alpha}(p_1) \psi_{j,\beta}(p_2)$
\label{eq:BIMSp1_EvsctDeFbF}
    \imath [\widehat{\Delta} \cdot \Gamma_\OLDDReg {}_{\psi\bar{\psi} c}^{ji,a}]^{(1)}_\text{div} &= \frac{\hbar g^3}{16 \pi^2 \epsilon} \frac{\xi C_2(G)}{2} {T_R}^a_{ij} (\widehat{\slashed{p_1}} \Proj{R} + \widehat{\slashed{p_2}} \Proj{L}) \, .
\end{align}
\end{subequations}
The sum of these 1PI contributions evaluated in this section constitutes the non-vanishing contribution to $[\widehat{\Delta} \cdot \Gamma_\OLDDReg]^{(1)}_\text{div}$:
\begin{multline}
	[\widehat{\Delta} \cdot \Gamma_\OLDDReg]^{(1)}_\text{div} =
		\frac{\hbar}{16 \pi^2 \epsilon} \left\{ g^2 \frac{\xi C_2(G)}{2} \widehat{\Delta} + g^2 \frac{S_2(R)}{3} \int \dInt[d]{x} (\overline{\partial}^\mu c_a + g f^{abc} \bar{G}^{b\,\mu} c_c) \widehat{\partial}^2 \bar{G}^a_\mu \right.\\
		\left. + \frac{2 Y_2(S)}{3} \int \dInt[d]{x} \imath g \theta^a_{mn} c^a \Phi_m \widehat{\partial}^2 \Phi_n
		\right\}
	\, ,
\end{multline}
and by comparing with \cref{eq:bdSingularCT1Loop} that provides the expression of $b_d{S_\text{sct}^{(1)}}$, we conclude that there exists a perfect cancellation between $b_d{S_\text{sct}^{(1)}}$ and $[\widehat{\Delta} \cdot \Gamma_\OLDDReg]^{(1)}_\text{div}$ as we expected.

\subsection{Bonneau Identities and the Evaluation of $\mathop{\text{LIM}}_{d \to 4}[\widehat{\Delta} \cdot \Gamma^{(1)}]_\text{fin}$}
\label{subsect:Deltahatbreaking}

This subsection  presents the evaluation of the finite
quantity appearing in \cref{eq:defsymmetryrestore}, i.e.\
$\mathop{\text{LIM}}_{d \to 4}[\widehat{\Delta} \cdot \Gamma^{(1)}]_\text{fin}$.
This is the central quantity which describes the one-loop symmetry breaking
caused by the BMHV scheme for $\gamma_5$. As mentioned around \cref{eq:QAPforSTI},
this calculation will provide a particularly efficient way to evaluate the
symmetry breaking. Indeed, this finite quantity accounts for the finite part
of the Slavnov-Taylor identity which, if we were using the direct method instead,
would be evaluated using products of 1PI Green's functions, including their
finite parts, which is in general a difficult matter. Here instead, only
UV-divergent parts of specific Green's functions will be required, as we will
see.

At first order in $\hbar$, our quantity of interest may be expressed as
\begin{equation}
	\mathop{\text{LIM}}_{d \to 4}[\widehat{\Delta} \cdot \Gamma^{(1)}]_\text{fin} =
	[N[\widehat{\Delta}] \cdot \Gamma_\text{Ren}]^{(1)} \, ,
\end{equation}
where the subscript ``Ren'' implies minimal subtraction and taking the
$\mathop{\text{LIM}}_{d \to 4}$.
Here $N[{\cal O}]$ denotes the Zimmermann-like definition
\cite{Zimmermann:1972te,Zimmermann:1972tv,Lowenstein:1971vf,Piguet:1980nr} of a
renormalized local operator (also called \emph{``normal product''}), defined as
an insertion of a local operator ${\cal O}$ and followed, in the context%
\footnote{
  The actual definition for a ``normal product'' depends on the chosen
  renormalization procedure:
  for example in BPHZ renormalization, where the renormalization is performed by
  subtracting the first terms of a Taylor expansion of loop integrands up to a
  given order (called ``degree'' of subtraction), different normal products are
  associated to the choice of the ``degree'' of subtraction \cite{Piguet:1980nr}.
}
of Dimensional Regularization and Renormalization, by a minimal subtraction
prescription \cite{Collins:1974da}.

Let us begin with further comments on how to evaluate
$[N[\widehat{\Delta}] \cdot \Gamma_\text{Ren}]^{(1)}$.
% $[\widehat{\Delta} \cdot \Gamma^{(1)}]_\text{fin}$.
At the one-loop level, it is reasonably straightforward to carry out a direct
computation, extending the computation of the divergent parts in the previous
subsection. However, it is useful to first discuss the structure of the
computation in more detail.

The BRST breaking vertex operator $\widehat{\Delta}$ in its local form is proportional to the evanescent metric: $\widehat{\Delta} = \hat{g}_{\mu\nu} \Delta^{\mu\nu}$, see \cref{eq:BRSTTreeBreaking}, where $\Delta^{\mu\nu}$ contains $\partial^\mu \gamma^\nu$ covariants, so that $\widehat{\Delta}$ can be re-expressed as: $\widehat{\Delta} = (g_{\mu\nu} - \bar{g}_{\mu\nu}) \Delta^{\mu\nu}$.
Finite contributions are generated once $\widehat{\Delta}$ is inserted into loop diagrams,
and the evanescent numerator combines with a $1/\epsilon$ singularity
to form a finite term that behaves schematically as $\epsilon/\epsilon$.

Hence, we can expect that the finite symmetry breaking can also be obtained
from extracting only the singular parts of suitable diagrams. Such a
relationship is provided by an identity due to Bonneau
\cite{Bonneau:1980zp,Bonneau:1979jx}.
The general form of this identity is very involved, and we refer to
\cite{Bonneau:1980zp,Bonneau:1979jx,Martin:1999cc} for it. Here we discuss its
essence and its form applied to our one-loop case. This will provide
valuable additional understanding of the symmetry breaking as well as
a reference for future two-loop calculations, where Bonneau's
identity will be even more useful.

The essential property contained in the Bonneau identity can be
explained with the help of the equation
\begin{equation}
\label{eq:expansion_N_Delta}
	N[\widehat{\Delta}(x)] = N[g_{\mu\nu} \Delta^{\mu\nu}(x)] - N[\bar{g}_{\mu\nu} \Delta^{\mu\nu}(x)] = N[g_{\mu\nu} \Delta^{\mu\nu}(x)] - \bar{g}_{\mu\nu} N[\Delta^{\mu\nu}(x)] \, .
\end{equation}
The first equation in \eqref{eq:expansion_N_Delta} makes explicit the
appearance of the evanescent metric, which is decomposed as
$g_{\mu\nu} - \bar{g}_{\mu\nu}$. The second equation highlights that pulling
the metric out of the minimal subtraction procedure is possible only for the
purely 4-dimensional metric, but not for the $d$-dimensional metric where doing
this operation would not commute with the minimal subtraction procedure, and
therefore \cref{eq:expansion_N_Delta} does not vanish.
Note that $N[\Delta^{\mu\nu}(x)]$ is a 4-dimensional object since it has been
submitted to the renormalization procedure, therefore its contraction with
$\bar{g}_{\mu\nu}$ is the same as its contraction with $g_{\mu\nu}$ from outside.

Using this notation, the one-loop version of the Bonneau identity then reads
\begin{equation}
\label{eq:BonneauIdEvansctOneLoop}
    [N[\widehat{\mathcal{O}}] \cdot \Gamma_\text{Ren}]^{(1)} =
	\mathop{\text{LIM}}_{d \to 4}\left(
		-  \text{r.s.p.} \left[\widecheck{\mathcal{O}} \cdot \Gamma\right]^{(1)}_{\check{g}=0}
	\right)
	\, .
\end{equation}
Here on the right-hand side ``$\text{r.s.p.}$'' means the residue of the
simple pole in $\nu = 4 - d = 2\epsilon$ of the 1PI Green's function under
consideration%
\footnote{
  I.e.\ since we evaluate the divergent parts of the 1PI Green's functions
  in $d = 4 - 2\epsilon$,  we will have to take a factor 2 into account.
}.
The Feynman rules corresponding to the operator $\widecheck{\mathcal{O}}$
are obtained from the ones for $\widehat{\mathcal{O}}$ by formally replacing
all the evanescent Lorentz structures by their corresponding $d$-dimensional
versions contracted%
\footnote{
  For example: $\widehat{p}^2 = p_\mu p_\nu \hat{g}^{\mu\nu} \to p_\mu p_\nu \check{g}^{\mu\nu} \equiv \widecheck{p}^2$, and so on...
}
with the symmetric ``metric''-tensor $\check{g}_{\mu\nu}$,
possessing the following properties:
\begin{align}
\label{eq:CheckedMetricProps}
    \check{g}_{\mu\nu} g^{\nu\rho} &= \check{g}_{\mu\nu} \hat{g}^{\nu\rho}
    = \check{g}_{\mu\;}^{\;\rho} \; ,&
    \check{g}_{\mu\nu} \bar{g}^{\nu\rho} &= 0 \; ,&
    \check{g}_{\mu\;}^{\;\mu} &= 1 \; .
\end{align}
This symbol can be understood as corresponding to the evanescent metric
$\hat{g}_{\mu\nu}$ such that its trace has been normalized to one.
This explains also the appearance of the minus sign on the right-hand-side
of \cref{eq:BonneauIdEvansctOneLoop}: its left-hand-side is proportional to
$\hat{g}_{\mu\nu}$ which satisfies $\hat{g}_{\mu\nu} \hat{g}^{\nu\mu} = -2\epsilon$.
The equality \cref{eq:BonneauIdEvansctOneLoop} implements the
intuition developed above: the finite part of the breaking can be
obtained by evaluating the UV singularity of suitable diagrams,
involving the object $\check{g}_{\mu\nu}$.

The significant advantage of using the Bonneau identity is that it
further simplifies the evaluation of the required
$\mathop{\text{LIM}}_{d \to 4}[\widehat{\Delta} \cdot \Gamma^{(1)}]_\text{fin} =
[N[\widehat{\Delta}] \cdot \Gamma_\text{Ren}]^{(1)}$
to an evaluation of
\begin{equation}
\label{eq:finitebreakingsimplified}
  \mathop{\text{LIM}}_{d \to 4} \left(
		- \text{r.s.p.}[\widecheck{\Delta} \cdot \Gamma]^{(1)}_{\check{g}=0} \right) \, ,
\end{equation}
i.e.\ we need to determine all UV-divergent 1PI 1-loop diagrams with an insertion
of $\widecheck{\Delta}$.
Clearly, at fixed loop order there is only a limited finite number of UV-singular
diagrams to be evaluated. This constitutes the main advantage of this method.
%% It starts from $n=2$ for the graphs to be 1PI, and ends at $n_\text{max}$ the maximal number of external lines for a 1PI graph with insertion of $\widehat{\mathcal{O}}$ to be superficially divergent.
% More precisely, $n_\text{max} = 4 - \sum_i \delta_{\phi_i} + (\delta_\mathcal{O} - 4) = \delta_\mathcal{O} - \sum_i \delta_{\phi_i}$, where $\delta_{\phi_i}$ is the canonical dimension of the field $\phi_i$ and $\delta_\mathcal{O}$ the canonical dimension of the inserted operator $\mathcal{O}$.
In the following, we will present an exhaustive list of all diagrams
contributing to the breaking and determine their values.

\subsubsection{1-loop vertices with insertion of $\widecheck{\Delta}$}

As presented above, we need to evaluate all the non-vanishing contributions
to the finite breaking of the Slavnov-Taylor identity at the 1-loop level,
i.e.\ all the non-vanishing contributions to
\cref{eq:finitebreakingsimplified}. This requires evaluating the contributions
to the breaking functional $[N[\widehat{\Delta}] \cdot \Gamma_\text{Ren}]^{(1)}$,
see \cref{eq:BonneauIdEvansctOneLoop,eq:finitebreakingsimplified}.

We now discuss how this quantity is evaluated in practice, at 1-loop level.
\cref{eq:finitebreakingsimplified} tells us we first need to evaluate
$[\widecheck{\Delta} \cdot \Gamma]^{(1)}_{\check{g}=0}$, i.e.\ all the 1PI
1-loop diagrams with an insertion of $\widecheck{\Delta}$, that also are
UV-divergent so as to give a non-zero contribution when taking their $\text{r.s.p.}$ %~.
As mentioned above, at the level of Feynman rules $\widecheck{\Delta}$ is obtained
from $\widehat{\Delta}$ by converting all occurrences of evanescent Lorentz symbols
inside it into contractions of their corresponding $d$-dimensional versions with
the $\check{g}_{\mu\nu}$ symbol.
Evaluation of the obtained diagrams is then performed using standard loop techniques,
and is followed by a \emph{complete} tensor contraction and simplification (including Dirac structures) so as to eliminate as many $\check{g}_{\mu\nu}$ symbols as possible, using the properties \cref{eq:CheckedMetricProps}.
Finally an $\epsilon$-expansion is performed in order to keep only the simple-pole terms.
The property $\check{g}_{\mu\;}^{\;\mu} = 1$ of the $\check{g}_{\mu\nu}$ symbol has the effect of selecting the contributions of interest originally coming from the evanescent operator $\widehat{\Delta}$, that would have otherwise been absorbed into the finite part if the $\check{g}_{\mu\nu}$ symbol was not used and the original evanescent metric $\hat{g}_{\mu\nu}$ was used instead.

At the end of the calculation the remaining $\check{g}_{\mu\nu}$ symbols that have not been already eliminated (signalling the contribution of higher-order evanescent quantities) have to be discarded: indeed, according to the Bonneau identity, these remaining contributions would be one $\hbar$-order higher.
Finally, the different Lorentz structures arising from the calculation of the Green's function can be obtained and their corresponding coefficients can be extracted out.

In the following, we provide the list of all these non-vanishing contributions.
For each contribution, we provide the associated Feynman diagram, its result,
and the corresponding contribution to the breaking functional
$[N[\widehat{\Delta}] \cdot \Gamma_\text{Ren}]^{(1)}$.
Besides, since the operators contained in this functional are fully expressed in
4 space-time dimensions, we will omit all the ``overlines'' that would otherwise
be present over all the Lorentz covariants (vectors, tensors, fields, to symbolize
their 4-dimensionality), so as to simplify the notation.
We are as well employing the same notations for the integrated field monomials as
in \cref{eq:RModelDReg_Action} (\cref{subsect:RModelDReg}), but now all defined
purely in 4 dimensions.

\noindent
\uline{$\widecheck{\Delta} c^a G^b_\mu$}:\\
\begin{minipage}{0.3\textwidth}
	\centering
	\includegraphics[scale=0.6]{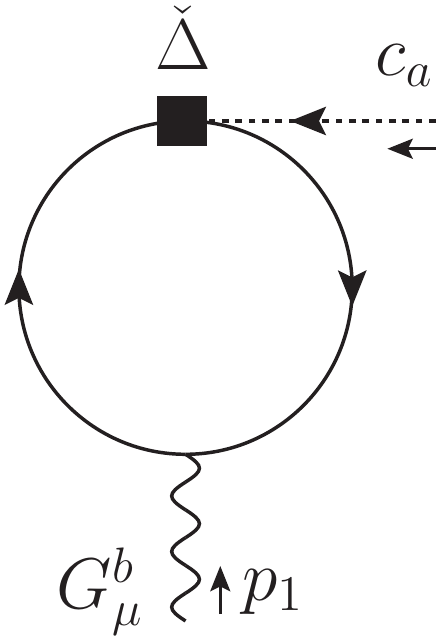}
\end{minipage}
\hfill
\begin{minipage}{0.7\textwidth}
% $\widecheck{\Delta}(q=0) c^a(-p_1) \to G^b_\mu(p_1)$
\begin{subequations}
\begin{equation}
    \imath [\widecheck{\Delta} \cdot \Gamma_\OLDDReg {}_{Gc}^{ba,\mu}]^{(1)}_\text{div} = \frac{-\hbar g^2}{16 \pi^2 \epsilon} \frac{S_2(R)}{6} \delta^{ab} \overline{p_1}^2 \overline{p_1}^\mu \, ,
\end{equation}
corresponding to the contribution
\begin{equation}
\label{eq:BIMSp1_BonDeG}
    [N[\widehat{\Delta}] \cdot \Gamma_\text{Ren}]^{(1)} \supset \frac{\hbar g^2}{16 \pi^2} \frac{S_2(R)}{3} \int \dInt[4]{x} (\partial^\mu c_a) (\partial^2 G^a_\mu) \, .
\end{equation}
\end{subequations}
\end{minipage}

\pagebreak

\noindent
\uline{$\widecheck{\Delta} c^a G^b_\mu G^c_\nu$}:\\
\begin{minipage}{0.3\textwidth}
	\centering
	\includegraphics[scale=0.6]{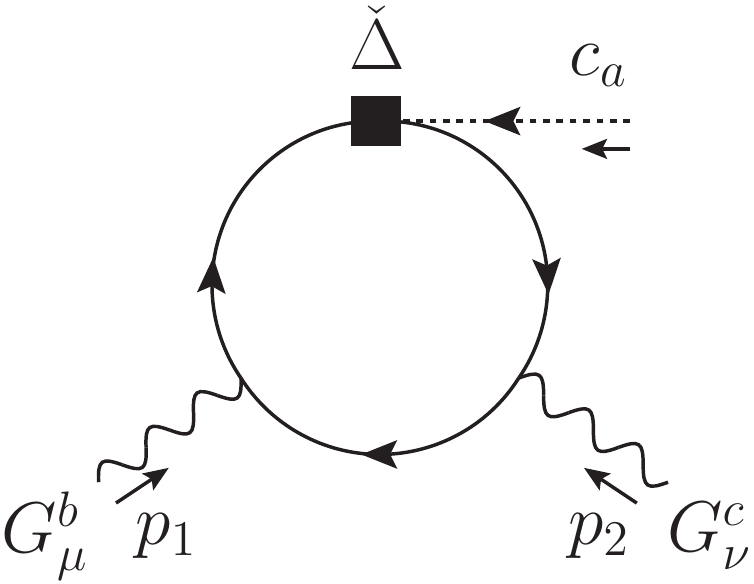}
	% $+ (p_1,\mu,b) \leftrightarrow (p_2,\nu,c)$ permutation.
\end{minipage}
\hfill
\begin{subequations}
\begin{minipage}{0.7\textwidth}
$+ (p_1,\mu,b) \leftrightarrow (p_2,\nu,c)$ permutation.
% $\widecheck{\Delta}(q=0) c^a(-p_1-p_2) \to G^b_\mu(p_1) G^c_\nu(p_2)$
\begin{multline}
    \imath [\widecheck{\Delta} \cdot \Gamma_\OLDDReg {}_{GGc}^{cba,\nu\mu}]^{(1)}_\text{div} =
        \frac{-\imath \hbar}{16 \pi^2 \epsilon} \frac{g^3}{6} \left[
            S_2(R) f^{abc} ((\overline{p_1}^2 - \overline{p_2}^2) \overline{g}^{\mu\nu} \right.\\
            \left.
            - 2 \overline{p_1}^\mu \overline{p_1}^\nu + 2 \overline{p_2}^\mu \overline{p_2}^\nu)
            + 2 d_R^{abc} \epsilon^{\mu\nu\rho\sigma} \overline{p_1}_\rho \overline{p_2}_\sigma
            \right] \, ,
\end{multline}
\end{minipage}
where we have defined the fully symmetric symbol $d_R^{abc} = \Tr[ {T_R}^a \{{T_R}^b , {T_R}^c\} ]$ for the R-representation.
This 1PI Green's function corresponds to the following contribution in the Bonneau identity and exhibits an anomalous contribution (second line):
\begin{multline}
\label{eq:BIMSp1_BonDeGG}
    [N[\widehat{\Delta}] \cdot \Gamma_\text{Ren}]^{(1)} \supset
        \frac{\hbar g^2}{16 \pi^2} \frac{S_2(R)}{3} \int \dInt[4]{x} g f^{abc} c_a G^b_\mu (\partial^2 g^{\mu\nu} - 2 \partial^\mu \partial^\nu) G^c_\nu \\
        - \frac{\hbar g^2}{16 \pi^2} \frac{d_R^{abc}}{3} \int \dInt[4]{x} g \epsilon^{\mu\nu\rho\sigma} c_a (\partial_\rho G^b_\mu) (\partial_\sigma G^c_\nu)
    \, .
\end{multline}
\end{subequations}

%\pagebreak

\noindent
\uline{$\widecheck{\Delta} c^a G^b_\mu G^c_\nu G^d_\rho$}:\\
\begin{minipage}{0.3\textwidth}
	\centering
	\includegraphics[scale=0.6]{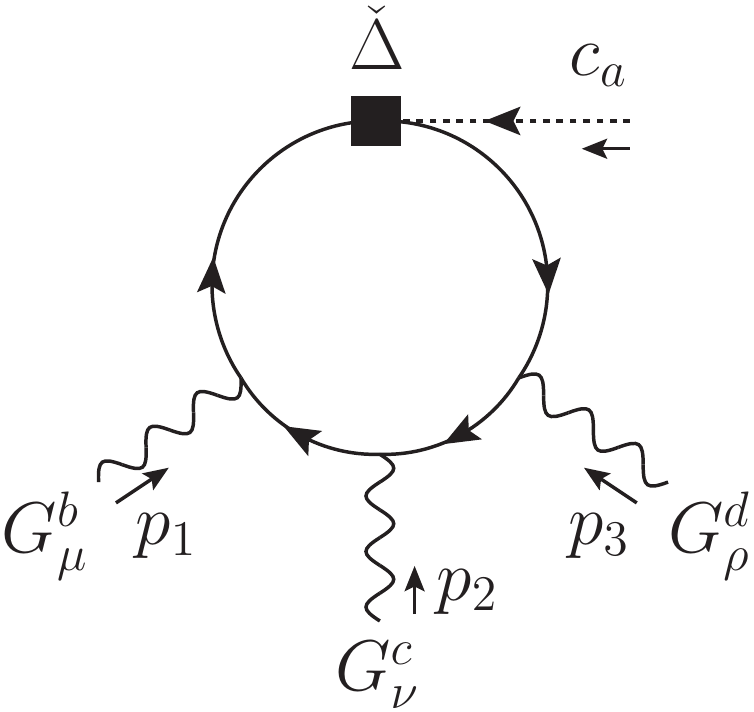}
\end{minipage}
\hfill
\begin{minipage}{0.7\textwidth}
	$+ \{ (p_1,\mu,b) \, , \, (p_2,\nu,c) \, , \, (p_3,\rho,d) \}$ permutations\footnotemark.
% $\widecheck{\Delta}(q=0) c^a(-p_1-p_2-p_3) \to G^b_\mu(p_1) G^c_\nu(p_2) G^d_\rho(p_3)$
\begin{multline}
    \imath [\widecheck{\Delta} \cdot \Gamma_\OLDDReg {}_{GGGc}^{dcba,\rho\nu\mu}]^{(1)}_\text{div} = \frac{-\hbar}{16 \pi^2 \epsilon} \frac{g^4}{6} \overline{(p_1 + p_2 + p_3)}_\sigma \left[ \right.\\
        \left.   \overline{g}^{\mu\nu} \overline{g}^{\rho\sigma} (\mathcal{A}_R^{abcd} + \mathcal{A}_R^{acbd})/2
               + \overline{g}^{\mu\rho} \overline{g}^{\nu\sigma} (\mathcal{A}_R^{abdc} + \mathcal{A}_R^{adbc})/2 \right.\\
        \left. + \overline{g}^{\mu\sigma} \overline{g}^{\nu\rho} (\mathcal{A}_R^{acdb} + \mathcal{A}_R^{adcb})/2
               - \mathcal{D}_R^{abcd} \epsilon^{\mu\nu\rho\sigma}
        \right]
    \, .
\end{multline}
\end{minipage}
\footnotetext{
	The third term of our calculation ($\propto \overline{g}^{\mu\sigma} \overline{g}^{\nu\rho}$) agrees with equation~(53) of \cite{Martin:1999cc}; however, an apparent discrepancy arises when comparing the first two terms ($\propto \overline{g}^{\mu\nu} \overline{g}^{\rho\sigma}$ and $\propto \overline{g}^{\mu\rho} \overline{g}^{\nu\sigma}$ with different group factors) with equation~(54) that tells that both $\overline{p_1}_\nu \overline{g}^{\mu\rho}$ and $\overline{p_1}_\rho \overline{g}^{\mu\nu}$ acquire the very same coefficient.
}
Introducing the notation $(T_R)^{a_1 \cdots a_n} = \Tr[{T_R}^{a_1} \cdots {T_R}^{a_n}]$ for the trace of a product of same group generators ${T_R}^a$, we have employed in the previous equation the group factor
\begin{equation}\label{eq:1PIDeltaPsiPsiGh}\begin{split}
    \mathcal{A}_R^{abcd} &= (T_R)^{abcd} - (T_R)^{abdc} + (T_R)^{acbd} - (T_R)^{acdb} + (T_R)^{adbc} + (T_R)^{adcb} \\
        &= (T_R)^{abcd} + (T_R)^{adcb} - S_2(R) f^{ace} f^{bde} = (T_R)^{acbd} + (T_R)^{adbc} - S_2(R) f^{abe} f^{cde} \\
        &= (T_R)^{abdc} + (T_R)^{acdb} - S_2(R) ( f^{abe} f^{cde} + f^{ace} f^{bde} ) \\
        &= \frac{1}{2}((T_R)^{abcd} + (T_R)^{adcb} + (T_R)^{acbd} + (T_R)^{adbc}) - \frac{S_2(R)}{2} (f^{abe} f^{cde} + f^{ace} f^{bde})
    \, ,
\end{split}\end{equation}
and we have defined the fully antisymmetric symbol%
\footnote{
	Here and in what follows, we employ the standard indicial notation for the (anti\nobreakdash-)symmetrization of tensor indices (or subset thereof): $T^{[a_1 \cdots a_n]} = \frac{1}{n!} \sum_\pi \sigma(\pi) T^{a_{\pi(1)}} \cdots T^{a_{\pi(n)}}$, and $T^{\{a_1 \cdots a_n\}} = \frac{1}{n!} \sum_\pi T^{a_{\pi(1)}} \cdots T^{a_{\pi(n)}}$.
}
$\mathcal{D}_R^{abcd} = (-\imath) 3! \Tr[ {T_R}^a {T_R}^{[b} {T_R}^c {T_R}^{d]} ] = \frac{1}{2}(d_R^{abe} f^{ecd} + d_R^{ace} f^{edb} + d_R^{ade} f^{ebc})$ for the R-representation, following the notations of Ref.\ \cite{Martin:1999cc}.
The 1PI Green's function \cref{eq:1PIDeltaPsiPsiGh} corresponds to the contribution
\begin{multline}
\label{eq:BIMSp1_BonDeGGG}
	[N[\widehat{\Delta}] \cdot \Gamma_\text{Ren}]^{(1)} \supset
		\frac{\hbar g^4}{16 \pi^2} \frac{\mathcal{A}_R^{abcd}}{6} \int \dInt[4]{x} c_a \partial^\nu \left( G^b_\mu G^{c\,\mu} G^d_\nu \right) \\
		- \frac{\hbar g^4}{16 \pi^2} \frac{\mathcal{D}_R^{abcd}}{3 \times 3!} \int \dInt[4]{x} c_a \epsilon^{\mu\nu\rho\sigma} \partial_\sigma \left( G^b_\mu G^c_\nu G^d_\rho \right)
	\, ,
\end{multline}
and also exhibits an anomaly (last term).

\noindent
\uline{$\widecheck{\Delta} c^a \Phi^m \Phi^n$}:\\
\begin{minipage}{0.3\textwidth}
	\centering
	\includegraphics[scale=0.6]{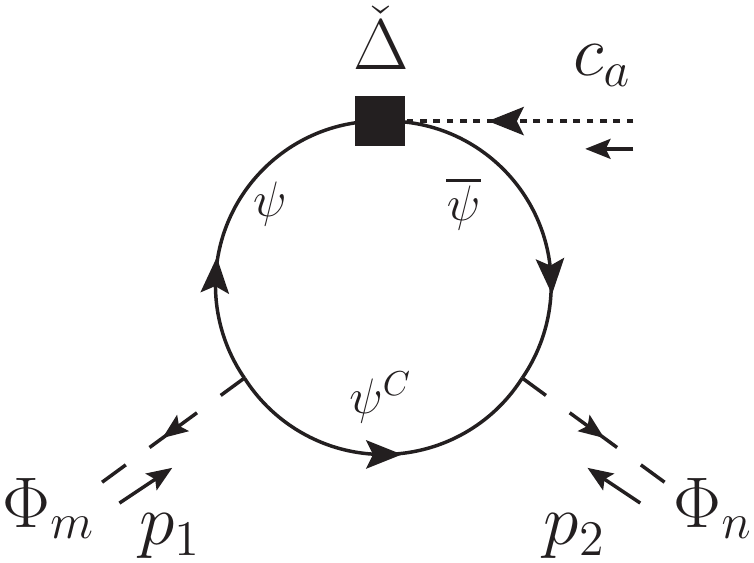}
\end{minipage}
\hfill
\begin{subequations}
\begin{minipage}{0.7\textwidth}
	$+ (p_1,m) \leftrightarrow (p_2,n)$ permutation.
% $\widecheck{\Delta}(q=0) c^a(-p_1-p_2) \to \Phi^m(p_1) \Phi^n(p_2)$
\begin{equation}
    \imath [\widecheck{\Delta} \cdot \Gamma_\OLDDReg {}_{\Phi\Phi c}^{nm,a}]^{(1)}_\text{div} = \frac{-\hbar g}{16 \pi^2 \epsilon} \frac{Y_2(S)}{6} \theta^a_{mn} (\overline{p_1}^2 - \overline{p_2}^2)
    \, ,
\end{equation}
\end{minipage}
corresponding to the contribution
\begin{equation}
\label{eq:BIMSp1_BonDeSS}
    [N[\widehat{\Delta}] \cdot \Gamma_\text{Ren}]^{(1)} \supset -\frac{\hbar}{16 \pi^2} \frac{Y_2(S)}{3} \int \dInt[4]{x} \imath g \theta^a_{mn} c^a \Phi_m \partial^2 \Phi_n
    \, .
\end{equation}
\end{subequations}

%\pagebreak

\noindent
\uline{$\widecheck{\Delta} c^a G^b_\mu \Phi^m \Phi^n$}:\\
\begin{figure}[h!]
	\centering
	\includegraphics[scale=0.6]{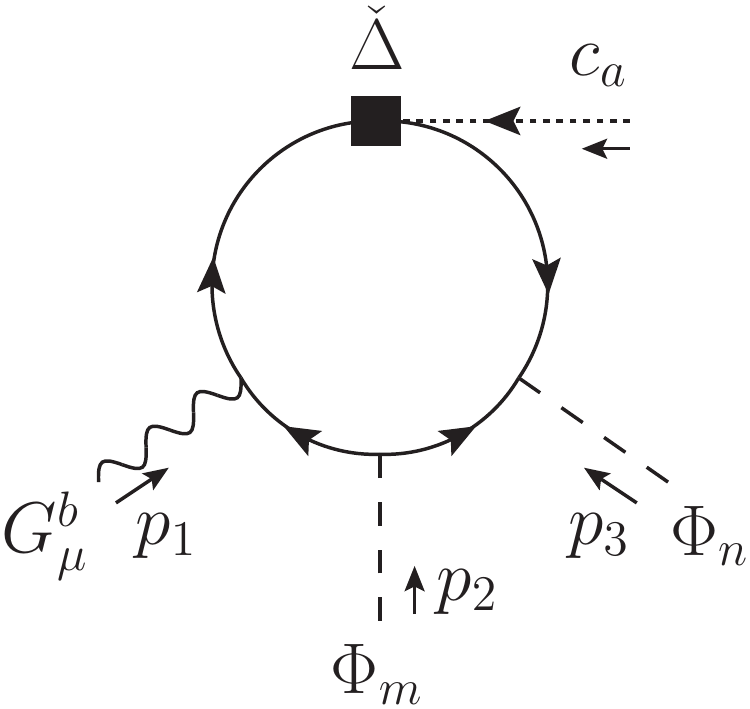}
	\includegraphics[scale=0.6]{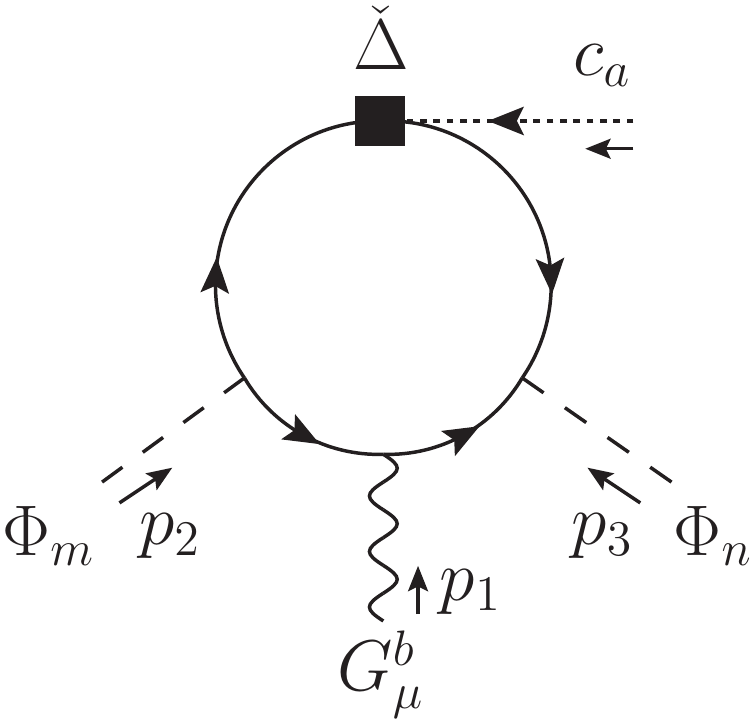}
	\includegraphics[scale=0.6]{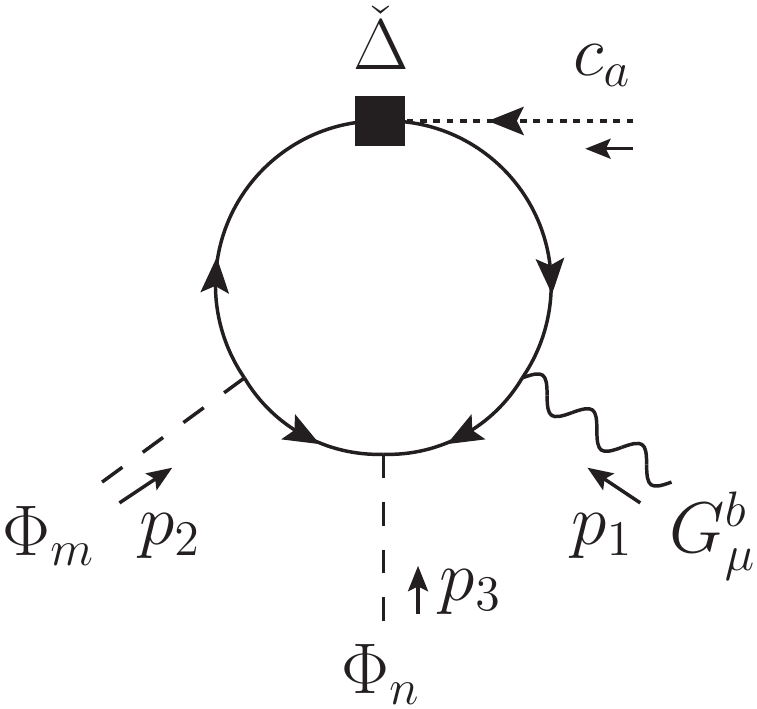} \\
	$+ (p_2,m) \leftrightarrow (p_3,n)$ permutation.
\end{figure}
% $\widecheck{\Delta}(q=0) c^a(-p_1-p_2) \to G^b_\mu(p_1) \Phi^m(p_2) \Phi^n(p_3)$
\begin{subequations}
\begin{equation}\label{eq:1PIDeltaGPhiPhiGh}\begin{split}
    \imath [\widecheck{\Delta} \cdot \Gamma_\OLDDReg {}_{\Phi\Phi G c}^{nm,ba,\mu}]^{(1)}_\text{div} &=
        \frac{\hbar g^2}{16 \pi^2 \epsilon} \frac{1}{6} \overline{(p_1 + p_2 + p_3)}^\mu
            \Tr\left[ 2 \{ {T_R}^a , {T_R}^b \} ( (Y_R^m)^* Y_R^n + (Y_R^n)^* Y_R^m ) \right.\\
&\left.\hspace{2cm} - {T_R}^a (Y_R^m)^* {T_{\overline{R}}}^b Y_R^n - {T_R}^a (Y_R^n)^* {T_{\overline{R}}}^b Y_R^m \right]
    \, ,
\end{split}\end{equation}
where, of course, the different ways of inserting the fields in the fermion loop, as well as the permutations of field legs of the same type, have to be considered.

The term $\Tr[\cdots]$ is equal to $(\mathcal{S}_R)^{ab}_{mn} \equiv ((\mathcal{C}_R)^{ab}_{mn} + (\mathcal{C}_R)^{ba}_{mn} + m \leftrightarrow n) / 2$, completely symmetric by exchanges $a \leftrightarrow b$ and $m \leftrightarrow n$, and $(\mathcal{C}_R)^{ab}_{mn} \equiv \Tr\left[ 2 \{ {T_R}^a , {T_R}^b \} (Y_R^m)^* Y_R^n - {T_R}^a (Y_R^m)^* {T_{\overline{R}}}^b Y_R^n \right]$.
Thus, the 1PI Green's function \cref{eq:1PIDeltaGPhiPhiGh} corresponds to the contribution
\begin{equation}
\label{eq:BIMSp1_BonDeGSS_SGS_SSG}
    [N[\widehat{\Delta}] \cdot \Gamma_\text{Ren}]^{(1)} \supset -\frac{\hbar}{16 \pi^2} \frac{(\mathcal{S}_R)^{ab}_{mn}}{3}
        \int \dInt[4]{x} \frac{g^2}{2} c_a \partial^\mu \left( G^b_\mu \Phi^m \Phi^n \right)
    \, .
\end{equation}
\end{subequations}
Besides, it is interesting to note that $\Tr\left[ {T_R}^a (Y_R^m)^* {T_{\overline{R}}}^b Y_R^n \right] = \Tr\left[ {T_{\overline{R}}}^a Y_R^n {T_R}^b (Y_R^m)^* \right]$, due to the symmetry properties of the Yukawa matrices and the definition of the generators in the conjugate representation.

\noindent
\uline{$\widecheck{\Delta} c^a \bar{\psi}_{i,\alpha} \psi_{j,\beta}$}:\\
\raisebox{-20pt}{\includegraphics[scale=0.55]{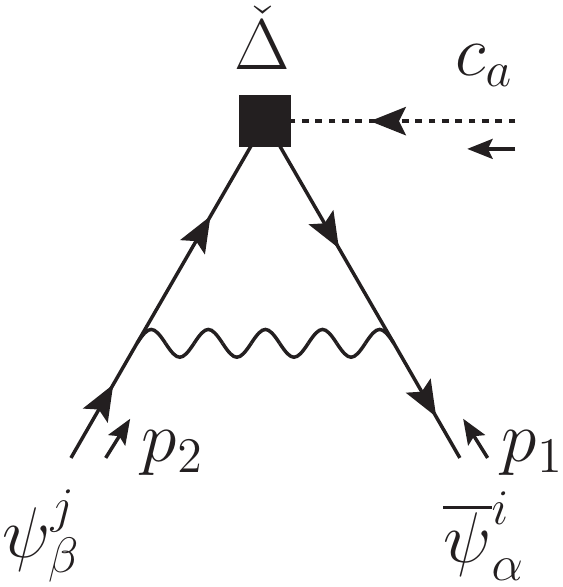}}
\raisebox{-20pt}{\includegraphics[scale=0.55]{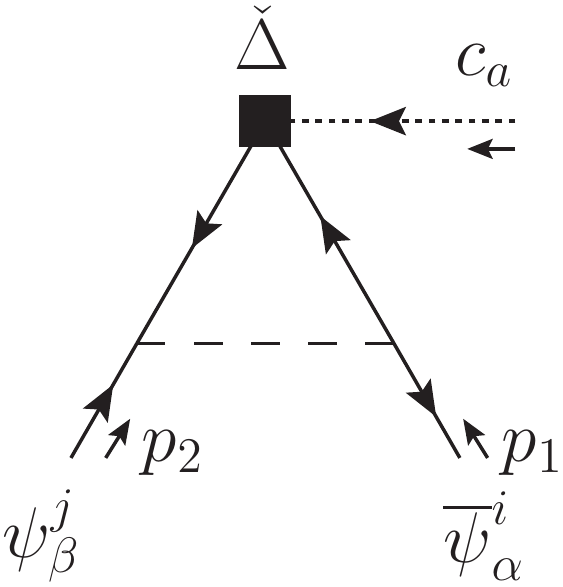}}
\raisebox{-20pt}{\includegraphics[scale=0.55]{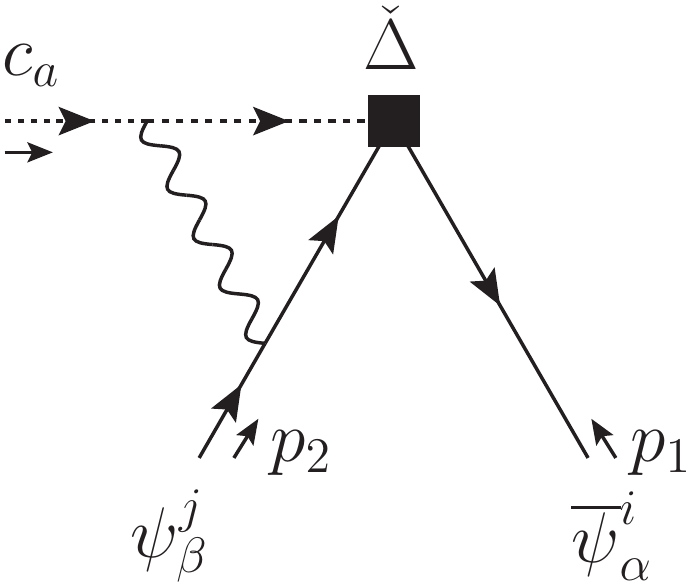}}
\raisebox{-20pt}{\includegraphics[scale=0.55]{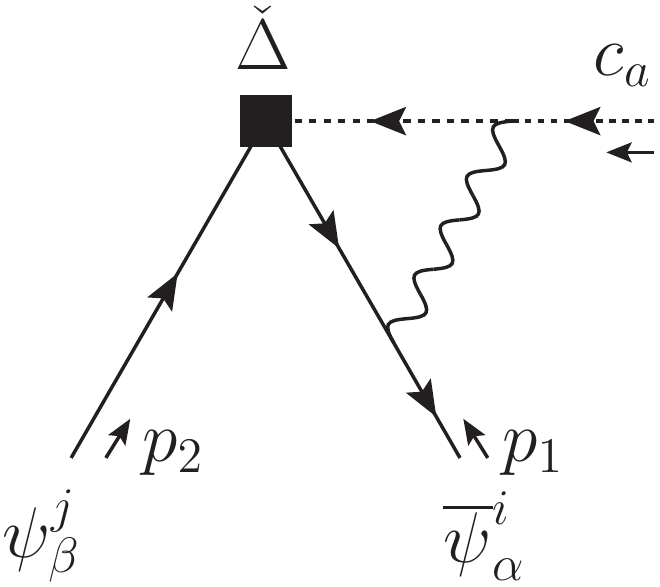}}
% $\widecheck{\Delta}(q=0) c^a(-p_1-p_2) \to \bar{\psi}_{i,\alpha}(p_1) \psi_{j,\beta}(p_2)$
\begin{subequations}
\begin{multline}
\label{eq:BIMSp1_BonDeFbF_Amp}
    \imath [\widecheck{\Delta} \cdot \Gamma_\OLDDReg {}_{\psi\bar{\psi} c}^{ji,a}]^{(1)}_\text{div} =
        \frac{\hbar g^3}{16 \pi^2 \epsilon} \left[ \frac{C_2(R) - C_2(G)/4}{2} + (\xi - 1) \frac{C_2(R)/6 - C_2(G)/4}{2} \right] {T_R}^a_{ij} \overline{\slashed{p_1} + \slashed{p_2}} \Proj{R} \\
        + \frac{\hbar g}{16 \pi^2 \epsilon} \frac{1}{4} ((Y_R^m)^* {T_{\overline{R}}}^a Y_R^m)_{ij} \overline{\slashed{p_1} + \slashed{p_2}} \Proj{R}
    \, .
\end{multline}
Note that here, contrary to the previous case when we inserted the evanescent $\widehat{\Delta}$ operator \cref{eq:BIMSp1_EvsctDeFbF}, the first two diagrams do not vanish, and the one with the scalar propagator provides the last scalar contribution in \cref{eq:BIMSp1_BonDeFbF_Amp}.
Using charge-conjugated fermionic legs, the scalar part becomes: $\frac{\hbar g}{16 \pi^2 \epsilon} (Y_R^m {T_R}^a (Y_R^m)^*)_{ji} \overline{\slashed{p_1} + \slashed{p_2}} \Proj{L} = -\frac{\hbar g}{16 \pi^2 \epsilon} ((Y_R^m)^* {T_{\overline{R}}}^a Y_R^m)_{ij} \overline{\slashed{p_1} + \slashed{p_2}} \Proj{L}$.
This 1PI Green's function corresponds to the contribution
\begin{multline}
\label{eq:BIMSp1_BonDeFbF}
    [N[\widehat{\Delta}] \cdot \Gamma_\text{Ren}]^{(1)} \supset
    -\frac{\hbar g}{16 \pi^2} \left\{ g^2 \left[ C_2(R) - \frac{C_2(G)}{4} + (\xi - 1) \left(\frac{C_2(R)}{6} - \frac{C_2(G)}{4}\right) \right] {T_R}^a_{ij} \right.\\
        \left. + \frac{1}{2} ((Y_R^m)^* {T_{\overline{R}}}^a Y_R^m)_{ij} \right\}
    \int \dInt[4]{x} c_a \partial_\mu (\overline{\psi}_i \gamma^\mu \Proj{R} \psi_j)
    \, .
\end{multline}
\end{subequations}

\noindent
\uline{$\widecheck{\Delta} c^a G^b_\mu \bar{\psi}_{i,\alpha} \psi_{j,\beta}$}:\\
\begin{figure}[h!]
	\centering
	\subfloat[Vanishing diagrams with fermion-scalar interactions.]{%
		\includegraphics[scale=0.55]{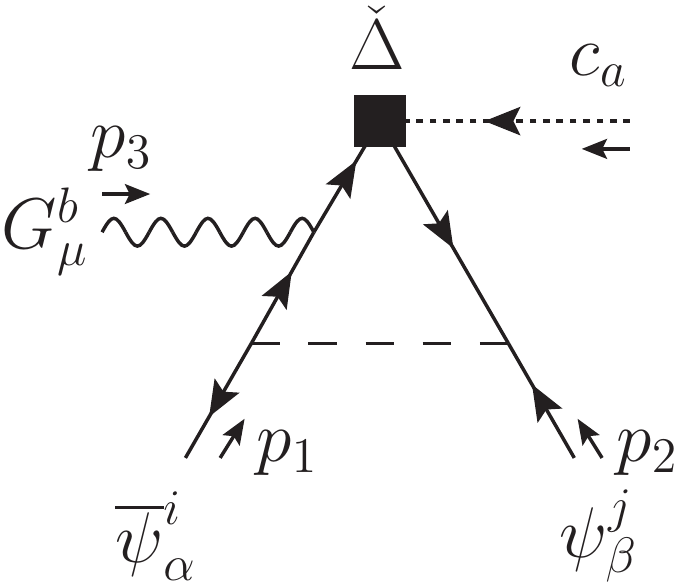}
		\includegraphics[scale=0.55]{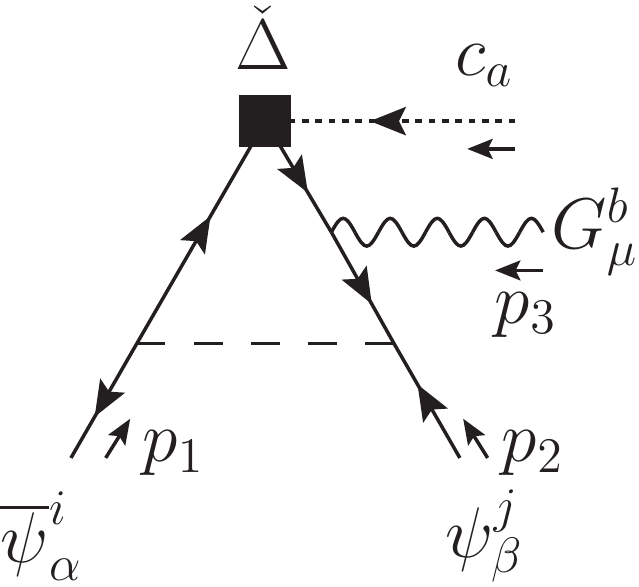}
		\includegraphics[scale=0.55]{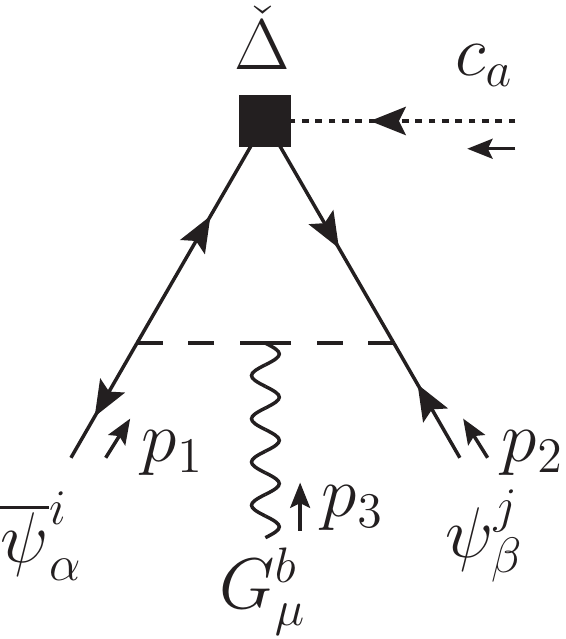}
	}
	\phantomcaption
\end{figure}
\begin{figure}[h!]
\ContinuedFloat
	\centering
	\subfloat[Vanishing diagrams with fermion-gauge boson interactions.]{%
		\includegraphics[scale=0.55]{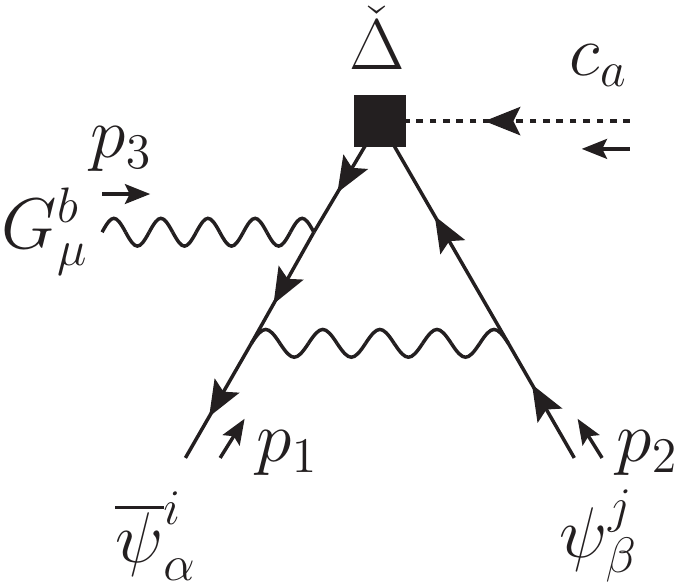}
		\includegraphics[scale=0.55]{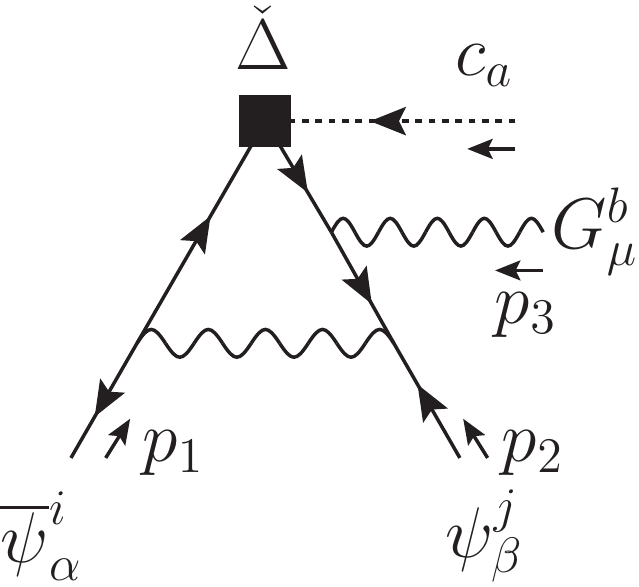}
		\includegraphics[scale=0.55]{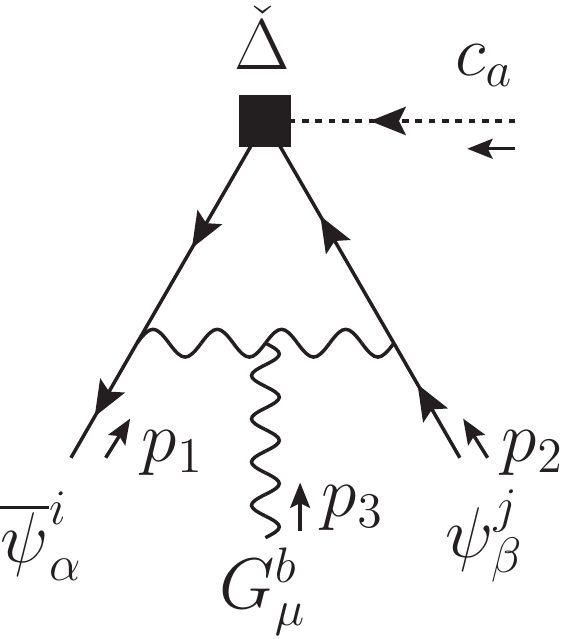}
	}
	\phantomcaption
\end{figure}
\begin{figure}[h!]
\ContinuedFloat
	\centering
	\subfloat[Diagrams cancelling with each other.]{%
		\includegraphics[scale=0.55]{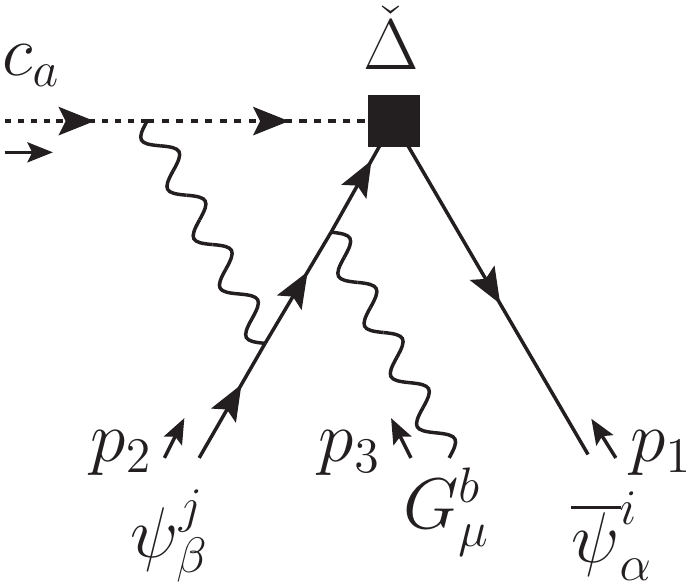}
		\includegraphics[scale=0.55]{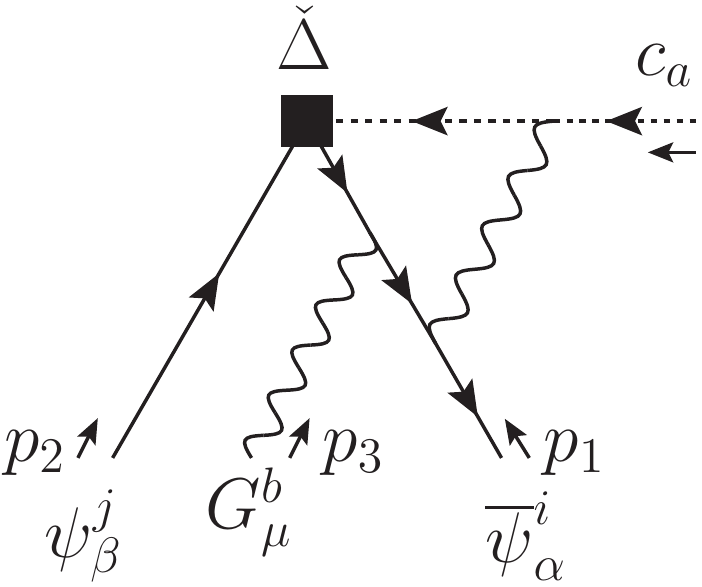}
	}
	\phantomcaption
\end{figure}
\begin{figure}[h!]
\ContinuedFloat
	\centering
	\subfloat[The four contributing diagrams; their group structures simplify considerably when summing the first two (and last two) diagrams together.]{%\shortstack{%
		\includegraphics[scale=0.55]{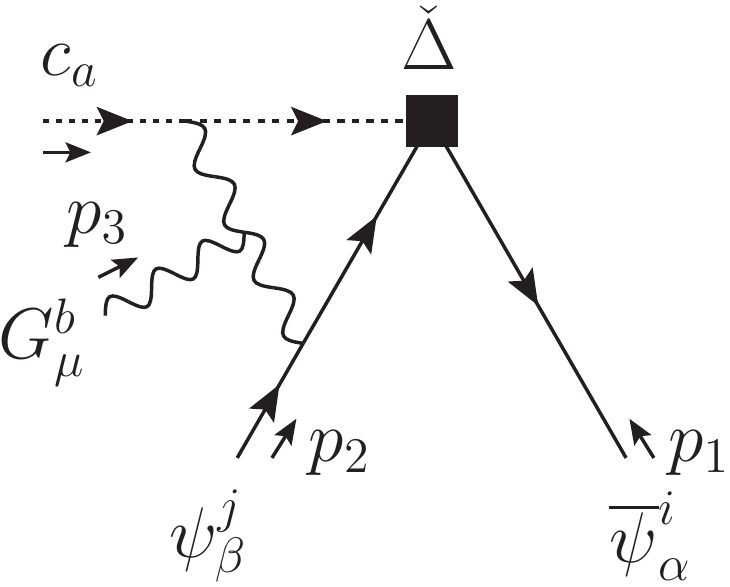}
		\includegraphics[scale=0.55]{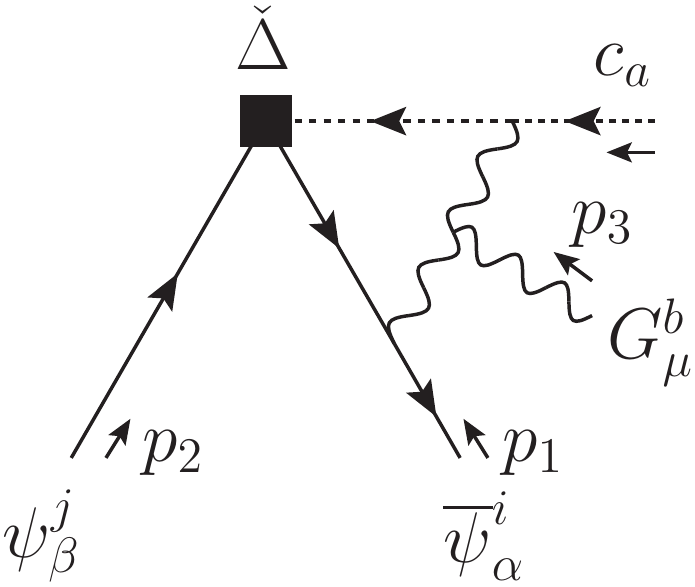}
		\includegraphics[scale=0.55]{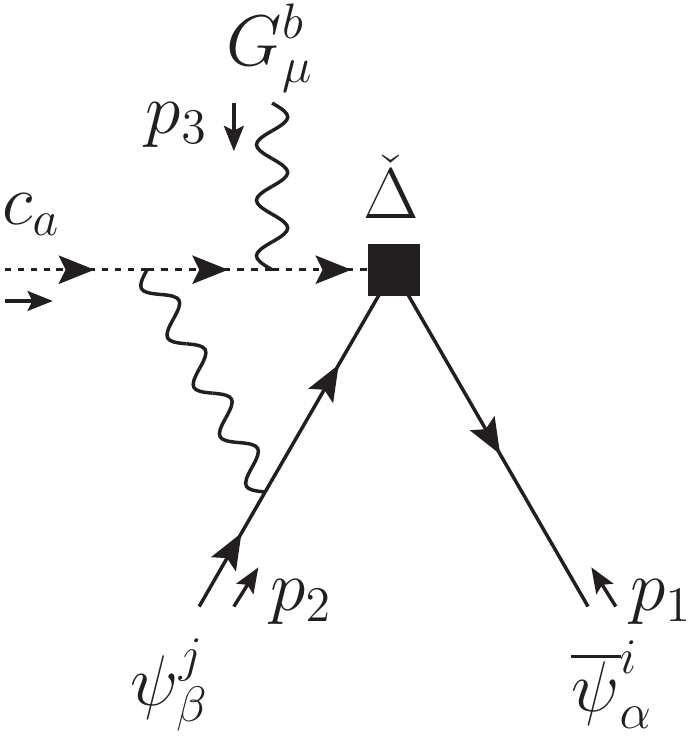}
		\includegraphics[scale=0.55]{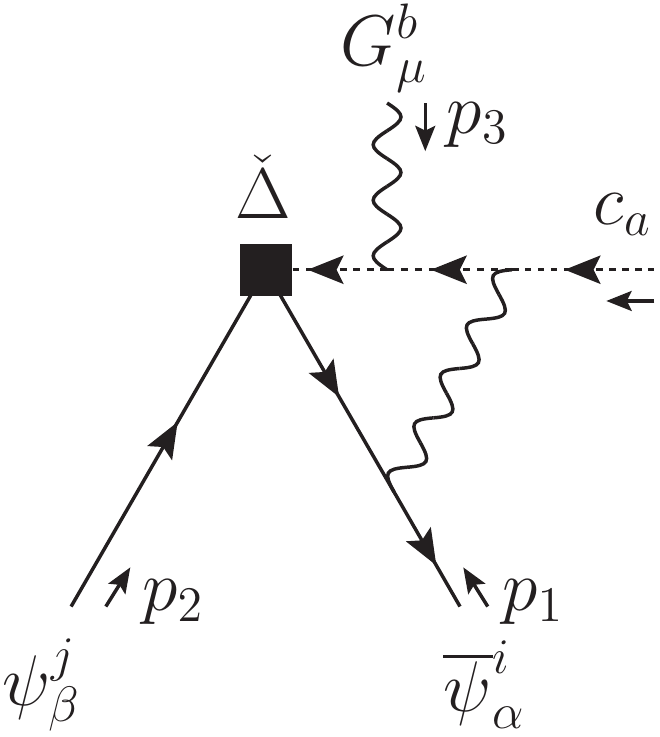}
	}
\end{figure}
% $\widecheck{\Delta}(q=0) c^a(-p_1-p_2-p_3) \to \bar{\psi}_{i,\alpha}(p_1) \psi_{j,\beta}(p_2) G^b_\mu(p_3)$
\begin{subequations}
\begin{equation}
    \imath [\widecheck{\Delta} \cdot \Gamma_\OLDDReg {}_{\psi\bar{\psi} G c}^{ji,ba,\mu}]^{(1)}_\text{div} =
          \frac{-\hbar g^4}{16 \pi^2 \epsilon} \frac{\xi C_2(G)}{8} \imath f^{abc} {T_R}^c_{ij} \overline{\gamma}^\mu \Proj{R}
        = \frac{-\hbar g^4}{16 \pi^2 \epsilon} \frac{\xi C_2(G)}{8} [{T_R}^a , {T_R}^b]_{ij} \overline{\gamma}^\mu \Proj{R}
    \, .
\end{equation}
Note that both the diagrams with the scalar propagators, and the diagrams with a gluon propagator connecting the fermions, are finite and thus do not contribute. Also, in our model there is no $GG\Phi$ vertex. The two diagrams with a gluon propagator connecting a fermion and the ghost leg cancel each other. The four remaining diagrams sum in pairs and their group structure simplify to get the simple result quoted above.
\\
This 1PI Green's function corresponds to the contribution
\begin{equation}
\label{eq:BIMSp1_BonDeFbFG}
    [N[\widehat{\Delta}] \cdot \Gamma_\text{Ren}]^{(1)} \supset
    \frac{-\imath \hbar g^2}{16 \pi^2} \frac{\xi C_2(G)}{4} \int \dInt[4]{x} \imath g^2 f^{abc} {T_R}^c_{ij} c_a \overline{\psi}_i \slashed{G}^b \Proj{R} \psi_j
    \, .
\end{equation}
\end{subequations}

\noindent
\uline{$\widecheck{\Delta} c^a \Phi^m \overline{\psi^C}_{i,\alpha} \psi_{j,\beta}$}:\\
\begin{figure}[h!]
	\centering
	\subfloat[Vanishing diagrams with fermion-scalar interactions.]{%
		\includegraphics[scale=0.55]{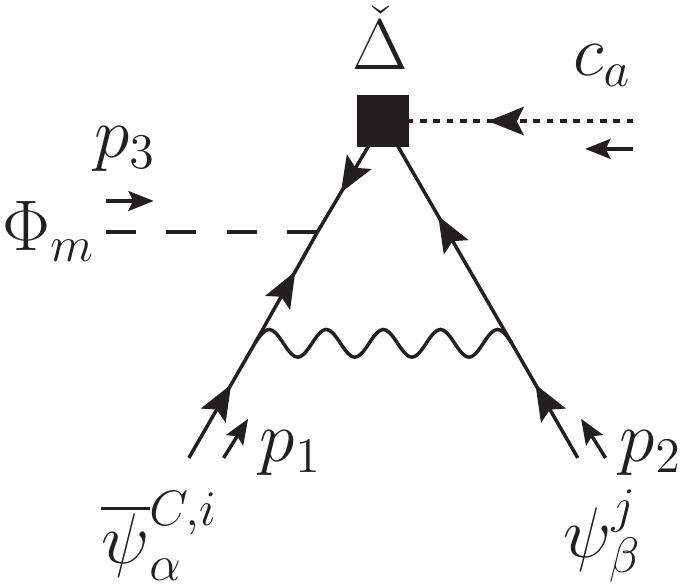}
		\includegraphics[scale=0.55]{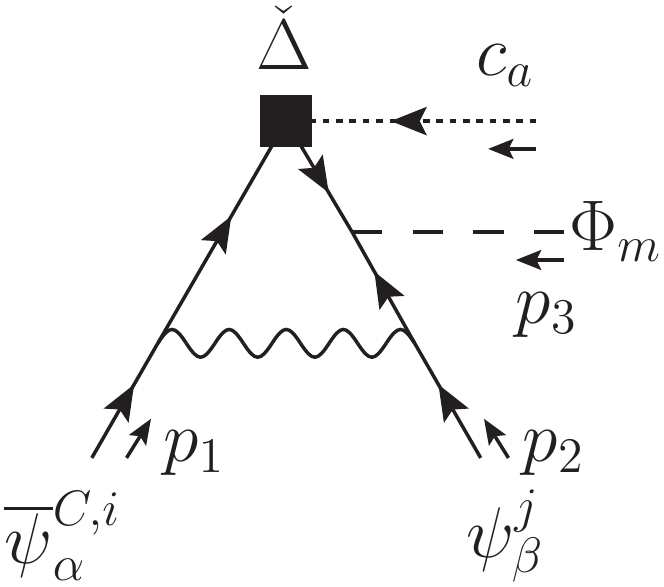}
	} \hfill
	\subfloat[Vanishing diagrams with fermion-gauge boson interactions.]{%
		\includegraphics[scale=0.55]{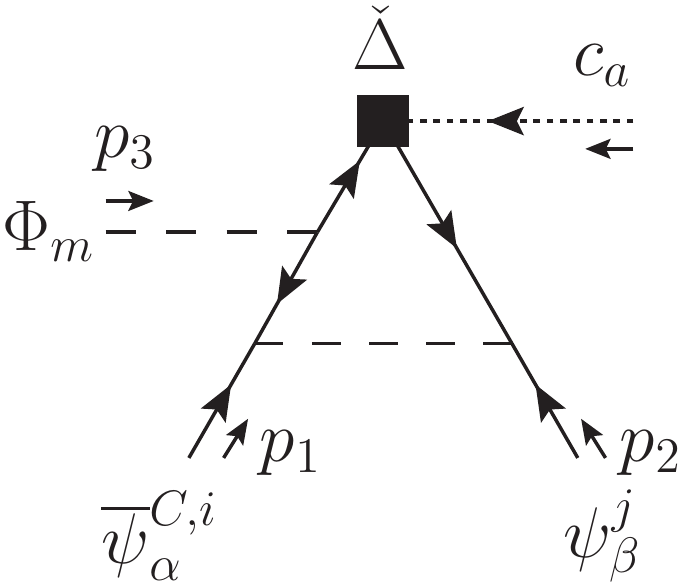}
		\includegraphics[scale=0.55]{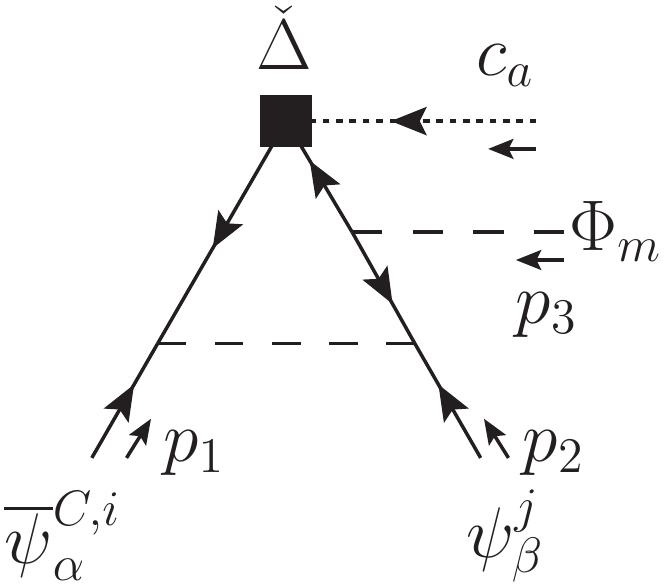}
	}
	\\
	\subfloat[Vanishing diagrams with fermion-gauge boson + fermion-scalar interactions.]{%
		\includegraphics[scale=0.55]{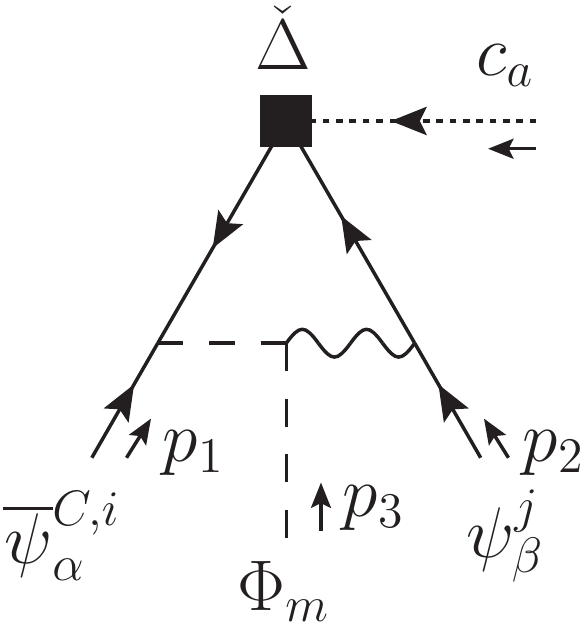}
		\includegraphics[scale=0.55]{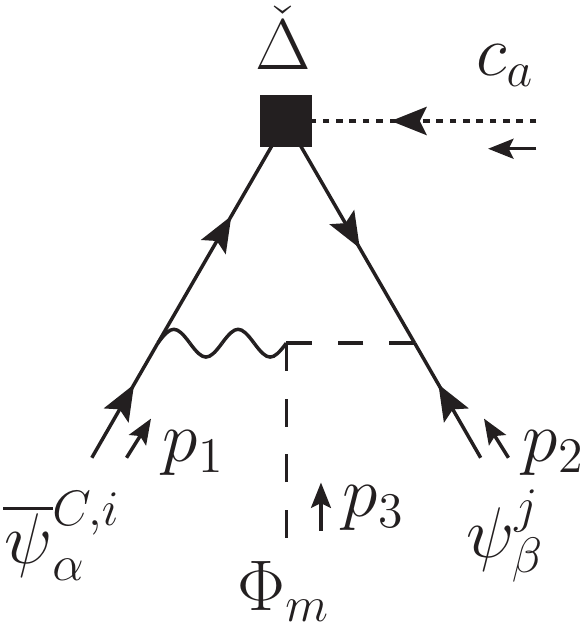}
	} \hfill
	\subfloat[Diagrams cancelling with each other.]{%
		\includegraphics[scale=0.55]{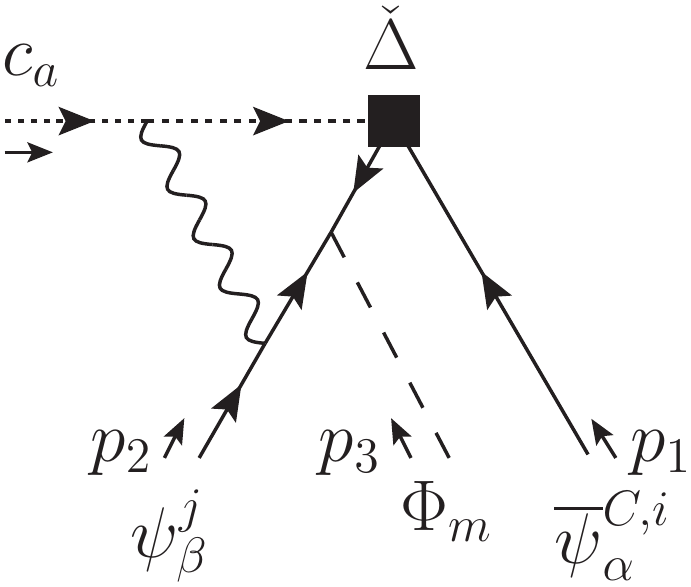}
		\includegraphics[scale=0.55]{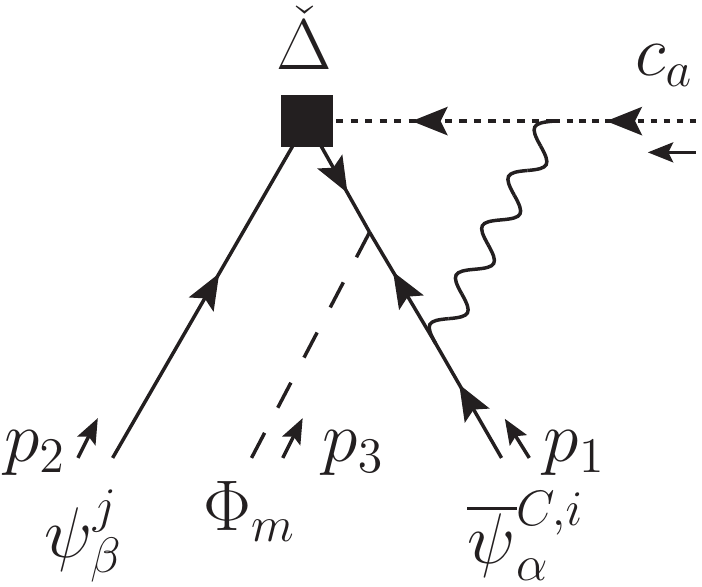}
	}
	\\
	\subfloat[The two contributing diagrams.]{%
		\includegraphics[scale=0.55]{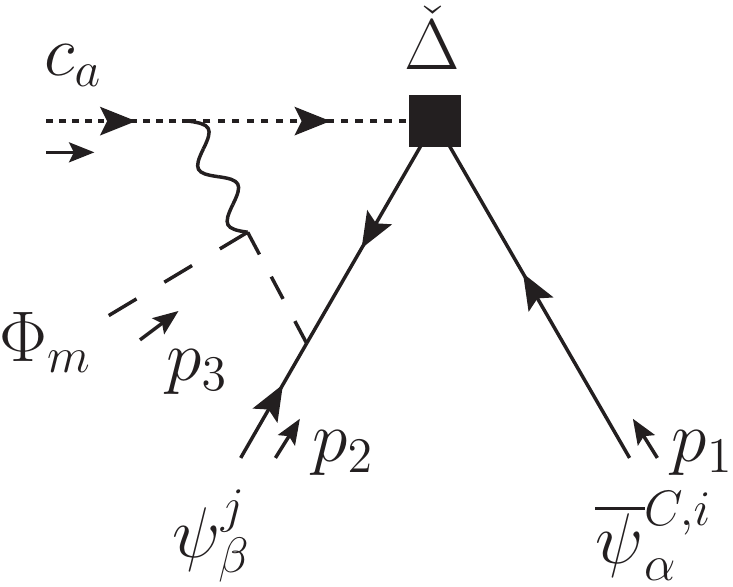}
		\includegraphics[scale=0.55]{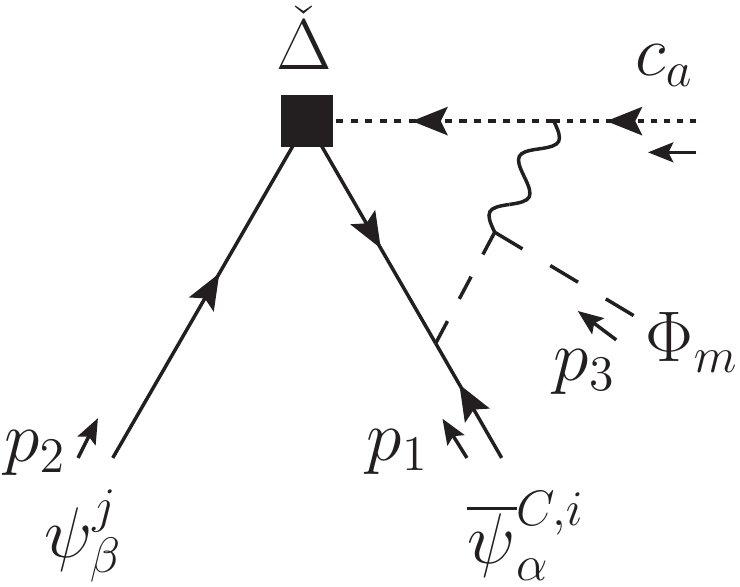}
	}
\end{figure}

\begin{subequations}
\begin{equation}
\label{eq:1PIDeltaPsiBCPsiPhiGh}
    \imath [\widecheck{\Delta} \cdot \Gamma_\OLDDReg {}_{\psi\overline{\psi^C} \Phi c}^{ji,m,a}]^{(1)}_\text{div} =
          \frac{\hbar g^3}{16 \pi^2 \epsilon} \frac{\xi C_2(G)}{8} (Y_R)^n_{ij} \theta^a_{nm} \Proj{R}
        = \frac{\hbar g^3}{16 \pi^2 \epsilon} \frac{\xi C_2(G)}{8} ( {T_{\overline{R}}}^a Y_R^m - Y_R^m {T_R}^a )_{ij} \Proj{R}
    \, .
\end{equation}
Similarly to the previous case $\widecheck{\Delta} c^a G^b_\mu \bar{\psi}_{i,\alpha} \psi_{j,\beta}$, the diagrams with scalar or gluonic propagators between the fermions, and also those with the vertex $G\Phi\Phi$ and scalar/gluon propagator between the fermions, are finite and thus do not contribute. Also, the two diagrams with a gluon propagator between a fermion and the ghost leg cancel each other. The two remaining diagrams form a pair whose total amplitude acquires a simpler group structure, after using the relation coming from the gauge-invariance of the Yukawa Lagrangian \cref{eq:GaugeInvarYuk}.
Thus, the 1PI Green's function \cref{eq:1PIDeltaPsiBCPsiPhiGh} corresponds to the contribution
\begin{equation}
\label{eq:BIMSp1_BonDeFbcFS}
    [N[\widehat{\Delta}] \cdot \Gamma_\text{Ren}]^{(1)} \supset
    \frac{\imath \hbar g^2}{16 \pi^2} \frac{\xi C_2(G)}{4} \int \dInt[4]{x} \frac{g}{2} (Y_R)^n_{ij} \theta^a_{nm} c_a \Phi^m \overline{\psi^C}_i \Proj{R} \psi_j
    \, .
\end{equation}
Associated with this term is the complex conjugate process \uline{$\widecheck{\Delta} c^a \Phi^m \overline{\psi}_{i,\alpha} \psi^C_{j,\beta}$} that generates a similar contribution to the Bonneau identity:
\begin{equation}
\label{eq:BIMSp1_BonDeFbFcS}
    [N[\widehat{\Delta}] \cdot \Gamma_\text{Ren}]^{(1)} \supset
    \frac{\imath \hbar g^2}{16 \pi^2} \frac{\xi C_2(G)}{4} \int \dInt[4]{x} \frac{g}{2} (Y_R)^{n\;*}_{ij} \theta^a_{nm} c_a \Phi^m \overline{\psi}_i \Proj{L} \psi^C_j
    \, .
\end{equation}
\end{subequations}

\subsubsection{1-loop vertices with insertion of one BRST-source-vertex and $\widecheck{\Delta}$}

At one-loop, and up to mass-dimension 4, the only 1PI diagrams containing a single insertion of $\widecheck{\Delta}$ and one BRST-source-vertex are those that only have one insertion of $\bar{R} s_d{\psi}$ or $R s_d{\overline{\psi}}$ BRST-source-vertex; these diagrams have mass-dimension four.
The reasons are as follows.

These diagrams should also have ghost number one since these are constituents of the Slavnov-Taylor identity.
The restriction on their mass-dimensions imposes that the sum of the mass-dimensions of their incoming and outgoing fields and derivatives, has to be smaller than or equal to four.
The BRST sources appear only as external fields and cannot be enclosed into loops, and their mass-dimensions are large (see \cref{tbl:fields_quantum_numbers}).
Furthermore, both the operator $\widecheck{\Delta}$ and any of the BRST-source-vertices contain only ghost fields, therefore all ghost lines from $\widecheck{\Delta}$ and any of the BRST-source-vertices give rise to an external ghost line.
% (since the only source of the antighost field is the $\bar{c}Gc$ vertex that leaves the ghost number unchanged).
Thus the mass-dimension and the ghost number constraints allow only the following operators: $\rho G cc$, $\rho \partial cc$, $\zeta ccc$, $\bar{R} \psi cc$, $R \overline{\psi} cc$ and $\mathcal{Y} \Phi cc$.
The operators $\rho G cc$, $\rho \partial cc$, $\zeta ccc$ and $\mathcal{Y} \Phi cc$ imply that the fermions from $\widecheck{\Delta}$ are enclosed into a loop, in which case one cannot form at one-loop level a 1PI diagram with the BRST-source-vertex. The remaining operators $\bar{R} \psi cc$ and $R \overline{\psi} cc$ may arise from one-loop contributions if one of the fermions of $\widecheck{\Delta}$ is contracted with a fermion from one of the operators $\bar{R} s_d{\psi}$ or $R s_d{\overline{\psi}}$.

\noindent
Only the following diagrams are therefore generated:
\\
\noindent
\uline{$\widecheck{\Delta} c^a c^b \bar{R}_{i,\alpha} \psi_{j,\beta}$}:\\
\begin{figure}[h!]
	\centering
	\includegraphics[scale=0.6]{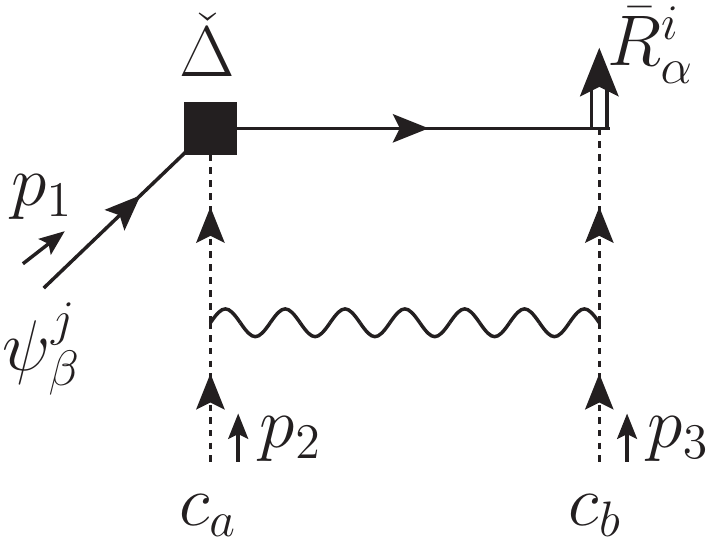}
	\includegraphics[scale=0.6]{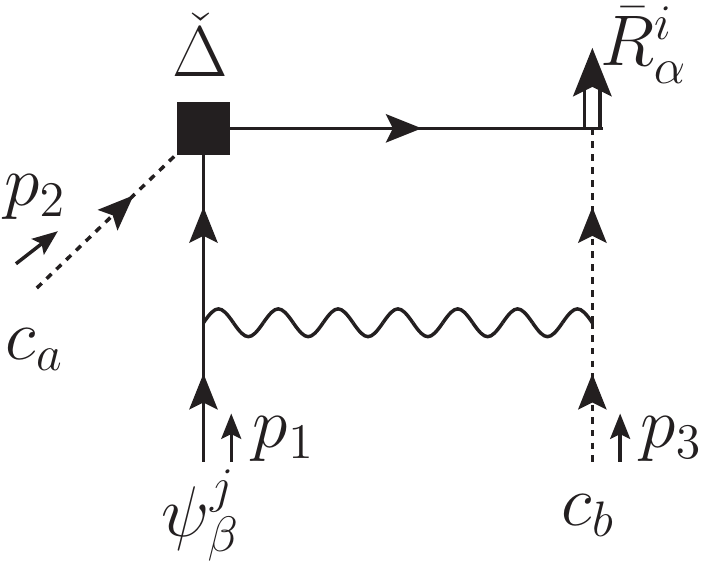} \\
	$+ (p_2,a) \leftrightarrow (p_3,b)$ permutation.
\end{figure}
% $\widecheck{\Delta}(q=0) c^a(p_2) c^b(p_3) \bar{R}_{i,\alpha}(-p_1-p_2-p_3) \psi_{j,\beta}(p_1)$
\begin{subequations}
\begin{equation}
    \imath [\widecheck{\Delta} \cdot \Gamma_\OLDDReg {}_{\psi \bar{R} c c}^{ji,ba}]^{(1)}_\text{div} =
          \frac{-\imath \hbar g^4}{16 \pi^2 \epsilon} \frac{\xi C_2(G)}{8} \imath f^{abc} {T_R}^c_{ij} \Proj{R}
        = \frac{-\imath \hbar g^4}{16 \pi^2 \epsilon} \frac{\xi C_2(G)}{8} [{T_R}^a , {T_R}^b]_{ij} \Proj{R}
    \, ,
\end{equation}
corresponding to the contribution
\begin{equation}
\label{eq:BIMSp1_BonDeFRbCC}
    [N[\widehat{\Delta}] \cdot \Gamma_\text{Ren}]^{(1)} \supset
    \frac{\hbar g^2}{16 \pi^2} \frac{\xi C_2(G)}{4} \int \dInt[4]{x} \imath \frac{g^2}{2} f^{abc} {T_R}^c_{ij} c^a c^b \bar{R}_i \Proj{R} \psi_j \, .
\end{equation}
\end{subequations}

\pagebreak

\noindent
\uline{$\widecheck{\Delta} c^a c^b \bar{\psi}_{i,\alpha} R_{j,\beta}$}:\\
\begin{figure}[h!]
	\centering
	\includegraphics[scale=0.6]{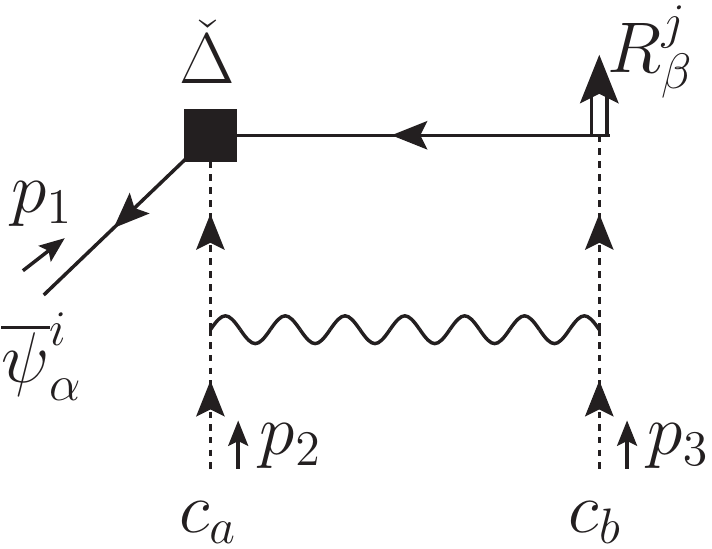}
	\includegraphics[scale=0.6]{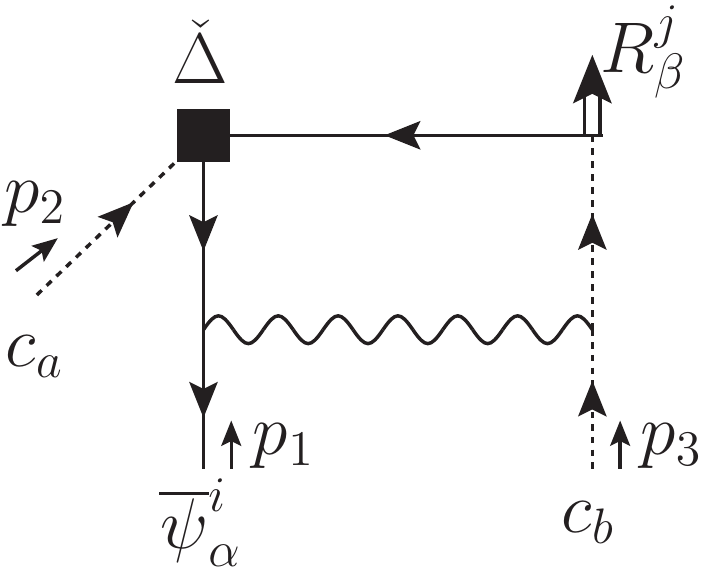} \\
	$+ (p_2,a) \leftrightarrow (p_3,b)$ permutation.
\end{figure}
% $\widecheck{\Delta}(q=0) c^a(p_2) c^b(p_3) \bar{\psi}_{i,\alpha}(p_1) R_{j,\beta}(-p_1-p_2-p_3)$
\begin{subequations}
\begin{equation}
    \imath [\widecheck{\Delta} \cdot \Gamma_\OLDDReg {}_{R \bar{\psi} c c}^{ji,ba}]^{(1)}_\text{div} =
          \frac{\imath \hbar g^4}{16 \pi^2 \epsilon} \frac{\xi C_2(G)}{8} \imath f^{abc} {T_R}^c_{ij} \Proj{L}
        = \frac{\imath \hbar g^4}{16 \pi^2 \epsilon} \frac{\xi C_2(G)}{8} [{T_R}^a , {T_R}^b]_{ij} \Proj{L}
    \, ,
\end{equation}
corresponding to the contribution
\begin{equation}
\label{eq:BIMSp1_BonDeFbRCC}
    [N[\widehat{\Delta}] \cdot \Gamma_\text{Ren}]^{(1)} \supset
    -\frac{\hbar g^2}{16 \pi^2} \frac{\xi C_2(G)}{4} \int \dInt[4]{x} \imath \frac{g^2}{2} f^{abc} {T_R}^c_{ij} c^a c^b \bar{\psi}_i \Proj{L} R_j \, .
\end{equation}
\end{subequations}
Note that only the diagrams with a gluon propagator connecting the two ghost lines do contribute, while those where the gluon propagator connects one ghost line with a fermion line do not.

\subsection{Finding the BRST-restoring finite 1-loop counterterms}
\label{subsect:fincts}

In this present section we evaluate the BRST-restoring finite 1-loop counterterms $S_\text{fct,restore}^{(1)}$.
From \cref{eq:defsymmetryrestore}, we see that these finite counterterms are defined such that their 4-dimensional linear BRST transformation $b{S_\text{fct,restore}^{(1)}}$ cancels $[N[\widehat{\Delta}] \cdot \Gamma_\text{Ren}]^{(1)}$, which has been evaluated in the previous \cref{subsect:Deltahatbreaking}.
We calculate these counterterms without imposing constraints on the fermion group representations and we also obtain the expression for the gauge anomalies as a by-product. These finite counterterms will be sufficient to restore the BRST invariance if the anomaly cancellation condition is met \cite{Piguet:1995er}.

In order to prepare our calculations and make them easier, it is reasonable to assume that $S_\text{fct,restore}^{(1)}$ will be a linear combination of all possible mass-dimension $\leq 4$ field monomials whose structure can appear while calculating 1-loop Feynman diagrams.
We therefore first evaluate all the linear BRST transformations of these monomials in \cref{subsect:bMonomialsEval},
then we combine these results and compare them in \cref{subsect:FiniteCTs} with the terms from $[N[\widehat{\Delta}] \cdot \Gamma_\text{Ren}]^{(1)}$ so as to find the finite counterterms $S_\text{fct,restore}^{(1)}$.

\subsubsection{Evaluation of linear BRST transformation for some field monomials}
\label{subsect:bMonomialsEval}

The following calculations are also performed in 4 dimensions, so we will again
omit all the ``overlines'' over all the Lorentz covariants so as to simplify the notation.
The notations for the integrated field monomials are the same as in
\cref{eq:RModelDReg_Action} (\cref{subsect:RModelDReg}), but now all defined
purely in 4 dimensions.
We obtain:

\begin{align}
	b \int \dInt[4]{x} \frac{1}{2} G^{a\,\mu} \partial^2 G^a_\mu &= \int \dInt[4]{x} (\partial^\mu c_a + g f^{abc} G^{b\,\mu} c_c) \partial^2 G^a_\mu \, ,
\end{align}
\begin{align}
  b S_{GG} &= b \int \dInt[4]{x} \frac{1}{2} G^a_\mu (g^{\mu\nu} \partial^2 - \partial^\mu \partial^\nu) G^a_\nu = - g f^{abc} \int \dInt[4]{x} c^a G^b_\mu (g^{\mu\nu} \partial^2 - \partial^\mu \partial^\nu) G^c_\nu \, ,
\end{align}
where we used the fact that $(\partial_\mu c^a)(g^{\mu\nu} \partial^2 - \partial^\mu \partial^\nu) G^a_\nu = 0$ when using integrations by parts.
\begin{align}
	b S_{GGGG} &= -\frac{g^2}{2} (f^{abe} f^{cde} + f^{ace} f^{bde}) \int \dInt[4]{x} c^a \partial_\nu \left( G^b_\mu G^{c\,\mu} G^{d\,\nu} \right) \, .
\end{align}
In this calculation, a term proportional to $\int \dInt[4]{x} c^f G^e_\mu G^{b\,\mu} G^c_\nu G^{d\,\nu}$ actually cancels. Indeed, its prefactor is given by: $(f^{acg} f^{bdg} + f^{adg} f^{bcg}) f^{aef}$, which vanishes after symmetrizing with respect to the group indices $e \leftrightarrow b$, $c\leftrightarrow d$, and the set $(e,b) \leftrightarrow (c,d)$.
Also, because $\frac{-1}{4} \int \dInt[4]{x} F^a_{\mu\nu} F^{a\,\mu\nu} = S_{GG} + S_{GGG} + S_{GGGG}$ is gauge-invariant, $b \int \dInt[4]{x} F^a_{\mu\nu} F^{a\,\mu\nu} = 0$ and we have:
\begin{align}
	b S_{GGG} &= - b S_{GG} - b S_{GGGG} \, .
\end{align}

\begin{multline}
	b (T_R)^{abcd} \int \dInt[4]{x} G^a_\mu G^{b\,\mu} G^c_\nu G^{d\,\nu} = \\
		-((T_R)^{abcd} + (T_R)^{acbd} + (T_R)^{adbc} + (T_R)^{adcb}) \int \dInt[4]{x} c^a \partial_\nu \left( G^b_\mu G^{c\,\mu} G^{d\,\nu} \right)
	\, .
\end{multline}
As before, a term proportional to $\int \dInt[4]{x} c^f G^e_\mu G^{b\,\mu} G^c_\nu G^{d\,\nu}$ cancels. Its prefactor is given by: $((T_R)^{abcd} + (T_R)^{acbd} + (T_R)^{adbc} + (T_R)^{adcb}) f^{aef}$ (using the shorthand notation $(T_R)^{abcd} \equiv \Tr[{T_R}^a \cdots {T_R}^d]$), and vanishes after symmetrization with respect to the group indices $e \leftrightarrow b$, $c\leftrightarrow d$, and the set $(e,b) \leftrightarrow (c,d)$.

\begin{align}
	b S_{\Phi\Phi} &= b \int \dInt[4]{x} \frac{-1}{2} \Phi_m \partial^2 \Phi_m = \int \dInt[4]{x} \imath g \theta^a_{mn} c^a \Phi_m \partial^2 \Phi_n \, ,
        \\
	b S_{\Phi GG \Phi} &= -\frac{g^2}{2} \{\theta^a,\theta^b\}_{mn} \int \dInt[4]{x} (\partial^\mu c^a) G^b_\mu \Phi^m \Phi^n \, ,
\end{align}
and because $\frac{1}{2} (D_\mu \Phi^m)^2 = S_{\Phi\Phi} + S_{\Phi G \Phi} + S_{\Phi GG \Phi}$ is gauge-invariant, $b (D_\mu \Phi^m)^2 = 0$ and we have:
\begin{align}
	b S_{\Phi G \Phi} &= - b S_{\Phi\Phi} - b S_{\Phi GG \Phi} \, .
\end{align}

For an arbitrary group symbol $\mathcal{C}^a_{mn}$,
\begin{equation}\begin{aligned}
	b \mathcal{C}^a_{mn} \int \dInt[4]{x} (\partial^\mu \Phi^m) G^a_\mu \Phi^n =\;&
		- \mathcal{C}^a_{mn} \int \dInt[4]{x} c^a (\partial^2 \Phi^m) \Phi^n \\
		& - \frac{1}{2} (\mathcal{C}^a_{mn} + \mathcal{C}^a_{nm}) \int \dInt[4]{x} c^a (\partial^\mu \Phi^m) (\partial_\mu \Phi^n) \\
		& + \imath g \left[ \imath f^{abc} \mathcal{C}^c_{nm} + \theta^a_{mo} (\mathcal{C}^b_{on} - \mathcal{C}^b_{no}) \right] \int \dInt[4]{x} c^a G^b_\mu \Phi^m (\partial^\mu \Phi^n) \\
		& + \frac{\imath g}{2} ( \theta^a_{mo} \mathcal{C}^b_{on} + \theta^a_{no} \mathcal{C}^b_{om} ) \int \dInt[4]{x} c^a (\partial^\mu G^b_\mu) \Phi^m \Phi^n
	\, ,
\end{aligned}\end{equation}
and, for an arbitrary group symbol $\mathcal{C}^{ab}_{mn}$,
\begin{equation}
	b \mathcal{C}^{ab}_{mn} \int \dInt[4]{x} G^a_\mu G^{b\,\mu} \Phi^m \Phi^n =
		- \mathcal{S}^{ab}_{mn} \int \dInt[4]{x} c^a \partial_\mu \left( G^{b\,\mu} \Phi^m \Phi^n \right)
	\, ,
\end{equation}
where $\mathcal{S}^{ab}_{mn} = (\mathcal{C}^{ab}_{mn} + \mathcal{C}^{ba}_{mn} + m \leftrightarrow n) / 2$, completely symmetric by exchanges $a \leftrightarrow b$ and $m \leftrightarrow n$.
In this calculation, a term proportional to the field monomial $\int \dInt[4]{x} c^a G^b_\mu G^{d\,\mu} \Phi^m \Phi^n$ actually cancels. Indeed, its prefactor is given by: $f^{acd} \mathcal{S}^{bc}_{mn} - \imath \theta^a_{mo} \mathcal{S}^{bd}_{on}$, and one can show that its contraction with the field monomial vanishes after symmetrizing with respect to the group indices $(b,d)$ and $(m,n)$.

We explicitly evaluate in addition the following 4-dimensional linear BRST transformations of the following fermionic operators, as these are the ones being involved in the definition of the finite counterterm action, which is naturally defined in 4 dimensions. (Note that if we were interested in their $d$-dimensional version, these would contain extra evanescent contributions.)
\begin{align}
	b S_{\overline{\psi}\psi} =\;&
		b \int \dInt[4]{x} \imath \overline{\psi}_i \slashed{\partial} \psi_i
		= g {T_R}^a_{ij} \int \dInt[4]{x} c^a \partial_\mu \left( \overline{\psi}_i \gamma^\mu \Proj{R} \psi_j \right)
	\, ,
	\\
	\begin{split}
	b ( S_{\bar{R} c \psi_R} + S_{R c \overline{\psi_R}} ) =\;&
		+ \imath \frac{g}{2} \theta^a_{nm} \int \dInt[4]{x} c^a \Phi^m \left( (Y_R)^n_{ij} \overline{\psi}^C_i \Proj{R} \psi_j + (Y_R)^{n\;*}_{ij} \overline{\psi}_i \Proj{L} \psi^C_j \right) \\
		& + \imath \frac{g^2}{2} f^{abc} {T_R}^c_{ij} \int \dInt[4]{x} c^a c^b \left( \bar{R}^i \Proj{R} \psi_j - \overline{\psi}_i \Proj{L} R^j \right) \\
		& + g^2 f^{abc} {T_R}^c_{ij} \int \dInt[4]{x} c^a \overline{\psi}_i \slashed{G}^a \Proj{R} \psi_j \\
		& + g {T_R}^a_{ij} \int \dInt[4]{x} c^a \partial_\mu \left( \overline{\psi}_i \gamma^\mu \Proj{R} \psi_j \right)
	\, .
	\end{split}
\end{align}

\subsubsection{Grouping all results together -- The Finite One-Loop Counterterms}
\label{subsect:FiniteCTs}

\noindent
The total contribution of \uline{$\widecheck{\Delta} c^a G^b_\mu + \widecheck{\Delta} c^a G^b_\mu G^c_\nu + \widecheck{\Delta} c^a G^b_\mu G^c_\nu G^d_\rho$} \cref{eq:BIMSp1_BonDeG,eq:BIMSp1_BonDeGG,eq:BIMSp1_BonDeGGG} is equal to:
\begin{equation}
    -\frac{\hbar g^2}{16 \pi^2} \left\{ \frac{S_2(R)}{6} b \left( 5 S_{GG} + S_{GGG} - \int \dInt[4]{x} G^{a\,\mu} \partial^2 G^a_\mu \right)
    + \frac{g^2}{12} (T_R)^{abcd} b \int \dInt[4]{x} G^a_\mu G^{b\,\mu} G^c_\nu G^{d\,\nu} \right\}
    \, ,
\end{equation}
together with relevant anomalies
\begin{equation}
    -\frac{\hbar g^2}{16 \pi^2} \left( \frac{S_2(R)}{3} d_R^{abc} \int \dInt[4]{x} g \epsilon^{\mu\nu\rho\sigma} c_a (\partial_\rho G^b_\mu) (\partial_\sigma G^c_\nu)
    + \frac{\mathcal{D}_R^{abcd}}{3 \times 3!} \int \dInt[4]{x} g^2 c_a \epsilon^{\mu\nu\rho\sigma} \partial_\sigma \left( G^b_\mu G^c_\nu G^d_\rho \right) \right)
    \, .
\end{equation}

\noindent
The contribution of \uline{$\widecheck{\Delta} c^a \Phi^m \Phi^n$} \cref{eq:BIMSp1_BonDeSS} is equal to:\\
\begin{equation}
    -\frac{\hbar}{16 \pi^2} \frac{Y_2(S)}{3} b{\overline{S_{\Phi\Phi}}} \, .
\end{equation}

\noindent
The contribution of \uline{$\widecheck{\Delta} c^a G^b_\mu \Phi^m \Phi^n$} \cref{eq:BIMSp1_BonDeGSS_SGS_SSG} is equal to:\\
\begin{equation}
    \frac{\hbar}{16 \pi^2} \frac{(\mathcal{C}_R)^{ab}_{mn}}{3}
        b \int \dInt[4]{x} \frac{g^2}{2} G^a_\mu G^{b\,\mu} \Phi^m \Phi^n
    \, ,
\end{equation}
with $(\mathcal{C}_R)^{ab}_{mn} \equiv \Tr\left[ 2 \{ {T_R}^a , {T_R}^b \} (Y_R^m)^* Y_R^n - {T_R}^a (Y_R^m)^* {T_{\overline{R}}}^b Y_R^n \right]$.

\noindent
The total contribution of \uline{$\widecheck{\Delta} c^a \bar{\psi}_{i,\alpha} \psi_{j,\beta} + \widecheck{\Delta} c^a G^b_\mu \bar{\psi}_{i,\alpha} \psi_{j,\beta} + \widecheck{\Delta} c^a \Phi^m \overline{\psi^C}_{i,\alpha} \psi_{j,\beta} + \widecheck{\Delta} c^a \Phi^m \overline{\psi}_{i,\alpha} \psi^C_{j,\beta}$ \\ $+ \widecheck{\Delta} c^a c^b \bar{R}_{i,\alpha} \psi_{j,\beta} + \widecheck{\Delta} c^a c^b \bar{\psi}_{i,\alpha} R_{j,\beta}$} \cref{eq:BIMSp1_BonDeFbF,eq:BIMSp1_BonDeFbFG,eq:BIMSp1_BonDeFbcFS,eq:BIMSp1_BonDeFbFcS,eq:BIMSp1_BonDeFRbCC,eq:BIMSp1_BonDeFbRCC} is equal to:\\
\begin{multline}
      \frac{-\hbar g^2}{16 \pi^2} \left(1 + \frac{\xi - 1}{6}\right) C_2(R) b{S_{\overline{\psi}\psi}}
    + \frac{\hbar}{16 \pi^2} \frac{\left( (Y_R^m)^* {T_{\overline{R}}}^a Y_R^m \right)_{ij}}{2} b \int \dInt[4]{x} g \overline{\psi}_i \slashed{G}^a \Proj{R} \psi_j \\
    + \frac{\hbar g^2}{16 \pi^2} \frac{\xi C_2(G)}{4} b{(S_{\bar{R} c \psi_R} + S_{R c \overline{\psi_R}})}
    \, .
\end{multline}

All in all, the BRST-restoring finite counterterms defined in 4 dimensions such as $b S_\text{fct,restore}^{(1)}$ cancels the contributions from $[N[\widehat{\Delta}] \cdot \Gamma_\text{Ren}]^{(1)}$, are:
\begin{equation}
\label{eq:FiniteCT1Loop}
\begin{split}
	S_{\text{fct,restore}}^{(1)} =\;&
	\frac{\hbar}{16 \pi^2} \left\{
		g^2 \frac{S_2(R)}{6} \left( 5 S_{GG} + S_{GGG} - \int \dInt[4]{x} G^{a\,\mu} \partial^2 G^a_\mu \right) + \frac{Y_2(S)}{3} \overline{S_{\Phi\Phi}} \right.\\
		&\left. + g^2 \frac{(T_R)^{abcd}}{3} \int \dInt[4]{x} \frac{g^2}{4} G^a_\mu G^{b\,\mu} G^c_\nu G^{d\,\nu}
		        - \frac{(\mathcal{C}_R)^{ab}_{mn}}{3} \int \dInt[4]{x} \frac{g^2}{2} G^a_\mu G^{b\,\mu} \Phi^m \Phi^n \right.\\
		&\left. + g^2 \left(1 + \frac{\xi - 1}{6}\right) C_2(R) S_{\overline{\psi}\psi}
		        - \frac{\left( (Y_R^m)^* {T_{\overline{R}}}^a Y_R^m \right)_{ij}}{2} \int \dInt[4]{x} g \overline{\psi}_i \slashed{G}^a \Proj{R} \psi_j \right.\\
		&\left. - g^2 \frac{\xi C_2(G)}{4} (S_{\bar{R} c \psi_R} + S_{R c \overline{\psi_R}})
		\right\}
	\, ,
\end{split}
\end{equation}
with $(\mathcal{C}_R)^{ab}_{mn} \equiv \Tr\left[ 2 \{ {T_R}^a , {T_R}^b \} (Y_R^m)^* Y_R^n - {T_R}^a (Y_R^m)^* {T_{\overline{R}}}^b Y_R^n \right]$, and the relevant (non-spurious) anomalies are:
\begin{equation}
\label{eq:Anomalies}
    -\frac{\hbar g^2}{16 \pi^2} \left( \frac{S_2(R)}{3} d_R^{abc} \int \dInt[4]{x} g \epsilon^{\mu\nu\rho\sigma} c_a (\partial_\rho G^b_\mu) (\partial_\sigma G^c_\nu)
    + \frac{\mathcal{D}_R^{abcd}}{3 \times 3!} \int \dInt[4]{x} g^2 c_a \epsilon^{\mu\nu\rho\sigma} \partial_\sigma \left( G^b_\mu G^c_\nu G^d_\rho \right) \right)
    \, ,
\end{equation}
with the fully symmetric symbol $d_R^{abc} = \Tr[ {T_R}^a \{{T_R}^b , {T_R}^c\} ]$,
and the fully antisymmetric symbol $\mathcal{D}_R^{abcd} = (-\imath) 3! \Tr[ {T_R}^a {T_R}^{[b} {T_R}^c {T_R}^{d]} ]$ for the R-representation.
In realistic renormalizable models, the fermionic content and the associated group representations are chosen so as to cancel these anomalies, i.e. by cancelling separately both $\sum_R S_2(R) d_R^{abc}$ (proportional to the usual triangle anomaly) and $\sum_R \mathcal{D}_R^{abcd}$.

This equation \cref{eq:FiniteCT1Loop} thus represents the main result of this paper.
If the anomalies \cref{eq:Anomalies} are canceled, these finite counterterms are \emph{necessary and sufficient} to restore the BRST symmetry at 1-loop level in the BMHV scheme.
They are necessary building blocks for a consistent 1-loop applications of the scheme, and they are vital ingredients in two-loop and higher-loop order calculations.
It should be noted that these finite counterterms, purely 4-dimensional and non-evanescent, are \emph{not} gauge-invariant!
They modify all the self-energies, as well as some specific interactions: the gauge-boson self-interactions, and the interactions between gauge-boson and scalars or fermions.

As previously mentioned in the remarks around \cref{eq:SFct_Struct},
one can also add, to these BRST-restoring finite counterterms, any other finite counterterms that are BRST-invariant, or even that are evanescent (because they will nonetheless vanish after taking the $\mathop{\text{LIM}}_{d \to 4}$), when being defined in $d$ dimensions. However, both of these will not contribute to BRST restoration; they will instead only correspond to a change of renormalization prescription for higher-order calculations, see discussion below \cref{eq:CT_structure} in \cref{sect:standardrenormalizationstructure}.
For example, the BRST-invariant finite counterterms could contain a linear combination of the $L_\varphi$ functionals defined in \cref{eq:Lfuncts_fields,eq:Lfuncts_couplings}.

\section{The Renormalization Group Equation in the Renormalized Model}
\label{sect:RGE}

In the present and the subsequent sections, we present the derivation of the renormalization group equation in the BMHV scheme. We focus particularly on the role of the extra counterterms specific to this scheme. Since the result will be equal to the known one, this serves as a check of the procedure and as an explanation how the additional counterterms can be treated. The present section uses methods from the abstract framework of algebraic renormalization theory, while the subsequent section proceeds in the more familiar way using renormalization constants. In both cases, we see that evanescent contributions play no role at the 1-loop level but will have an influence at higher orders. Hence, these sections provide important background information for future multi-loop applications.

As we have shown in the previous sections, the set of operators in the tree-level action is not the same set that exists at the one-loop level when using the BMHV dimensional renormalization scheme. Due to the presence of evanescent operators and finite non-evanescent counterterms needed to restore the BRST symmetry, the formalism of multiplicative renormalization (with bare fields, bare coupling constants and $Z$-factors) will not straightforwardly lead to the true renormalization group equation, that involves only fields and parameters of the original 4-dimensional tree-level action (see also discussion in Ref.\ \cite{Martin:1999cc}).
This will be briefly overviewed in \cref{sect:MultRenorm}.

Instead if we start with the dimensionally renormalized 1PI functional $\Gamma_\text{Ren}$, see \cref{eq:definitionGammaRen},
and we use the Quantum Action Principle and the Bonneau identities, the formalism of bare objects and $Z$-factors can be avoided.
From now on we take this effective action to be anomaly free, i.e. the anomalies described by \cref{eq:Anomalies} are cancelled.

\subsection{Basis of Insertions}

In the context of the algebraic renormalization framework, it can be shown \cite{Piguet:1995er} that the renormalization group equation corresponds to the expansion of the operator insertion
\begin{equation}
\label{eq:eqinsert}
	\mu \frac{\partial}{\partial \mu}\Gamma_\text{Ren} = \mathcal{O} \cdot \Gamma_\text{Ren}	
\end{equation}
in a suitable basis of operators of ultraviolet dimension 4, ghost number 0, with contracted Lorentz indices but free gauge indices (later contracted with  group factors from the associated coefficients).
The basis is compounded of operators that respect the same symmetries as the functional $\mu {\partial \Gamma_\text{Ren}}/{\partial \mu}$, and they are, generally speaking, operators comprising derivatives with respect to the parameters of the theory, and field-counting operators,
\begin{equation}
\label{eq:RGExc}
	\mu \frac{\partial \Gamma_\text{Ren}}{\partial \mu} = \Big(-\sum_{g} \beta_g \frac{\partial}{\partial g} + \sum_{\phi} \gamma_\phi N_{\phi} \Big) \Gamma_\text{Ren} \, .
\end{equation}
As we will see in \cref{subsect:mu_DGamma_Dmu}, evaluating \eqref{eq:eqinsert} and \eqref{eq:RGExc}
independently will result in a system of equations, overdetermined and solvable by direct comparison of their coefficients.

Let us now specialize these generally valid facts to the model discussed in our paper. Our basis will have the same symmetries  as $\Gamma_\text{Ren}$, so it should respect the following equations \cite{Piguet:1995er}:
\begin{align}
\label{eq:STIInvarianceEqs}
	\mu \frac{\partial \mathcal{S}(\Gamma_\text{Ren})}{\partial \mu} &= \mathcal{S}_{\Gamma_\text{Ren}} \, \mu \frac{\partial \Gamma_\text{Ren}}{\partial \mu} = 0 \, , &
	\frac{\delta}{\delta B} \, \mu \frac{\partial \Gamma_\text{Ren}}{\partial \mu} &= 0 \, , &
	\mathcal{G} \, \mu \frac{\partial \Gamma_\text{Ren}}{\partial \mu} &= 0 \, ,
\end{align}
i.e. respectively the BRST equation, the gauge-fixing condition and the ghost equation \cite{Piguet:1995er} (with $\mathcal{G} \equiv {\delta}/{\delta \bar{c}_a} + \partial^\mu {\delta}/{\delta \rho_a^\mu} \equiv {\delta}/{\delta \widetilde{\rho}_a^\mu}$), and where
\begin{equation*}
\label{eq:3eq}
\begin{split}
	&\mathcal{S}_{\Gamma_\text{Ren}} = \int \dInt[4]{x} \left( \frac{\delta \Gamma_\text{Ren}}{\delta \rho_a^\mu} \frac{\delta}{\delta G^a_\mu} + \frac{\delta \Gamma_\text{Ren}}{\delta G^a_{\mu}} \frac{\delta}{\delta \rho_a^\mu} + \frac{\delta \Gamma_\text{Ren}}{\delta \zeta_a} \frac{\delta}{\delta c^a} + \frac{\delta \Gamma_\text{Ren}}{\delta c^a} \frac{\delta}{\delta \zeta_a} + B^a \frac{\delta \Gamma_\text{Ren}}{\delta \bar{c}_a} \right.\\
	  &\left. + \frac{\delta \Gamma_\text{Ren}}{\delta \mathcal{Y}^m}\frac{\delta}{\delta \Phi_m} + \frac{\delta \Gamma_\text{Ren}}{\delta \Phi_m}\frac{\delta}{\delta \mathcal{Y}^m}
	          + \frac{\delta \Gamma_\text{Ren}}{\delta \bar{R}^i} \frac{\delta}{\delta \psi_i} + \frac{\delta \Gamma_\text{Ren}}{\delta \psi_i} \frac{\delta}{\delta \bar{R}^i} + \frac{\delta \Gamma_\text{Ren}}{\delta R^i} \frac{\delta}{\delta \overline{\psi}_i} + \frac{\delta \Gamma_\text{Ren}}{\delta \overline{\psi}_i} \frac{\delta}{\delta R^i} \right)
	\, ,
\end{split}
\end{equation*}
is the linearized BRST operator of our model. The basis that respects those equations is constructed from its classical approximation, % in the sense of \cref{eq:clbasis},
by employing the functionals $L_G$, $L_c$, $L_{\Phi}$, $L_{\psi_R}$ that are $b$-invariant in 4 dimensions and whose definitions have been introduced in \cref{sect:standardrenormalizationstructure}, \cref{eq:Lfuncts_fields}.
% (see also \cref{app:LinearBRST}).
These functionals can be expressed as linear combinations of field-counting operators for $d=4$ acting on the tree-level action: $L_\varphi \equiv \mathcal{N}_\varphi S_0$ for $\varphi = G, c, \Phi, \psi_R$, as well as the operators $L_g$, ${L_{Y_R}}^m_{ij}$ and $L_{\lambda_{mnop}}$ defined by differentiating the action with respect to the coupling parameters of the theory, \cref{eq:Lfuncts_couplings}.

A quantum extension of this classical basis is constructed \cite{Piguet:1995er} by the action on $\Gamma_\text{Ren}$ of the symmetric differential operators  we have just introduced (see Ref.\ \cite{Martin:1999cc} for the details), and up to order $\hbar^n$ the following equation holds:
\begin{equation}
	\Big[ \mu \frac{\partial}{\partial \mu} + \beta \; g \frac{\partial}{\partial g} + (\beta_Y)_{ij}^m \frac{\partial}{\partial Y_{ij}^m} + {\beta_\lambda}_{mnop} \frac{\partial}{\partial \lambda_{mnop}} - \gamma_G \mathcal{N}_G - \gamma_c \mathcal{N}_{c} - \gamma_\Phi \mathcal{N}_{\Phi} - \gamma_\psi \mathcal{N}_{\psi}^R \Big] \Gamma_\text{Ren} = 0 \, .
\end{equation}
This is the renormalization group equation of our theory. Now, thanks to the consequence of the Quantum Action Principle that any differential operator contained in our quantum basis can be expressed as insertions of normal products in $\Gamma_\text{Ren}$, and the fact that the first non-vanishing contribution to these expansions is of order $\hbar$, at one-loop level we have:
\begin{equation}
\label{eq:RGEBfinal}
	\mu \frac{\partial \Gamma_\text{Ren}}{\partial \mu}
	\stackrel{\mathcal{O}(\hbar)}{=}
	-\beta^{(1)} g \frac{\partial S_0^{(4D)}}{\partial g} - (\beta_Y^{(1)})_{ij}^m \frac{\partial S_0^{(4D)}}{\partial Y_{ij}^m} - {\beta_\lambda}_{mnop}^{(1)} \frac{\partial S_0^{(4D)}}{\partial \lambda_{mnop}} + \sum_{\phi}\gamma_\phi^{(1)} \mathcal{N}_{\phi} S_0^{(4D)} \, ,
\end{equation}
where $S_0^{(4D)}$ symbolizes the 4-dimensional restriction of the tree-level action of our model, \cref{eq:TreeLevel4DAction}.
The RHS of equation \eqref{eq:RGEBfinal} is the first constituent needed in the construction of our system of the renormalization group equations.

\subsection{Evaluation of $\mu {\partial \Gamma_\text{Ren}}/{\partial \mu}$}
\label{subsect:mu_DGamma_Dmu}

The first non-trivial contribution to the functional $\mu {\partial \Gamma_\text{Ren}}/{\partial \mu}$ is always of order $\hbar$, since the tree-level action does not depend on the renormalization scale $\mu$. The problem of expressing $\mu {\partial \Gamma_\text{Ren}}/{\partial \mu}$ as an insertion of normal product operators into $\Gamma_\text{Ren}$ (keep in mind that $\mu$ is not a parameter of the action) was solved by Bonneau \cite{Bonneau:1980zp} and generalized by Martin \cite{Martin:1999cc} due to the presence of different types of fields and external sources, evanescent contributions and finite counterterms. Its \emph{restriction to one-loop ($\hbar$) order} reads:
\begin{equation}
%\label{eq:BonneauRGEallx}
\label{eq:BonneauRGEOneLoop}
    \mu \frac{\partial}{\partial \mu}  \Gamma_\text{Ren} = N\big[\text{r.s.p.} \, \Gamma_{\text{DReg}}^{(1)}\big] \cdot \Gamma_{\text{Ren}} \, ,
\end{equation}
where we recall that ``$\text{r.s.p.} \, \Gamma_{\text{DReg}}^{(1)}$'' means
the residue of the simple pole in $\nu = 4 - d = 2\epsilon$ of all the 1PI
Green's functions described by the dimensionally-regularized effective action
at $\hbar$ order, $\Gamma_{\text{DReg}}$.

Notice that, since the singular parts of Feynman diagrams contributing to 1PI Green's functions are local polynomials in external momenta expressed in $d$, 4 and/or $\epsilon$ (i.e. evanescent) dimensions, the results generally contain evanescent contributions.

In order to handle these evanescent contributions, we will recall the results of the so-called Bonneau identities that have been first employed in \cref{subsect:Deltahatbreaking} in the specific one-loop case, \cref{eq:BonneauIdEvansctOneLoop}.
The Bonneau identities \cite{Bonneau:1980zp,Bonneau:1979jx} form a linear system whose unique solution provides an expansion of any anomalous (e.g. evanescent) operator in terms of a quantum basis of standard insertions. More precisely, any anomalous normal product can be re-expressed as a linear combination of standard and evanescent monomial normal products \cite{Martin:1999cc}, taking at any loop order the form:
\begin{equation}
\label{eq:BonneauIdentity_AllOrders}
	N[\hat{g}_{\mu\nu}\mathcal{O^{\mu \nu}}](x) \cdot \Gamma_\text{Ren} = \sum_i \bar{\alpha}_i N[\bar{\mathcal{M}}^i](x) \cdot \Gamma_\text{Ren} + \sum_j \hat{\alpha}_j N[\hat{\mathcal{M}}^j](x) \cdot \Gamma_\text{Ren} \, ,
\end{equation}
where the $\bar{\alpha}_i$, $\hat{\alpha}_j$ coefficients are evaluated similarly to those presented in \cref{subsect:Deltahatbreaking}, \cref{eq:BonneauIdEvansctOneLoop}, i.e. as $\text{r.s.p.}$'s in $4-d$ of specific 1PI Green's functions. This engenders a $\hbar$ power expansion for these coefficients, thus showing that evanescent operators generate $\hbar$-order contributions.
As shown by the latter term in \cref{eq:BonneauIdentity_AllOrders}, the Bonneau identities can also generate extra evanescent operators, but ponderated by additional $\hbar$-sized coefficients $\hat{\alpha}_j$. Such terms can be further reduced to pure standard operators by recursively applying the Bonneau identities, so that the anomalous normal product ultimately reduces to
\begin{equation}
\label{Bonneau_reduction}
	N[\hat{g}_{\mu\nu}\mathcal{O^{\mu \nu}}](x) \cdot \Gamma_\text{Ren} = \sum_i q_i N[\bar{\mathcal{M}}^i](x) \cdot \Gamma_\text{Ren} \, ,
\end{equation}
where $q_i$ are formal series in $\hbar$, having no order $\hbar^0$ contribution due to the $\text{r.s.p.}$ extractions on the calculations of 1PI functions entering into the definitions of the $\alpha_i$ coefficients.
This equation \eqref{Bonneau_reduction} thus holds \emph{for a fixed $\hbar$ order}, after reapplying a finite number of times the Bonneau identities.
Fortunately, at lowest order in $\hbar$ the linear system is trivial and decoupled, i.e. loops with anomalous insertions can be transformed in sum of tree-level diagrams with insertions of standard operators.

In general, for the calculation of the coefficient $q_i$ at order $\hbar^{n}$, we need the coefficients $\alpha_i$ up to order $\hbar^{n}$ and $\hat{\alpha}_j$ up to order $\hbar^{n-1}$, since the evanescent operators count for an order $\hbar$ higher, according to the Bonneau identities, what is of crucial importance in particular for the calculation at one-loop level.
If we now use these general results for specific operator  $\mu {\partial \Gamma_\text{Ren}}/{\partial \mu}$, the expansion
\begin{equation}
\label{eq:RGEBb}
    \mu \frac{\partial \Gamma_\text{Ren}}{\partial \mu} = \sum_i \bar{r}_i N[\bar{\mathcal{W}_i}] \cdot \Gamma_\text{Ren} +
    \sum_{j} \hat{r}_j N[\hat{\mathcal{W}_j}] \cdot \Gamma_\text{Ren}
\end{equation}
holds. Thanks to the Bonneau identities \cref{Bonneau_reduction}, the last term of this expansion can be re-expressed as
\begin{equation}
\label{eq:RGEBbc}
	N[\hat{\mathcal{W}_j}] \cdot \Gamma_\text{Ren} = \sum_i c_{ji} N[\bar{\mathcal{W}_i}] \cdot \Gamma_\text{Ren} \, ,
\end{equation}
where the $c_{ji}$ are formal expansions in $\hbar$, having no order $\hbar^0$ contribution due to the $\text{r.s.p.}$ extractions. This results in the expansion
\begin{equation}
\label{eq:RGEB}
    \mu \frac{\partial \Gamma_\text{Ren}}{\partial \mu} = \sum_i r_i N[\bar{\mathcal{W}_i}] \cdot \Gamma_\text{Ren} \, ,
\end{equation}
and at $\hbar$ order, this equation reduces to:
\begin{equation}
\label{eq:RGEBbx}
	\mu \frac{\partial \Gamma_\text{Ren}}{\partial \mu} = \sum_i r_i N[\bar{\mathcal{W}_i}] \cdot \Gamma_\text{Ren} = \sum_i(\bar{r}_i + {\textstyle\sum_j} \hat{r}_j c_{ji}) N[\bar{\mathcal{W}_i}] \cdot \Gamma_\text{Ren} \stackrel{\mathcal{O}(\hbar)}{=} \sum_i \bar{r}_i \bar{\mathcal{W}_i} \, ,
\end{equation}
where in the last step, the non-zero contributions at lowest $\hbar$ order come from the coefficients $\bar{r}_i$ only and thus, \emph{evanescent contributions do not affect one-loop level RGEs}.
In addition, the corresponding field product insertions $N[\bar{\mathcal{W}_i}] \cdot \Gamma_\text{Ren}$ are tree-level $\hbar^0$ insertions, simply equal to $\bar{\mathcal{W}_i}$.
The general algorithm for calculating of the $\bar{r}_i$ and $\hat{r}_j$ coefficients at any order is explained in \cite{Martin:1999cc}.

Now, there is a question of choice of basis for the set of 4-dimensional monomials $\bar{\mathcal{W}}_i$. Fortunately, any such basis of renormalized insertions is completely characterized by the corresponding classical basis \cite{Piguet:1995er}. If
\begin{equation}
\label{eq:clbasis}
	\big\{\Delta^p\cdot\Gamma = \Delta_\text{class}^p + \mathcal{O}(\hbar) \mid p=1,2,\dots ; \text{dim}(\Delta^p)\leq d\big\}
\end{equation}
is the set of insertions whose classical approximations form a basis for classical insertions up to dimension $d$, then the same set is a basis for the quantum insertions bounded by $d$. This means that a convenient
choice for the set of monomials are the field operators that are contained in the tree-level action $S_0$.

The insertion in \cref{eq:eqinsert} then can be chosen as a linear combination of operators from the 4-dimensional action,
\begin{equation}
	\mu \frac{\partial \Gamma_\text{Ren}}{\partial \mu} = \sum_{i \in f.b.} \sum_{a_i} \overline{c}_{\phi_1 \phi_2 \dots}^{(1),a_i} \, \overline{S^{0,a_i}_{\phi_1 \phi_2 \dots}},
\end{equation}
where $f.b.$ denotes the full basis of field operators $(\phi_1 \phi_2 \dots)$ in the tree-level action $S_0$. Thus, using our notation for Green's functions, and in regards to \cref{eq:BonneauRGEOneLoop} at one-loop ($\hbar$) order only, each contribution in the above equation takes the form
\begin{multline}
	N[\text{r.s.p.} (-i) \langle \widetilde{\phi_n}(p_n) \cdots \widetilde{\phi_1}(p_1) \rangle^\text{\,1PI} {\textstyle\prod_i} \phi_i] \cdot \Gamma \\
	= \text{r.s.p.} N[{\textstyle\prod_i} \phi_i \Gamma_{\phi_n \cdots \phi_1}(\bar{p}_1, \dots, \bar{p}_n)] \cdot \Gamma %\\
	\stackrel{\mathcal{O}(\hbar)}{\subset} \text{r.s.p.}(-S_\text{sct}^{(1),4D}) \equiv -2\epsilon S_\text{sct}^{(1),4D} \, ,
\end{multline}
where $(1),4D$ denotes $\hbar$ order and 4-dimensional space, respectively. Therefore, at $\hbar$ order, the Renormalization Group equation acquires the simple form
\begin{equation}
\label{eq:RGEa}
	\mu \frac{\partial \Gamma_\text{Ren}}{\partial \mu} \equiv -2\epsilon S_\text{sct}^{(1),4D} \, ,
\end{equation}
where $S_\text{sct}^{(1),4D}$ is just equal to \cref{eq:SingularCT1Loop,eq:SingularCT1Loop_NoScalarContrib,eq:SingularCT1Loop_ScalarContrib} but projected onto 4 dimensions only (thus there are no appearance of evanescent operators).
We again emphasize that the absence of any evanescent contribution is a one-loop effect only.

\subsection{Solution of the System}
\label{subsect:RGE_AlgRen}

By direct comparison of \eqref{eq:RGEBfinal} with \eqref{eq:RGEa} we obtain the following system of equations:
\begin{align}
S_{GG} &\rightarrow 2\gamma_G^{(1)} = \frac{-2\hbar}{16 \pi^2} g^2 \frac{(13-3\xi) C_2(G) - 4 S_2(R) - S_2(S)}{6}
\, ,
\\
S_{GGG} &\rightarrow -\beta^{(1)}+ 3\gamma_G^{(1)} = \frac{-2\hbar}{16 \pi^2} g^2 \frac{(17-9\xi) C_2(G) - 8 S_2(R) - 2 S_2(S)}{12}
\, ,
\\
S_{GGGG} &\rightarrow -2\beta^{(1)}+ 4\gamma_G^{(1)} = \frac{-2\hbar}{16 \pi^2} g^2 \frac{2(2-3\xi) C_2(G) - 4 S_2(R) - S_2(S)}{6}
\, ,
\\
S_{\bar{\psi}\psi_R} &\rightarrow 2\gamma_\psi^{(1)} = \frac{2\hbar}{16 \pi^2} \left(g^2 \xi C_2(R) + \frac{Y_2(R)}{2}\right)
\, ,
\\
S_{\overline{\psi} G \psi_R} &\rightarrow  -\beta^{(1)} + \gamma_G^{(1)} + 2\gamma_\psi^{(1)} = \frac{2\hbar}{16 \pi^2} \left(g^2 \frac{(3+\xi) C_2(G) + 4 \xi C_2(R)}{4} + \frac{Y_2(R)}{2}\right)
\, ,
\\
S_{\Phi \Phi} &\rightarrow 2\gamma_\Phi^{(1)} = \frac{-2\hbar}{16 \pi^2} \left(g^2 (3-\xi) C_2(S) - Y_2(S)\right)
\, ,
\\
S_{\Phi G \Phi} &\rightarrow -\beta^{(1)} + \gamma_G^{(1)} + 2\gamma_\Phi^{(1)} = \frac{-2\hbar}{16 \pi^2} \left(g^2 \left((3-\xi) C_2(S) -\frac{3+\xi}{4} C_2(G)\right) - Y_2(S)\right)
\, ,
\\
S_{\Phi GG \Phi} &\rightarrow -2\beta^{(1)} + 2\gamma_G^{(1)} + 2\gamma_\Phi^{(1)} = \frac{-2\hbar}{16 \pi^2} \left(g^2 \left((3-\xi) C_2(S) - \frac{3+\xi}{2} C_2(G)\right) - Y_2(S)\right)
\, ,
\\
S_{\Phi^4_{mnop}} &\rightarrow -{\beta_\lambda}_{mnop}^{(1)} + 4\gamma^{(1)}_{\Phi}\lambda_{mnop} = \frac{-2\hbar}{16 \pi^2}
\frac{1}{2} (3 g^4 A - g^2 \xi \Lambda^S - 4 H + \Lambda^2)_{mnop}
\, ,
\\
\begin{split}
	S_{\overline{\psi_R}^C_i \Phi^m {\psi_R}_j} &\rightarrow - (\beta_Y^{(1)})_{ij}^m + (Y_R)^m_{ij} (\gamma^{(1)}_{\Phi}+2\gamma_\psi^{(1)}) \\
		&= \frac{-2\hbar}{16 \pi^2} \left( (Y_R^n (Y_R^m)^* Y_R^n) -g^2 \frac{2 C_2(R) (3+\xi) - C_2(S) (3-\xi)}{2} Y_R^m \right)_{ij}
	\, ,
\end{split}
\\
S_{\bar{c}c} & , S_{\rho c} \rightarrow -\gamma_G^{(1)}+\gamma_c^{(1)} = \frac{-2\hbar}{16 \pi^2} g^2 \frac{3-\xi}{4} C_2(G)
\, ,
\\
S_{\bar{c}Gc} & , S_{\rho G c} , S_{\zeta c c} , S_{\bar{R} c \psi_R} , S_{R c \overline{\psi_R}} , S_{\mathcal{Y} c \Phi}
    \rightarrow -\beta^{(1)}+\gamma_c^{(1)} = \frac{2\hbar}{16 \pi^2} g^2 \frac{\xi C_2(G)}{2}
\, .
\end{align}
This is an overdetermined system of equations that provides the following solutions for the $\beta$-functions and anomalous dimensions at one-loop level:
\begin{align}
\beta &= \frac{\hbar}{16 \pi^2} g^2\left(\frac{-22 C_2(G) + 4 S_2(R) + S_2(S)}{6}\right)
\, ,
\\
{\beta_\lambda}_{mnop} &= \frac{\hbar}{16 \pi^2} (3 g^4 A_{mnop} - 4 H_{mnop} + \Lambda^2_{mnop}+\Lambda^{Y}_{mnop}-3 g^2\Lambda^{S}_{mnop})
\, ,
\\
\begin{split}
{\beta_Y}_{ij}^m &= \frac{\hbar}{16 \pi^2} \left(\vphantom{\frac{1}{1}} 2(Y_R^n (Y_R^m)^* Y_R^n)_{ij} - 3 g^2\{C_2(R),Y_R^m\}_{ij} + (Y_R)^m_{ij} Y_2(S) \right.\\
	&\left.\quad + \frac{1}{2} ((Y_R)^m_{ij} Y_2(R) + Y_2(\bar{R}) (Y_R)^m_{ij}) \right)
	\, ,
\end{split}
\\
\gamma_G &= \frac{\hbar}{16 \pi^2} g^2 \frac{(3\xi-13) C_2(G) + 4 S_2(R) + S_2(S)}{6}
\, ,
\\
\gamma_\psi &= \frac{\hbar}{16 \pi^2} \frac{2 g^2 \xi C_2(R) + Y_2(R)}{2}
\, ,
\\
\gamma_\Phi &= \frac{\hbar}{16 \pi^2} \left(g^2 (\xi-3) C_2(S) + Y_2(S)\right)
\, ,
\\
\gamma_c &= \frac{\hbar}{16 \pi^2} g^2\frac{(6\xi-22) C_2(G) + 4 S_2(R) + S_2(S)}{6}
\, ,
\end{align}

\section{Comparison with the Standard Multiplicative Renormalization Approach in the BMHV Scheme}
\label{sect:MultRenorm}

In this section, we explain the derivation of the RGE using the
standard approach based on divergences of renormalization
constants. In the BMHV scheme there are extra divergences for
evanescent operators, and we focus particularly on their role in the
derivation.

The standard textbook approach to deriving RGEs in the context of DReg
was developed in Ref.\ \cite{tHooft:1973mfk} and applied e.g.\ in
Refs.\ \cite{Machacek:1983tz,Machacek:1983fi,Machacek:1984zw}. It starts from the observation that the
bare action (i.e.\ the sum of tree-level and counterterm action) can
be written in terms of bare fields and parameters which depend on the
$\overline{\text{MS}}$-renormalization scale $\mu$. For a generic bare
parameter $g_i$ in a massless theory,
and in the $\overline{\text{MS}}$-renormalization scheme, this may be written as
\begin{align}
	g_{i,\text{bare}} &= \mu^{\rho_i\epsilon}\left(g_i + \delta g_i\right)
	\, , &
	\delta g_i &=\sum_{n=1}^{\infty} \frac{a_i^{(n)}}{\epsilon^n} \, ,
\end{align}
where $\rho_i$ is a constant, $g_i$ the renormalized parameter and
$\delta g_i$ the renormalization constant, which is a pure divergence.
The coefficients $a_i^{(n)}$ depend on the parameters of the theory, but depend
on $\mu$ only implicitly through the $\mu$-dependence of these parameters.
The corresponding $\beta$ function
defined as $\beta_i(\epsilon) \equiv {\partial g_i}/{\partial \ln\mu}$ is
then obtained as
\begin{equation}
  \beta_i(\epsilon) = -\rho_i\epsilon g_i - \rho_i a_i^{(1)} +
    \sum_{k} \rho_k g_k \frac{\partial a_i^{(1)}}{\partial g_k} \, ,
\end{equation}
where the sum runs over all parameters $g_k$ of the theory. Similarly, the anomalous dimension is obtained from the renormalization constant associated with an irreducible self-energy Green's function, which has the expansion
\begin{equation}
	Z_{\phi} = 1 + \sum_{n=1}^{\infty} \frac{a_{\phi}^{(n)}}{\epsilon^{n}} \, ,
\end{equation}
and, assuming equal renormalization of the fields in self-energy Green's function, is equal to
\begin{equation}
	\gamma_{\phi}(\epsilon) = \frac{1}{2} \mu \frac{d}{d\mu} \ln Z_{\phi} \, .
	%= -C_{\Gamma}^{(1)} \equiv -\epsilon %\delta Z_{\Gamma}^{(1)} \, ,
\end{equation}
Proceeding similarly for all fields of the theory one obtains
the generic RGE
\begin{equation}
	\mu \frac{\partial}{\partial \mu} \Gamma_\text{DReg} = \Big(-\sum_{k} \beta_k(\epsilon) \frac{\partial}{\partial g_k} + \sum_{\phi} \gamma_\phi(\epsilon) N_{\phi} \Big) \Gamma_\text{DReg} \, .
\end{equation}
This equation holds even for $\epsilon\ne0$. An important detail is
that at this level the $\beta$ and $\gamma$ functions are
$\epsilon$-dependent and have the structure
\begin{equation}
  \beta_i(\epsilon),\gamma_i(\epsilon) =
  \mathcal{O}(\epsilon)\times(\text{tree-level})
  +\mathcal{O}(\epsilon^0)\times(\text{$\ge1$-loop level})
  \, .
\end{equation}

\subsection{On the Influence of the Evanescent Counterterms}

In principle, all of these remarks apply to the BMHV scheme. However, in this
scheme, the action contains evanescent divergent counterterms, see
\cref{eq:Ssctevan}. These have no tree-level counterpart. In order to
apply the method in the BMHV context we can amend the tree-level
action by additional terms, such that for each term in
$S_{\text{sct,evan}}^{(1)}$ there is a new tree-level parameter,
i.e.\ writing
\begin{equation}
  S_{0,\text{amend}} = S_{0,\text{inv}}+S_{0,\text{evan}}
  + S_{0,\text{evan,add}}
\end{equation}
instead of \cref{eq:S0Def2}. The parameters in the new part
$S_{0,\text{evan,add}}$ will generically be denoted as $\hat{g}_i$.
Likewise we can amend the renormalization transformation
\cref{eq:rentransform} by
\begin{align}
  \label{eq:rentransformevan}
  \hat{g}_i & \to \hat{g}_i + \delta\hat{g}_i\,.
\end{align}
As a result we can obtain all singular counterterms, including the
evanescent ones, from a renormalization transformation, as
\begin{equation}
	S_{0,\text{amend}}
	\stackrel{\text{Eqs.\ (\ref{eq:rentransform},\ref{eq:rentransformevan})}}{\longrightarrow}
	S_{0,\text{amend}}+S_{\text{ct,inv}} + S_{\text{ct,evan}}
\end{equation}
in place of \cref{eq:Sctinv}. Via the logic described above, this leads to
an RGE with the generic structure
\begin{align}
\label{eq:GenericRGEDReg}
  \mu \frac{\partial}{\partial \mu} \Gamma_\text{DReg} =
        \Big(-\sum_{k} \beta_k(\epsilon) \frac{\partial}{\partial g_k}
        -\sum_{k} \hat{\beta}_k(\epsilon) \frac{\partial}{\partial \hat{g}_k}
        + \sum_{\phi} \gamma_\phi(\epsilon) N_{\phi} \Big) \Gamma_\text{DReg} \, ,
\end{align}
where the second sum on the right-hand side is over all parameters
$\hat{g}_k$ of the evanescent additional action
$S_{0,\text{evan,add}}$.

In the following, we need to discuss the influence of these additional
``evanescent'' parameters $\hat{g}_k$. We first remark that such or
similar parameters have been discussed in various contexts
before. Ref.\ \cite{Schubert:1988ke} considered the same problem as
the present section, but in the context of a non-gauge theory, and
discussed the influence of such parameters on the RGE. In the context
of regularization by dimensional reduction (DRed), evanescent
quantities do not correspond to $\gamma_5$ but to the extra $(4-d)$
degrees of freedom of the gauge fields (the so-called
``$\epsilon$-scalars''). Accordingly, the impact of the
$\epsilon$-scalar mass term on the 2-loop RGE of softly broken
supersymmetric gauge theories has been discussed in
Ref.\ \cite{Jack:1994rk}. Finally, in applications of DRed
to non-supersymmetric QCD, the evanescent coupling $\alpha_e$ between
the $\epsilon$-scalar and quarks appears. The need for treating this coupling
and its $\beta$ function as independent has been explained first in
Ref.\ \cite{Jack:1994bn}, for a further overview and references see
\cite{Gnendiger:2017pys}.

We now provide the following remarks:
\begin{itemize}[leftmargin=*]
\item
  Our original formulation of the theory in
  \crefrange{sect:RModel}{sect:Rmodel1LoopSCT} corresponds to setting the
  evanescent parameters $\hat{g}_k=0$ at tree-level. This is
  compatible with the RGE in $\epsilon\ne0$ only at one particular
  renormalization scale $\mu$. At other scales $\mu'$, the RGE generates
  non-vanishing tree-level values $\hat{g}_k(\mu')\ne0$.
\item
  Non-vanishing $\hat{g}_i$ enter the theory in three ways up to the
  1-loop level:
  \begin{enumerate}
  \item at tree-level in the purely evanescent part;
    \item at the 1-loop level in finite contributions to standard
      (non-evanescent) Green's functions;
      \item at the 1-loop level in $1/\epsilon$ contributions to
        evanescent Green's functions, and in finite contributions to
        $\hat{\beta}$ functions of evanescent parameters.
  \end{enumerate}
\item
  Hence, applying the $\mathop{\text{LIM}}_{d \to 4}$ operation at the
  1-loop level to the generic RGE \eqref{eq:GenericRGEDReg}, i.e.\ its
  renormalized limit, leads to:
  \begin{enumerate}
  \item the derivative
    $\frac{\partial}{\partial \hat{g}_k}\Gamma_\text{DReg}$ reduces to a finite,
    pure 1-loop quantity;
  \item
    all coefficients $\beta_k(\epsilon)$, $\gamma_\phi(\epsilon)$, and $\hat{\beta}_k(\epsilon)$
    vanish at tree-level and become quantities of 1-loop order; in the $\epsilon \to 0$
	limit we denote $\beta_k(0) \equiv \beta_k \, , \, \gamma_\phi(0) \equiv \gamma_\phi \,$;
  \item
    the coefficients $\beta_k$ and $\gamma_\phi$ corresponding to
    non-evanescent operators are independent of the evanescent $\hat{g}_k$.
  \end{enumerate}
\end{itemize}
In view of these comments, the renormalized limit at the 1-loop level of \cref{eq:GenericRGEDReg} leads to the RGE
\begin{align}
  \mu \frac{\partial \Gamma_\text{Ren}}{\partial \mu} =
        \Big(-\sum_{k} \beta_k \frac{\partial}{\partial g_k}
        + \sum_{\phi} \gamma_\phi N_{\phi} \Big) \Gamma_\text{Ren} \, ,
\end{align}
where both sides are understood to be evaluated up to the 1-loop
level. The dependence on evanescent parameters $\hat{g}_i$ has dropped
out, and the non-evanescent coefficients $\beta_k$, $\gamma_\phi$ may
be evaluated by setting the $\hat{g}_i=0$.
This shows that the correct 1-loop RGE in the BMHV context may be
obtained by the simple recipe of applying the usual procedure of
Refs.\ \cite{tHooft:1973mfk,Machacek:1983fi}
from the divergences of renormalization constants, ignoring the
additional evanescent objects contained
in the amended tree-level action, and instead taking only the theory
as defined in \crefrange{sect:RModel}{sect:Rmodel1LoopSCT}.

On the other hand, this analysis also shows that starting from the 2-loop level,
the situation will be more involved. E.g.\ the term
$\hat{\beta}_k\frac{\partial}{\partial \hat{g}_k}\Gamma_\text{DReg}$
can be expected to provide finite, non-vanishing 2-loop contributions,
and the $\beta_i$, $\gamma_\phi$ coefficients might depend on the
evanescent parameters $\hat{g}_i$. Both effects have appeared in the
contexts of Refs.\ \cite{Schubert:1988ke,Schubert:1993wg,Jack:1994rk,Jack:1994bn}
mentioned above, and additional calculations are required to replace
the dependence on the $\hat{g}_i$ by modifications of the $\beta_i$,
$\gamma_\phi$.

\subsection{Full system of Renormalization Group Equations}

According to the previous discussion, the 1-loop RGE can be obtained
from the divergences of the renormalization constants by ignoring all
evanescent contributions. In this way we obtain schematically
\begin{align}
\delta g^{(1)}/g&\to&
	\beta &= \frac{\hbar}{16 \pi^2} g^2\left(\frac{-22 C_2(G) + 4 S_2(R) + S_2(S)}{6}\right)
\, ,
\\
\delta \lambda_{mnop}^{(1)} &\to&
	{\beta_\lambda}_{mnop} &= \frac{\hbar}{16 \pi^2} (3 g^4 A_{mnop} - 4 H_{mnop} + \Lambda^2_{mnop} + \Lambda^{Y}_{mnop} - 3 g^2\Lambda^{S}_{mnop})
\, ,
\\
\delta (Y_R)^{m,(1)}_{ij} &\to&
\begin{split}
	{\beta_Y}_{ij}^m &= \frac{\hbar}{16 \pi^2} \left(\vphantom{\frac{1}{1}}
		2(Y_R^n (Y_R^m)^* Y_R^n)_{ij} - 3 g^2\{C_2(R),Y_R^m\}_{ij} + (Y_R)^m_{ij} Y_2(S) \right.\\
		&\left.\quad + \frac{1}{2} ((Y_R)^m_{ij} Y_2(R) + Y_2(\bar{R}) (Y_R)^m_{ij}) \right)
	\, ,
\end{split}
\\
\delta Z_G^{(1)} &\to&
	\gamma_G &= \frac{\hbar}{16 \pi^2} g^2 \frac{(3\xi-13) C_2(G) + 4 S_2(R) + S_2(S)}{6} = -\gamma_{\bar{c}} = -\gamma_{\rho}
\, ,
\\
\delta Z_{\psi_L}^{(1)} &\to&
	\gamma_\psi &= \frac{\hbar}{16 \pi^2} \frac{2 g^2 \xi C_2(R) + Y_2(R)}{2} = -\gamma_R
\, ,
\\
\delta Z_{\Phi}^{(1)} &\to&
	\gamma_\Phi &= \frac{\hbar}{16 \pi^2} \left(g^2 (\xi-3) C_2(S) + Y_2(S)\right) = -\gamma_{\mathcal{Y}}
\, ,
\\
\delta Z_c^{(1)} &\to&
	\gamma_c &= \frac{\hbar}{16 \pi^2} g^2 \frac{(6\xi-22) C_2(G) + 4 S_2(R) + S_2(S)}{6} = -\gamma_{\zeta}
\, .
\end{align}
This result is the same as the one obtained in \cref{sect:RGE},
demonstrating that both methods may be applied to obtain the correct
1-loop RGE in the BMHV scheme.

\section{The Left-Handed (L) Model}
\label{sect:LModel}

In this section, we indicate how our previous results adapt for a model including only left-handed fermions.
We define the Left-Handed (L) model to be the same as the Right-Handed (R) model studied so far, except now with the usage of purely left-handed fermions $\psi_L \equiv \Proj{L} \psi$: the gauge, scalar, and gauge-scalars sectors remain unchanged, while only the fermion kinetic and Yukawa terms get modified. Our aim is to know how our results derived so far change when considering these left-handed fermions.

It is possible to construct a mapping between the L-model and the R-model: indeed, using the charge-conjugation construction from \cref{subsect:ChargeConjDd}, the charge-conjugate of a left-handed fermion is a right-handed fermion:
\begin{equation}
	{\psi_L}^C = C \overline{\psi_L}^T = C (\overline{\psi} \Proj{R})^T = C \Proj{R}^T \overline{\psi}^T = C \Proj{R}^T C^{-1} C \overline{\psi}^T = \Proj{R} C \overline{\psi}^T = \Proj{R} \psi^C \equiv \Proj{R} \widehat{\psi} \equiv \widehat{\psi}_R \, ,
\end{equation}
with the definition $\widehat{\psi} \equiv \psi^C$.

The same discussion as in \cref{subsect:RModelDReg} holds and we can promote this L-model to $d$ dimensions.
The left-handed fermion-kinetic term is:
\begin{equation}
	\mathcal{L}_\text{fermions} = \imath \overline{\psi}_i \slashed{\partial} {\psi}_i + g {T_L}^a_{ij} \overline{\psi_L}_i \slashed{G}^a {\psi_L}_j \, ,
\end{equation}
where ${T_L}$ is the generator for their corresponding representation.
Since the kinetic term is a scalar function, it is also equal to its transpose in spinor space, and thus we obtain:
\begin{equation}\begin{split}
	\mathcal{L}_\text{fermions} &= \imath (\overline{\psi}_i \slashed{\partial} {\psi}_i)^T + g {T_L}^a_{ij} (\overline{\psi_L}_i \slashed{G}^a {\psi_L}_j)^T
		 = -\imath {\psi}_i^T \overset{\leftarrow}{\slashed{\partial}^T} \overline{\psi}_i^T - g {T_L}^a_{ij} G^a_\mu {\psi_L}_j^T (\gamma^\mu)^T \overline{\psi_L}_i^T \\
		&= -\imath {\psi}_i^T C^{-1} C \overset{\leftarrow}{\slashed{\partial}^T} C^{-1} C \overline{\psi}_i^T - g {T_L}^a_{ij} G^a_\mu {\psi_L}_j^T C^{-1} C (\gamma^\mu)^T C^{-1} C \overline{\psi_L}_i^T \\
		&= \imath \overline{\psi}^C_i C \overset{\leftarrow}{\slashed{\partial}^T} C^{-1} \psi^C_i + g {T_L}^a_{ij} G^a_\mu \overline{\psi_L}^C_j C (\gamma^\mu)^T C^{-1} {\psi_L}^C_i \\
		&= -\imath \overline{\widehat{\psi}}_i \overset{\leftarrow}{\slashed{\partial}} \widehat{\psi}_i + g (-{T_L}^a_{ij}) G^a_\mu \overline{\widehat{\psi}_R}_j \gamma^\mu {{}\widehat{\psi}_R}_i
		 = \imath \overline{\widehat{\psi}}_i \slashed{\partial} \widehat{\psi}_i + g {T_{\overline{L}}}^a_{ij} \overline{\widehat{\psi}_R}_i \slashed{G}^a {{}\widehat{\psi}_R}_j \, ,
\end{split}\end{equation}
where in the second equality we used the anticommutativity of the fermion fields, in the second line we inserted $\mathbb{1} = C^{-1} C$ and used the properties of the charge-conjugation as defined in \cref{subsect:ChargeConjDd}, and in the last line we used an integration by parts (supposing the absence of surface terms) to rewrite the pure kinetic (first) term, and defined in the interaction term ${T_{\overline{L}}}^a_{ij} = -{T_L}^a_{ji}$ corresponding to the complex-conjugated representation of the left-handed fermions. Posing ${T_{\widehat{R}}}^a_{ij} \equiv {T_{\overline{L}}}^a_{ij}$, we see that this conjugated $L$-representation corresponds to the representation for the associated right-handed fermions.

Let us now turn to the Yukawa term, which is a real number and therefore equals to its hermitian conjugate:
\begin{equation}\begin{split}
	2 \times \mathcal{L}_\text{Yukawa}
		= -(Y_L)^m_{ij} \Phi_m \overline{\psi_L}^C_i {\psi_L}_j -(Y_L)^{m\;*}_{ij} \Phi^\dagger_m \overline{\psi_L}_i {\psi_L}^C_j
		= -(Y_L)^{m\;*}_{ij} \Phi^\dagger_m \overline{\widehat{\psi}_R}^C_i {{}\widehat{\psi}_R}_j + \hc \, ,
\end{split}\end{equation}
and we can define $(Y_{\widehat{R}})^m_{ij} \equiv (Y_L)^{m\;*}_{ij}$ the corresponding Yukawa matrix for the associated right-handed fermions, which is just the complex conjugate of the one for the left-handed fermions.

External sources for the fermion fields need to be introduced in the L-model due to the BRST quantization procedure:
\begin{equation}\begin{split}
	S_{\bar{L} c \psi_L} &= \bar{L}^i s_d{\psi_i} = \imath g \bar{L}^i c^a {T_L}^a_{ij} {\psi_L}_j \equiv \imath g \bar{L}^i c^a {T_L}^a_{ij} \Proj{L} \psi_j \, , \\
	S_{L c \overline{\psi_L}} &= s_d{\overline{\psi}_i} L^i = \imath g \overline{\psi_L}_j c^a {T_L}^a_{ji} L^i \equiv \imath g \overline{\psi}_j \Proj{L} c^a {T_L}^a_{ji} L^i
	\, .
\end{split}\end{equation}
Since these are scalar functions, we can take their transpose, and use the fact that $L$ and $\bar{L}$ are \emph{commuting} fermions (their ghost number $=-1$) to obtain:
\begin{equation}\begin{split}
	S_{\bar{L} c \psi_L} &= \imath g c^a {T_L}^a_{ij} (\bar{L}^i {\psi_L}_j)^T = \imath g c^a {T_L}^a_{ij} {\psi_L}^T_j \bar{L}^T_i = \imath g c^a {T_L}^a_{ij} {\psi_L}^T_j C^{-1} C \bar{L}^T_i \\
		&= \imath g \overline{\psi_L}^C_j c^a (-{T_L}^a_{ij}) (-C \bar{L}^T_i) = \imath g \overline{\widehat{\psi}_R}_j c^a {T_R}^a_{ji} \widehat{R}_i
		\equiv S_{\widehat{R} c \overline{\widehat{\psi}_R}} \, ,
\end{split}\end{equation}
where we have employed the notations introduced above and have defined the external source $\widehat{R}_i$ for the corresponding right-handed fermions: $\widehat{R}_i \equiv -C \bar{L}^T_i = -{L^C}_i$.
Similarly we obtain for the other source term:
\begin{equation}\begin{split}
	S_{L c \overline{\psi_L}} &= \imath g (\overline{\psi_L}_j L^i)^T c^a {T_L}^a_{ji} = \imath g {L^T}^i \overline{\psi_L}^T_j c^a {T_L}^a_{ji} = \imath g {L^T}^i C^{-1} C \overline{\psi_L}^T_j c^a {T_L}^a_{ji} \\
		&= \imath g (-{T_L}^a_{ji}) {L^T}^i C^{-1} c^a {\psi_L}^C_j = \imath g {T_R}^a_{ij} \overline{\widehat{R}}^i c^a {{}\widehat{\psi}_R}_j
		\equiv S_{\overline{\widehat{R}} c \widehat{\psi}_R} \, ,
\end{split}\end{equation}
where we used that $\overline{\widehat{R}}_i = -\overline{L^C}_i = L^T_i C^{-1}$, stemming from the properties of $C$.

These calculations demonstrate that we can establish a one-to-one mapping between a left-handed model with fermions $\psi$ ($\psi_L \equiv \Proj{L} \psi$) defined in a left-representation of the considered gauge group with generators $T_L$ that couple to scalar fields with the Yukawa interaction $(Y_L)^m_{ij}$, and a right-handed model with fermions $\widehat{\psi}$ related via charge-conjugation: $\widehat{\psi} \equiv \psi^C$ ($\widehat{\psi}_R \equiv \Proj{R} \widehat{\psi}$), in a right-representation ${T_{\widehat{R}}}^a_{ij} \equiv {T_{\overline{L}}}^a_{ij}$ that couple to the scalar fields with the Yukawa interaction $(Y_{\widehat{R}})^m_{ij} \equiv (Y_L)^{m\;*}_{ij}$. Therefore, all of our calculations derived so far in this work apply to the left-handed model as well.

We are thus able to evaluate the tree-level breaking of the BRST symmetry by the action of this Left-Handed model, similarly to what has been done in \cref{subsect:BRST_breaking_RModel}. We find that the breaking $\widehat{\Delta} = s_d S_0$ is given by:
\begin{equation}
%%\label{eq:BRSTTreeBreaking}
    \widehat{\Delta} = \int \dInt[d]{x}
        (g {T_L}^a_{ij}) c^a \left\{
            \overline{\psi}_i \left(\overset{\leftarrow}{\widehat{\slashed{\partial}}} \Proj{L} + \overset{\rightarrow}{\widehat{\slashed{\partial}}} \Proj{R}\right) \psi_j
        \right\}
    \equiv \int \dInt[d]{x} \widehat{\Delta}(x)
    \, ,
\end{equation}
generating a corresponding Feynman rule:
\\
% c^a(-p_1-p_2) \bar{\psi}_{i,\alpha}(p_1) \psi_{j,\beta}(p_2)
\begin{equation}
\begin{tabular}{rl}
	\raisebox{-40pt}{\includegraphics[scale=0.6]{diagrams/evanescent_vertices/BRST_breaking.pdf}} &
	$\begin{aligned}
		&= \frac{g}{2} {T_L}^a_{ij} \left((\widehat{\slashed{p_1}} + \widehat{\slashed{p_2}}) - (\widehat{\slashed{p_1}} - \widehat{\slashed{p_2}}) \gamma_5 \right)_{\alpha\beta} \\
		&= g {T_L}^a_{ij} \left(\widehat{\slashed{p_1}} \Proj{L} + \widehat{\slashed{p_2}} \Proj{R} \right)_{\alpha\beta}
		\, ,
	\end{aligned}$
\end{tabular}
\end{equation}
and the one corresponding to the charge-conjugated fermions:
\begin{equation}
    \widehat{\Delta} = \int \dInt[d]{x}
        (g {T_{\overline{L}}}^a_{ij}) c^a \left\{
        \overline{\psi^C}_i \left(\overset{\leftarrow}{\widehat{\slashed{\partial}}} \Proj{R} + \overset{\rightarrow}{\widehat{\slashed{\partial}}} \Proj{L}\right) \psi^C_j
        \right\}
% = \int \dInt[d]{x}
        % \left(\frac{g}{2} {T_{\overline{L}}}^a_{ij}\right) c^a \left\{
        % \partial_\mu (\overline{\psi^C}_i \widehat{\gamma}^\mu \psi^C_j)
        % - \overline{\psi^C}_i \overset{\leftrightarrow}{\widehat{\slashed{\partial}}} \gamma_5 \psi^C_j
        % \right\}
    \, ,
\end{equation}
generating the Feynman rule:
\\
% c^a(-p_1-p_2) \overline{\psi^C}_{i,\alpha}(p_1) \psi^C_{j,\beta}(p_2)
\begin{equation}
\begin{tabular}{rl}
	\raisebox{-40pt}{\includegraphics[scale=0.6]{diagrams/evanescent_vertices/BRST_breaking_CC.pdf}} &
	$\begin{aligned}
		&= \frac{g}{2} {T_{\overline{L}}}^a_{ij} \left((\widehat{\slashed{p_1}} + \widehat{\slashed{p_2}}) + (\widehat{\slashed{p_1}} - \widehat{\slashed{p_2}}) \gamma_5 \right)_{\alpha\beta} \\
		&= g {T_{\overline{L}}}^a_{ij} \left(\widehat{\slashed{p_1}} \Proj{R} + \widehat{\slashed{p_2}} \Proj{L} \right)_{\alpha\beta}
		\, .
	\end{aligned}$
\end{tabular}
\end{equation}
where the difference with the previous result is in the appearance of the generator ${T_{\overline{L}}}^a$ for the fermionic conjugate representation $L$.

The group invariants related to the scalar-fields representation $C_2(S) , S_2(S) , Y_2(S)$ and those defined in \cref{eq:quartic_scalar_invariants}: $A_{mnop} , H_{mnop} , \Lambda^2_{mnop} , \Lambda^S_{mnop}$ all remain the same. The group invariants $C_2(L) , S_2(L) , Y_2(L)$ of the left-representation are actually equal to those of the corresponding right-representation: $C_2(L) \mathbb{1} = {T_L}^a {T_L}^a = (-{T_L}^{a\,T})(-{T_L}^{a\,T}) = {T_{\overline{L}}}^a {T_{\overline{L}}}^a = C_2(\overline{L}) \mathbb{1} \equiv C_2(\widehat{R}) \mathbb{1}$; $S_2(L) \delta^{ab} = \Tr({T_L}^a {T_L}^b) = \Tr((-{T_L}^{b\,T})(-{T_L}^{a\,T})) = \Tr({T_{\overline{L}}}^b {T_{\overline{L}}}^a) = S_2(\overline{L}) \delta^{ab} = \Tr({T_{\widehat{R}}}^a {T_{\widehat{R}}}^b) = S_2(\widehat{R}) \delta^{ab}$; and $Y_2(L) \mathbb{1} = (Y_L^m)^* Y_L^m = Y_{\widehat{R}}^m (Y_{\widehat{R}}^m)^* = (Y_{\widehat{R}}^m)^* Y_{\widehat{R}}^m \equiv Y_2(\widehat{R}) \mathbb{1}$ by using the symmetry of the Yukawa matrices.

The singular counterterms $S_\text{sct}^{(1)} = S_\text{sct}^{(1)\,\substack{\text{No Scalar} \\ \text{contrib.}}} + S_\text{sct}^{(1)\,\substack{\text{Scalar} \\ \text{contrib.}}}$ are then obtained, and are the same as in \cref{eq:SingularCT1Loop,eq:SingularCT1Loop_NoScalarContrib,eq:SingularCT1Loop_ScalarContrib}, except for the replacements:
\begin{subequations}
\label{eq:RtoL_Changes}
\begin{align}
	S_2(R) \to S_2(L) \, , &&
	C_2(R) \to C_2(L) \, , &&
	Y_2(R) \to Y_2(L) \, , &&
	Y_R^m \to Y_L^m \, , \\
	\overline{S_{\bar{\psi}\psi_R}} \to \overline{S_{\bar{\psi}\psi_L}} \, , &&
	\overline{S_{\overline{\psi} G \psi_R}} \to \overline{S_{\overline{\psi} G \psi_L}} \, , &&
	S_{\overline{\psi_R}^C_i \Phi^m {\psi_R}_j} \to S_{\overline{\psi_L}^C_i \Phi^m {\psi_L}_j} \, , \\
	S_{\bar{R} c \psi_R} \to S_{\bar{L} c \psi_L} \, , &&
	S_{R c \overline{\psi_R}} \to S_{L c \overline{\psi_L}} \, .
\end{align}
\end{subequations}
Again, we can make contact to the usual renormalization transformation, and
express the singular counterterms as follows:
\begin{equation}
%\label{eq:Ssctinvevan}
	S_\text{sct}^{(1)} = S_{\text{sct,inv}}^{(1)} + S_{\text{sct,evan}}^{(1)} \, .
\end{equation}
The invariant counterterms $S_{\text{sct,inv}}^{(1)}$ acquire the same form as
those from \cref{eq:Sctinvstructure}, in terms of the functionals $L_\varphi$, and
with the changes: % see also \cref{eq:SingularCT1Loop_Lphi}
\begin{align}
	{\delta Z_{\psi_R}} \overline{L_{\psi_R}} \to {\delta Z_{\psi_L}} \overline{L_{\psi_L}} \, , &&
	{\delta (Y_R)^m_{ij}} {L_{Y_R}}^m_{ij} \to {\delta (Y_L)^m_{ij}} {L_{Y_L}}^m_{ij} \, ,
\end{align}
and the corresponding $\delta Z_\varphi$, $\delta g_i$ renormalization constants
are again the same as their counterparts \crefrange{eq:RenConstFirst}{eq:RenConstLast},
but with the coefficients changed according to \cref{eq:RtoL_Changes}.
The purely evanescent counterterms $S_{\text{sct,evan}}^{(1)}$
% appearing in \cref{eq:Ssctinvevan}
\cref{eq:Ssctevan,eq:SsctevanTildeOps}
are also expressed in the same way, with the substitution $S_2(R) \to S_2(L)$.

Therefore, following the explanations given in \cref{sect:RGE,sect:MultRenorm},
the resulting renormalization group equations for the Left-handed model are the
very same ones as those for the Right-handed model, with the obvious changes $R \leftrightarrow L$.

The BRST-restoring finite counterterms \cref{eq:FiniteCT1Loop} now read:
\begin{equation}\begin{split}
	S_{\text{fct,restore}}^{(1)} =\;&
	\frac{\hbar}{16 \pi^2} \left\{
		g^2 \frac{S_2(L)}{6} \left( 5 S_{GG} + S_{GGG} - \int \dInt[4]{x} G^{a\,\mu} \partial^2 G^a_\mu \right) + \frac{Y_2(S)}{3} \overline{S_{\Phi\Phi}} \right.\\
		&\left. + g^2 \frac{(T_L)^{abcd}}{3} \int \dInt[4]{x} \frac{g^2}{4} G^a_\mu G^{b\,\mu} G^c_\nu G^{d\,\nu}
		        - \frac{(\mathcal{C}_L)^{ab}_{mn}}{3} \int \dInt[4]{x} \frac{g^2}{2} G^a_\mu G^{b\,\mu} \Phi^m \Phi^n \right.\\
		&\left. + g^2 \left(1 + \frac{\xi - 1}{6}\right) C_2(L) S_{\overline{\psi}\psi}
		        - \frac{\left( (Y_L^m)^* {T_{\overline{L}}}^a Y_L^m \right)_{ij}}{2} \int \dInt[4]{x} g \overline{\psi}_i \slashed{G}^a \Proj{L} \psi_j \right.\\
		&\left. - g^2 \frac{\xi C_2(G)}{4} (S_{\bar{L} c \psi_L} + S_{L c \overline{\psi_L}})
		\right\}
	\, ,
\end{split}\end{equation}
where we have used the following group factors:
\begin{subequations}
\begin{align}
\begin{split}
	(T_{\widehat{R}})^{a_1 \cdots a_n} &= \Tr[{T_{\overline{L}}}^{a_1} \cdots {T_{\overline{L}}}^{a_n}]
		 = \Tr\left[ (-{T_L}^{a_1\,T}) \cdots (-{T_L}^{a_n\,T}) \right] \\
		&= (-1)^n \Tr[{T_L}^{a_n} \cdots {T_L}^{a_1}]
		 = (-1)^n (T_L)^{a_n \cdots a_1}
	\, ,
\end{split} \\
	(\mathcal{C}_L)^{ab}_{mn} &\equiv \Tr\left[ 2 \{ {T_L}^a , {T_L}^b \} (Y_L^n)^* Y_L^m - {T_L}^a (Y_L^n)^* {T_{\overline{L}}}^b Y_L^m \right]
	\, ,
\end{align}
\end{subequations}
% (note that the following relation holds: $\Tr\left[ {T_L}^a (Y_L^n)^* {T_{\overline{L}}}^b Y_L^m \right] = \Tr\left[ {T_{\overline{L}}}^a Y_L^m {T_L}^b (Y_L^n)^* \right]$).
Again, this expression is formally completely unchanged with respect to \cref{eq:FiniteCT1Loop}, with the only change $R \leftrightarrow L$.
However, the relevant (non-spurious) anomalies \cref{eq:Anomalies} now become:
\begin{equation}
    \boldsymbol{\pmb{+}} \frac{\hbar g^2}{16 \pi^2} \left( \frac{S_2(L)}{3} d_L^{abc} \int \dInt[4]{x} g \epsilon^{\mu\nu\rho\sigma} c_a (\partial_\rho G^b_\mu) (\partial_\sigma G^c_\nu)
    + \frac{\mathcal{D}_L^{abcd}}{3 \times 3!} \int \dInt[4]{x} g^2 c_a \epsilon^{\mu\nu\rho\sigma} \partial_\sigma \left( G^b_\mu G^c_\nu G^d_\rho \right) \right)
    \, ,
\end{equation}
where the group factors are the fully symmetric symbol $d_L^{abc} = \Tr[ {T_L}^a \{{T_L}^b , {T_L}^c\} ]$ and $\mathcal{D}_L^{abcd} = \frac{1}{2}(d_L^{abe} f^{ecd} + d_L^{ace} f^{edb} + d_L^{ade} f^{ebc})$ for the L-representation.
The change of sign in front of the equation, with respect to the one in \cref{eq:Anomalies}, comes from the fact that these group factors for the L-representation are related to the corresponding ones in the corresponding right-handed model by: $d_L^{abc} = -d_{\widehat{R}}^{abc}$ and $\mathcal{D}_L^{abcd} = -\mathcal{D}_{\widehat{R}}^{abcd}$.
This has phenomenological consequences for model-building: relevant anomalies can be cancelled in a given model if ones includes both right-handed and left-handed fermions whose representations are the complex-conjugate of the other.

\section{Conclusions}
\label{sect:Concl}

The present paper starts a systematic study of the BMHV scheme for
$\gamma_5$ and its application to chiral gauge theories such as the
electroweak Standard Model. Our motivation is the increasing need for
high-precision predictions including electroweak corrections at the
(multi\nobreakdash-)loop level. Many alternative $\gamma_5$ schemes have been
proposed and used in the literature. The BMHV scheme is singled out by
its mathematical rigor. It is the only scheme for which mathematical
and quantum field theoretical consistency as well as useful theorems
like the ones by Breitenlohner/Maison and Bonneau are fully
established at all orders. Its understanding is thus not only
important for practical BMHV calculations but also as a point of
reference and benchmark for the study of alternative $\gamma_5$
schemes.

In the present paper, we have investigated a chiral gauge theory at
the one-loop level. The theory includes massless chiral fermions and scalars,
for simplicity restricting to irreducible representations and a simple gauge
group. We have focused on the special, BMHV-specific aspects of
renormalization and counterterms. Our results and conclusions can be
summarized as follows.
\begin{itemize}[leftmargin=*]
\item
  In \cref{sect:RModel} we explained in detail the setup of the
  BMHV scheme on the level of the $d$-dimensional tree-level action and the
  resulting breaking of BRST invariance (similarly to Ref.\ \cite{Martin:1999cc}
  for a theory without scalars). The breaking of BRST invariance is localized
  in one single term, the evanescent part of the fermion kinetic term;
  the breaking has been expressed in a set of Feynman rules in
  \cref{subsect:BRST_breaking_RModel}.

\item
  \cref{sect:standardrenormalizationstructure} provided a
  detailed overview of the different renormalization and counterterm
  structure in the BMHV scheme compared to the usual case where
  counterterms can be generated by a renormalization
  transformation. Even in the BMHV scheme, a large part of the
  counterterms can be generated by the usual renormalization
  transformation, but there are several additional, BMHV-specific new
  counterterm structures.

\item
  \cref{sect:Rmodel1LoopSCT} presented the results for the
  singular, i.e.\ UV-divergent 1-loop counterterms. Most of the
  counterterms follow the usual pattern and can be written in terms of
  field and parameter renormalization constants, see
  \cref{eq:Ssctinvevan} and the following equations. However, there
  are extra, evanescent singular counterterms. In line with the
  general definitions of the BMHV scheme
  \cite{Breitenlohner:1975qe,Breitenlohner:1977hr,Breitenlohner:1975hg,Breitenlohner:1976te}
  as well as comparable known results in the context of dimensional reduction
  \cite{Gnendiger:2017pys,Jack:1993ws,Jack:1994bn,Harlander:2006rj, Harlander:2006xq,
    Kilgore:2011ta, Kilgore:2012tb,Broggio:2015ata,Broggio:2015dga}
  such counterterms are necessary at higher order to ensure unitarity
  and finiteness. Most of the evanescent counterterms are still BRST
  invariant (despite being evanescent), but there are two non-BRST
  invariant evanescent counterterms, related to the scalar and vector
  self-energies, respectively.

\item
  \cref{sect:BRSTrestoration} corresponds to the central complication
  of the BMHV scheme --- the breaking of gauge and BRST
  invariance. The breaking already present in the tree-level action
  implies a violation of Slavnov-Taylor identities at the 1-loop
  level, and special, symmetry-restoring counterterms have to be
  found.  We have explained in detail the role and the structure of
  these counterterms and described various possible ways of how these
  counterterms may be determined. Our calculation is based on the
  regularized quantum action principle and the
  so-called Bonneau identities (this combination of tools has also
  been used in Ref.\ \cite{Martin:1999cc});
  in this way, the computation is
  simplified to the evaluation of only UV-divergent parts of specific
  Feynman diagrams. While not strictly necessary at the 1-loop level,
  we expect that this method will lead to significant simplifications
  at the 2-loop level, which is why we use and explain it here in
  detail.

  The final result for the symmetry-restoring counterterms is given in
  \cref{eq:FiniteCT1Loop}. There is some freedom in this choice, since
  invariant or evanescent counterterms may be changed. Our choice is
  particularly simple, and is constructed to the largest possible
  extent from objects which appear already in the tree-level
  action. The terms in the symmetry-restoring counterterm action
  correspond to finite contributions to self-energies of scalars,
  fermions and gauge bosons and finite contributions to a subset of
  the interaction terms between scalars, fermions and gauge bosons.

\item
  The symmetry-restoring counterterms may be changed by
  adding/changing evanescent terms, corresponding to defining a
  counterterm action $S_{\text{fct,evan}}$, see
  \cref{eq:CT_structure}. However, all renormalized 1-loop quantities
  are blind to this choice, hence we do not discuss this option in the
  present paper. It will be relevant for a 2-loop application of the
  BMHV scheme, and a 2-loop calculation might be simplified by an
  optimized choice of $S_{\text{fct,evan}}$ at the 1-loop level.

\item
  \cref{sect:RGE} and \cref{sect:MultRenorm} are devoted to the
  derivation of the RGE in the context of the BMHV scheme. We
  demonstrate in two different ways that despite the extra,
  BMHV-specific counterterms the 1-loop RGE is unchanged compared to
  the familiar case of using a symmetry-invariant
  regularization. However, both the more abstract derivation using
  Bonneau identities and the textbook method based on divergent
  renormalization constants show that this statement relies on
  specific simplifications which occur at the 1-loop level. Therefore,
  it will be interesting and nontrivial to investigate the same
  situation at the 2-loop level.
\end{itemize}
As an outlook, several future extensions are of interest. First, the
results can be slightly extended and specialized to the case of the
electroweak SM, which has a non-semisimple gauge group, reducible
representations and both right-handed and left-handed (see
\cref{sect:LModel}) chiral fermions. This is work in progress. Second,
the results can be extended to higher loop orders; specifically, a
knowledge of the required symmetry-restoring 2-loop counterterms will
open up the possibility of 2-loop calculations in the BMHV scheme, and
the determination of the 2-loop RGE will provide important information
on the interpretation and the relationship between BMHV and other
calculations.

%% \section*{Acknowledgements}
\acknowledgments

We thank Jiangyang You for detailed discussions on algebraic renormalization techniques in general and specifically for discussions about the work \cite{Martin:1999cc} by C.P.~Martin and D.~Sanchez-Ruiz.
H.Bélusca-Maïto thanks FeynCalc's co-author Vladyslav Shtabovenko for various help and support provided with the FeynCalc software package.

The authors highly acknowledge the financial support from the Croatian Science Foundation (HRZZ) under the project ``PRECIOUS'' (``Precise Computations of Physical Observables in Supersymmetric Models'') number \verb|HRZZ-IP-2016-06-7460|, as well as the hospitality of the Institut für Kern- und Teilchenphysik, TU Dresden, where part of this work has been pursued. In addition A.Ilakovac highly acknowledges the financial support of the previous Croatian Science Foundation project \verb|HRZZ-IP-2013-11-8799|.

% \clearpage
\newpage

% % \bibliographystyle{utphys}
% \bibliographystyle{JHEPmod}
% \bibliography{references}

\providecommand{\href}[2]{#2}\begingroup\raggedright\endgroup

\end{document}